\documentclass{article}

\usepackage{PRIMEarxiv}

\usepackage[utf8]{inputenc} %
\usepackage[T1]{fontenc}    %
\usepackage{hyperref}       %
\usepackage{url}            %
\usepackage{booktabs}       %
\usepackage{amsfonts}       %
\usepackage{nicefrac}       %
\usepackage{microtype}      %
\usepackage{lipsum}
\usepackage{fancyhdr}       %
\usepackage{graphicx}       %
\graphicspath{{figures/}}     %

\usepackage{placeins}
\usepackage{soul,color}
\usepackage[table,xcdraw]{xcolor}
\usepackage[normalem]{ulem}
\useunder{\uline}{\ul}{}
\usepackage{longtable}

\title{CSPBench: a benchmark and critical evaluation of Crystal Structure Prediction 
}

\author{
  Lai Wei, Sadman Sadeed Omee\\
 Department of Computer Science and Engineering\\
  University of South Carolina\\
  Columbia, SC 29201 \\
    \And
 Rongzhi Dong, Nihang Fu\\
 Department of Computer Science and Engineering\\
  University of South Carolina\\
  Columbia, SC 29201 \\  
   \And
    Yuqi Song\\
   Department of Computer Science\\
University of Southern Maine\\
Portland, Maine, 04101\\
  \And
 Edirisuriya M. D. Siriwardane\\
 Department of Physics\\
  University of Colombo\\
  Sri Lanka \\
  \And
  Meiling Xu \\
 School of Physics and Electronic Engineering\\
  Jiangsu Normal University\\
  Xuzhou, China \\
  \And
  Chris Wolverton\\
 Department of Materials Science and Engineering\\
  Northwestern University\\
  Chicago, USA \\  
  \And 
    Jianjun Hu* \\
 Department of Computer Science and Engineering\\
  University of South Carolina\\
  Columbia, SC 29201 \\
  \texttt{jianjunh@cse.sc.edu} \\
}

\begin{document}
\maketitle

\begin{abstract}
Crystal structure prediction (CSP) is now increasingly used in discovering novel materials with applications in diverse industries. However, despite decades of developments and significant progress in this area, there lacks a set of well-defined benchmark dataset, quantitative performance metrics, and studies that evaluate the status of the field. We aim to fill this gap by introducing a CSP benchmark suite with 180 test structures along with our recently implemented CSP performance metric set. We benchmark a collection of 13 state-of-the-art (SOTA) CSP algorithms including template-based CSP algorithms, conventional CSP algorithms based on DFT calculations and global search such as CALYPSO, CSP algorithms based on machine learning (ML) potentials and global search, and distance matrix based CSP algorithms. Our results demonstrate that the performance of the current CSP algorithms is far from being satisfactory. Most algorithms cannot even identify the structures with the correct space groups except for the template-based algorithms when applied to test structures with similar templates. We also find that the ML potential based CSP algorithms are now able to achieve competitive performances compared to the DFT-based algorithms. These CSP algorithms' performance is strongly determined by the quality of the neural potentials as well as the global optimization algorithms. Our benchmark suite comes with a comprehensive open-source codebase and 180 well-selected benchmark crystal structures, making it convenient to evaluate the advantages and disadvantages of CSP algorithms from future studies. All the code and benchmark data are available at \url{https://github.com/usccolumbia/cspbenchmark}

\end{abstract}


\section{Introduction}
\vspace{-0.4cm}
The critical assessment of protein structure prediction (CASP) and advancements like Alphafold have significantly propelled research in predicting protein structures \cite{kryshtafovych2021critical, jumper2021highly}. Similarly, crystal structure prediction (CSP) methods have gained attention for organic molecules \cite{bowskill2021crystal}. However, the focus on crystal structure prediction within the domain of inorganic materials is steadily growing and proving to be vital for discovering new materials across diverse industries. Understanding the crystal structure of a material holds immense importance as it significantly influences its physical, chemical, and mechanical properties. Traditionally, experimental techniques such as Density Functional Theory (DFT) calculations, coupled with global search algorithms or tailored experiments, have been utilized to determine crystal structures. While successful for many materials, these methods are often time-consuming, costly, and particularly challenging when dealing with novel or intricate compounds. Therefore, the applications and quantitative metrics within CSP are becoming increasingly indispensable for advancing research in inorganic materials and guiding their practical utilization.

Nowadays, a plethora of approaches for crystal structure prediction exists, including evolutionary algorithms \cite{glass2006uspex, trimarchi2009predicting, liu2021copex}, data mining \cite{hofmann2003crystal, fischer2006predicting} and machine learning \cite{GN-OA, oliynyk2017disentangling}. Current methods of evaluating structures are mainly based on manual structural inspection, comparison with experimentally observed structures, comparison of energy or enthalpy values, success rate analysis, and computation of distances between structures. Nevertheless, the absence of a quantitative approach for evaluating predicted structures remains a challenge, hindering the ability to confidently ascertain their reliability and guide experimental validation. As the field progresses, developing robust quantitative evaluation methods is essential to unlock new frontiers in materials research and development. In this paper, we conducted a large scale benchmark study on the main CSP algorithms selected from Table \ref{tab:softwares}, including CSP algorithms with ML potentials including ab initio CSP \cite{CrySPY, XtalOpt}, GN-OA (Graph Networks for crystal structure Optimization with Atomistic potentials) with ML-potentials \cite{GN-OA}, AGOX with M3GNet potential \cite{christiansen2022atomistic}, Random structure search in Atomistic Global Optimization X (AGOX), Basin-hopping \cite{wales1997global}, Parallel tempering, Local GPR basin-hopping \cite{kofke2002acceptance} , Evolutionary algorithm, the Bayesian Optimization GOFEE \cite{bisbo2022global}, template-based CSP \cite{wei2022tcsp,griesemer2021high}, DL-based CSP algorithms. Ab initio methods involve the calculation of the electronic structure and total energy of a crystal system using quantum mechanical principles. CrySPY and XtalOpt are two widely used algorithms for crystal structure prediction based on the ab initio approach that employ optimization algorithms to search for stable crystal structures with low energies. AGOX employs a machine-learned potential called M3GNet, which is designed to accurately describe the energy landscape of materials, explore the vast configuration space of crystal structures, and identify stable candidates. The Basin-hopping algorithm involves a stochastic exploration of the potential energy landscape, efficiently searching for the global minimum by jumping between these basins (local energy minima). DL-based crystal structure prediction algorithms utilize neural networks and deep learning techniques to learn representations of atomic structures and predict their energies or stability.  GN-OA is an algorithm that utilizes machine learning potentials to predict crystal structures. It employs graph networks, which can represent atomic structures efficiently, and atomistic potentials learned from data to optimize the crystal structures. We also compared the performances of the non-DFT-based CSP algorithms with the leading DFT-based CSP algorithms including CALYPSO \cite{CALYPSO} and USPEX \cite{glass2006uspex}. 
We then analyze and evaluate these CSP algorithms and utilize them to generate target structures. By calculating metrics through our quantitative evaluation method, we can examine the performance of each algorithm. 
In our paper, we conduct a comprehensive analysis and evaluation of leading crystal structure prediction (CSP) algorithms. To assess their performance, we calculate a set of quantitative metrics for CSP benchmarking, which serve as objective measures to gauge the algorithms' accuracy, efficiency, and reliability in predicting crystal structures. Through this evaluation process, we aim to provide a clear picture of the capabilities of each algorithm and how they fare in comparison to one another. The results obtained from the benchmarking analysis shed light on their performance in terms of predicting known crystal structures and identifying novel structures not present in the training dataset. Our quantitative evaluation metrics ensure that the analysis is conducted in a fair and unbiased manner, allowing for meaningful comparisons. Overall, this thorough analysis and evaluation are fundamental to advancing the state-of-the-art in CSP and accelerating the discovery of new materials with tailored properties for diverse applications.

Another category of modern CSP algorithms combines global search with machine learning potential functions for structure search. In early attempts, these algorithms mainly used specialized ML potential functions that only covered one or a few element types: in \cite{dolgirev2016machine}, ML potentials for four systems: Al, C, He and Xe were trained for CSP using the USPEX algorithm. A follow-up study \cite{podryabinkin2019accelerating} used active learning to develop a ML potential and apply it to the CSP of carbon allotropes, sodium structures under pressure, and boron allotropes. CALYPSO algorithm has also been combined with machine learning potentials for structure prediction of Boron (B) clusters \cite{tong2018accelerating}, 24-atom cubic boron phases \cite{yang2021hard} and gallium nitride (GaN) phase simulation with 4096-atom \cite{tong2021machine}. However, all these ML potentials are not universal enough to cover elements of the whole or a majority of the periodic table.

Recent ML potentials have been developed to cover a large portion of the periodic table. Takamoto et al. \cite{takamoto2022teanet} developed TeaNet, a 16-layer graph convolution network with a residual network (ResNet) architecture and recurrent GCN weights initialization, for the simulation of metals, and amorphous SiO$_2$ structures. Their universal model can cover 18 elements initially and was later extended to 45 elements \cite{takamoto2022towards}. Their neural network potential has been shown to speed up the simulation of lithium diffusion in LiFeSO$_4$F, molecular adsorption in metal-organic frameworks, an order–disorder transition of Cu-Au alloys, and material discovery for a Fischer-Tropsch catalyst. Choudhary et al. developed a graph neural network ML potential and combined it with a genetic algorithm for crystal structure prediction of alloys \cite{choudhary2023unified}.

\section{Method}
\label{sec:headings}

\subsection{Summary of main category of CSP algorithms}

\begin{table}[]
\centering
\caption{A summary of the main CSP softwares. MLP: machine learning potentials; MOGA: multi-objective genetic algorithm; }
\begin{tabular}{|l|r|l|l|l|l|}
\hline
\textbf{Algorithm} & \multicolumn{1}{l|}{\textbf{Year}}                  & \textbf{Category}                                                                & \textbf{Open-source} & \textbf{URL link} & \textbf{Program Lang}                                                                                \\ \hline
USPEX \cite{glass2006uspex}              & 2006                                                & De novo (DFT)                                                                    & No                   & {\color[HTML]{0000EE} {\href{https://uspex-team.org/en/uspex/overview}{link}}}  & Matlab                                       \\ \hline
CALYPSO \cite{CALYPSO}            & 2010                                                & De novo (DFT)                                                                    & No                   & {\color[HTML]{0000EE} {\href{http://www.calypso.cn/}{link}}}   & Python                                                        \\ \hline
ParetoCSP \cite{omee2024crystal}              & 2024                                                & MOGA+MLP                                                                   & Yes                   & {\color[HTML]{0000EE} {\href{https://github.com/sadmanomee/ParetoCSP}{link}}}  & Python                                       \\ \hline
GNOA \cite{GN-OA}             & 2022                                                & BO/PSO  + MLP                                                                      & Yes                  & {\color[HTML]{0000EE} {\href{ http://www.comates.group/links?software=gn\_oa}{link}}}  & Python                                 \\ \hline
TCSP \cite{wei2022tcsp}              & 2022                                                & Template                                                                         & Yes                   & {\color[HTML]{0000EE} {\href{ http://materialsatlas.org/crystalstructure}{link}}}   & Python                                    \\ \hline
CSPML \cite{CSPML}             & 2022                                                & Template                                                                         & Yes                  & {\color[HTML]{0000EE} {\href{ https://github.com/Minoru938/CSPML}{link}}}   & Python                                            \\ \hline
GATor \cite{curtis2018gator}            & 2018 & GA + FHI potential                                                                 & Yes                  & {\color[HTML]{0000EE} {\href{ https://www.noamarom.com/software/gator/}{link}}}            & Python                             \\ \hline
AIRSS \cite{pickard2006high, pickard2011ab}            & {\color[HTML]{333333} 2011}                         & Random + DFT or pair Potential                                                     & Yes                  & {\color[HTML]{0000EE} {\href{ https://airss-docs.github.io/}{link}}} & Fortran                                                   \\ \hline
GOFEE \cite{bisbo2020global}            & 2020                                                & \cellcolor[HTML]{FFFFFF}{\color[HTML]{333333} ActiveLearning  +  Gaussian Pot.} & Yes                  & {\color[HTML]{0000EE} {\href{ http://grendel-www.cscaa.dk/mkb/}{link}}} & Python                                               \\ \hline
AGOX \cite{christiansen2022atomistic}             & 2022                                                & Search  +  Gaussian Potential                                                        & Yes                  & {\color[HTML]{0000EE} {\href{ https://gitlab.com/agox/agox}{link}}}  & Python                                                  \\ \hline
GASP \cite{tipton2012gasp}              & 2007                                                & GA + DFT                                                                           & Yes                  & {\color[HTML]{0000EE} {\href{ https://github.com/henniggroup/gasp}{link}}} & Java                                            \\ \hline
M3GNet \cite{chen2022universal}            & 2022                                                & Relax with MLP                                                                 & Yes                  & {\color[HTML]{0000EE} {\href{ https://github.com/materialsvirtuallab/m3gnet}{link}}} & Python                                   \\ \hline
ASLA \cite{mortensen2020atomistic}              & 2020                                                & NN + RL                                                                            & No                   & {\color[HTML]{0000EE} {\href{ https://journals.aps.org/prb/abstract/10.1103/PhysRevB.102.075427} {link}}}      & N/A          \\ \hline
CrySPY \cite{yamashita2021cryspy}            & 2023                                                & GA/BO + DFT                                                                        & Yes                  & {\color[HTML]{0000EE} {\href{ https://tomoki-yamashita.github.io/CrySPY\_doc/tutorial/random/\#running-cryspy}{link}}} & Python\\ \hline
XtalOpt \cite{lonie2011xtalopt}           &  2011 & GA + DFT                                                                           & Yes                  & {\color[HTML]{0000EE} {\href{ http://xtalopt.github.io/download.html}{link}}} & C++                                         \\ \hline
AlphaCrystal \cite{hu2021alphacrystal,song2024alphacrystal}      & 2023                                                & GA + DL                                                                            & Yes                  & {\color[HTML]{0000EE} {\href{ https://github.com/usccolumbia/AlphaCrystal}{link}}}                             & Python        \\ \hline
\end{tabular}
\label{tab:softwares}
\end{table}

\subsubsection{ab initio CSP}
There are several open-source CSP codes based on combining search algorithms with DFT energy calculation, including CrySPY \cite{yamashita2021cryspy}, XtalOpt \cite{lonie2011xtalopt}, GASP \cite{tipton2012gasp} , AIRSS \cite{pickard2006high, pickard2011ab}. However, the most widely used and well-established leading software for de novo CSP 
are GA based USPEX and particle swarm optimization (PSO) based CALYPSO \cite{CALYPSO}. Due to the computational costs associated with DFT-based CALYPSO, we selected a subset of 23 structures from the test dataset for prediction. 
Despite their closed-source code, their binary programs can be easily obtained and both come with several advanced search techniques such as symmetry handling, crowding niche, and so on. In this algorithm, structures within the first population are randomly generated while adhering to proper physical constraints, such as the interatomic distances and crystal symmetry. Similar crystal structures are then removed using structure characterization techniques, including bond characterization metrics and coordination characterization function to streamline the search. Once all structures of each population are established, local optimizations are performed using DFT-based methods to locate the local minima. Structural evolution is further carried out using swarm intelligence algorithms such as particle swarm optimization or artificial bee colony. New structures are generated based on the information gathered from the previous generation. By combining random structure generation, local optimizations, and swarm intelligence algorithms, the CALYPSO method efficiently explores the PES, increasing the chances of locating the global energy minimum. In summary, the general idea is to iteratively generate and optimize structures to navigate the complex energy landscape.

Due to the demanding computational costs for DFT calculations, we allocated 3,000 DFT energy calculations in all their experimental runs for different benchmark test samples.
Here, the structural relaxations are performed using the Vienna Ab initio simulations package (VASP) \cite{kresse1996efficient} by considering the Perdew-Burke-Ernzerhof generalized gradient approximation \cite{perdew1996generalized} for the exchange-correlation function and the projector-augmented-wave potentials \cite{blochl1994projector} for the electron-ion interactions. 
VASP allows for geometry optimization using different optimization algorithms, such as the conjugate gradient method (CG) and quasi-Newton RMM-DIIS algorithm. 
The VASP running parameters mainly involve the plane-wave cutoff energy (the maximum kinetic energy for the electronic wavefunctions), Monkhorst–Pack k meshes(sampling in the Brillouin zone), energy and force convergence precisions. By gradually optimizing the structure by adjusting these corresponding parameters, the optimization process can be accelerated while still obtaining reliable and accurate results. This approach can help save overall time in structure prediction by efficiently exploring the configuration space and converging toward the optimal structure. 
The relevant parameter settings for DFT calculations of CALYPSO  during the optimization process for its predictions are listed in detail in Table S7 of the supplementary file.

\subsubsection{GNOA with ML-potentials}

Our approach involves GN-OA algorithms \cite{GN-OA}, a machine-learning method for crystal structure prediction. In this framework, a graph network (GN) is employed to establish a correlation model between crystal structures, and an optimization algorithm (OA) is utilized to accelerate the search for the crystal structure with the lowest formation enthalpy. In this work, we evaluate the CSP algorithms based on ML potentials. Two graph neural network potentials have been tested here including the MEGNet \cite{chen2019graph} and M3GNet \cite{chen2022universal},  which has been combined with random search (RAS), Bayesian optimization (BO), and Particle Swarm Optimization (PSO) for crystal structure prediction as implemented in their GNOA package \cite{yin2022search}.

\subsubsection{AGOX with M3GNet potential}
We adopt Atomistic Global Optimization X (AGOX)~\cite{christiansen2022atomistic}, a customizable and efficient global structural optimization framework that has six search global optimization algorithms implemented. AGOX uses the effective medium theory (EMT) potential~\cite{jacobsen1987interatomic} to optimize and relax the generated candidate structures by default. For better comparison with other algorithms, we replace the simple EMT potential with the more powerful M3GNet~\cite{chen2022universal} inter-atomic potential. M3GNet is based on graph neural networks and explicitly incorporates many-body interactions and is much faster than DFT-based energy calculations~\cite{hohenberg1964inhomogeneous,sham1966one}. After completing the optimized structure search using each algorithm, the final optimized structure is further relaxed using M3GNet. An overview of the AGOX framework is shown in Figure~\ref{fig:agox_framework}. In our work, we use three different search algorithms: Basin-hopping (BH), parallel tempering (PT), and random search (RSS). The global search algorithms are described below:

\begin{figure}[!htb]
  \centering
  \includegraphics[width=\linewidth]{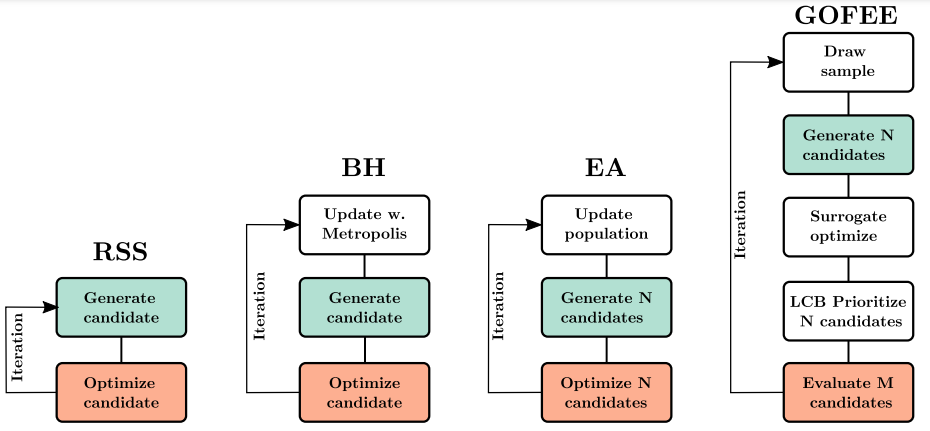}
  \caption{An overview of the global optimization algorithms included in the AGOX framework.}
  \label{fig:agox_framework}
\end{figure}

\paragraph{Random structure search:}
Random structure search (RSS) is the simplest algorithm in AGOX. In each iteration, it generates a candidate at random and optimizes it locally. The relaxed candidate is then stored in the database.

\paragraph{Basin-hopping:}
Basin-hopping (BH)~\cite{wales1997global} is a method of exploring the configuration space by performing a series of jumps from one potential energy surface (PES) minimum to another, turning the potential energy surface into a network of interpenetrating stairs. A sampler is employed to maintain track of a candidate that has already been reviewed and provides information for the creation of a new candidate. The process begins by rattling a prior candidate to produce a new candidate. Next, the generated candidate is relaxed locally. The Metropolis criterion is next examined to decide whether the new candidate is approved as the generation's beginning point. The probability of acceptance of the Metropolis criterion is defined by the following equation:
\begin{equation}
    A = min\{1, exp[\beta(E_{k - 1} - E_k)]\}
\end{equation}
where $\beta = 1/k_BT$ with $k_B$ the Boltzmann's constant, and $E_k$ is the energy of the structure found in iteration $k$.

\paragraph{Parallel tempering}
Simultaneous basin-hopping searches are conducted across different temperatures in the parallel tempering (PT) method, as described by Kofke et al.~\cite{kofke2002acceptance}. This approach promotes exploration at elevated temperatures and exploitation at lower ones, ensuring that structures adaptable to varying temperatures are swapped to prevent stagnation. In this setup, multiple workers with different processors each perform a basin-hopping search at a specific temperature, utilizing a single database.
The following equation calculates the probability of the structure swap between workers with adjacent temperatures every $N_t$ episode:
\begin{equation}
    A = min\{1, exp[(\beta_i - \beta_j)(E_i - E_j)]\}
\end{equation}
where $\beta = 1/k_BT_i$ with $k_B$ the Boltzmann's constant.

\paragraph{Local GPR basin-hopping}
The basin-hopping search is enhanced by the use of a local Gaussian process regression (GPR) model~\cite{bisbo2020efficient} implemented inside the AGOX framework. The Local GPR model uses a radial basis function (RBF) kernel~\cite{vert2004primer} and the smooth overlap of atomic positions (SOAP)~\cite{bartok2013representing} descriptor to perform the basin-hopping search.

\paragraph{Evolutionary algorithm}
Biological evolution theories serve as the foundation for evolutionary algorithms (EAs). The first step is to generate a population of potential solutions, and then each one is evaluated using a fitness function to determine how effective it is. With time, the population evolves and finds better solutions. With each iteration of EA, a population of candidates is maintained and used as input to generate a new candidate. After that, the newly generated candidate is relaxed. The sampler would keep a population of structurally different candidates in an EA so that they may serve as the parents of future candidates. The algorithm presented in \cite{vilhelmsen2014genetic} is used to select the population.

\paragraph{GOFEE: Bayesian Optimization}

GOFEE is a Bayesian search algorithm that Bisbo and Hammer~\cite{bisbo2022global} developed as an effective technique for locating low-energy structures in computationally expensive energy landscapes, termed global optimization with first-principles energy expressions. A combination of an evolutionary search strategy and an actively learned surrogate model of the energy space is deployed in GOFEE. This facilitates answering a lot more structural queries than the target potential would allow. However, a significantly smaller number of evaluations are performed utilizing the target potential on the structures that the surrogate model thought were most promising. These evaluations serve as training data to further refine the surrogate model. In GOFEE, a set of candidate structures is first locally optimized using a computationally inexpensive surrogate potential. Subsequently, a lower confidence bound acquisition function selects candidates for evaluation with the true potential. Each episode of GOFEE generates $N$ candidates, which are locally optimized using a GPR potential, or more precisely in the so-called lower-confidence-bound expression, defined by the following equation:
\begin{equation}
    E(x) = \hat{E}(x) - k\sigma (x)
\end{equation}
where $\hat{E}$ and $\sigma$ are the predicted energy and uncertainty of the GPR model for the structure represented by $x$.

\subsubsection{ParetoCSP}
ParetoCSP~\cite{omee2024crystal} based on the idea of the GN-OA algorithm~\cite{GN-OA} with two major upgrades including the multi-objective GA search algorithm and the use of M3GNet potential for energy calculation. GN-OA has been proven from previous research that incorporating symmetry constraint expedites CSP. Similar to the GN-OA approach, our method also considers crystal structure prediction with symmetry constraints. We incorporate two additional structural features, namely crystal symmetry $S$ and the occupancy of Wyckoff position $W_i$ for each atom $i$. These features are selected from a collection of $229$ space groups and associated $1506$ Wyckoff positions. The method begins by selecting a symmetry $S$ from the range of $P2$ to $P230$, followed by generating lattice parameters $L$ within the chosen symmetry. Next, a combination of Wyckoff positions $\{W_i\}$ is selected to fulfill the specified number of atoms in the cell. The atomic coordinates $\{R_i\}$ are then determined based on the chosen Wyckoff positions $\{W_i\}$ and lattice parameters $L$. To generate crystal structures, we need to tune the $S$, $\{W_i\}$, $L$, and $\{R_i\}$ variables. 

By selecting different combinations of $S$, ${W_i}$, $L$, and ${R_i}$, one can generate a comprehensive array of possible crystal structures for the given ${c_i}$. In theory, determining the energy of these various structures and selecting the one with the least energy should be the optimal crystal arrangement. However, exhaustively enumerating all these structures becomes practically infeasible due to the staggering number of potential combinations. To address this complexity, a more practical approach involves iteratively sampling candidate structures from the design space, under the assumption that one of the sampled structures will emerge as the most stable and optimal solution. Consequently, an optimization strategy is adopted to guide this search process towards identifying the structure with the lowest energy. In particular, a genetic algorithm, NSGA-III~\cite{seada2015u}, improved by incorporating AFPO~\cite{schmidt2010age} to enhance its performance and robustness, is utilized.

It starts by generating n random crystals and assigning them an age of 1, where n denotes the population size. One complete
generation then goes through the following steps: calculating the energy of the structures and fitness, selecting parents, performing
genetic operations, and updating the age. After a certain threshold of G generations, the lowest energy structure from the multidimensional Pareto front is chosen and further relaxed and symmetrized to obtain the final optimal structure. The genetic encoding is
shown in the lower right corner of the flowchart. It contains lattice parameters $a$, $b$, $c$, $\alpha$, $\beta$, and $\gamma$, the space group $S$, the wyckoff
position combination $W_i$, and the atomic coordinates $R_i$ of atom indexed by $i$.

\subsubsection{template-based CSP}

CSPML \cite{CSPML} is a machine learning-based crystal structure prediction algorithm that uses metric learning \cite{Metric} to automate the selection of template structures from a stable structure database with high chemical replaceability to the probable structure for a given chemical composition. For a given formula, CSPML first restricts the candidates to structures with the same compositional ratio and then uses XenonPy \cite{XenonPy} to calculate the compositional descriptor of the query formula and templates; only templates ranked as the top five can be considered candidate structures. For the 38 query compositions selected from the Materials Project database \cite{jain2013commentary}, 35 out of them have candidates with probabilities greater than 0.5, and 18 out of them have ranked the best template structure that is most similar to the true structure in the top five.

TCSP \cite{wei2022tcsp} is a template-based crystal structure prediction algorithm. For a given formula, TCSP first narrows down the candidates to structures with the same prototype and then uses Element’s mover distance (ElMD) \cite{ElMD} to measure the compositional similarity between the query formula and the compositions of all possible template structures. We implement BERTOS \cite{fu2023composition} in TCSP, which achieves over 96.82\% accuracy for all-element oxidation states prediction on the Inorganic Crystal Structure Database (ICSD), to leverage its significant capabilities to enhance the accuracy of predicting oxidation states in the searching template element process of TCSP. Templates with identical oxidation states are then added to the final template list. If no such templates are found, the top five candidate structures are taken as the final templates. We apply the M3GNet potential to optimize generated structures in this work.

\subsubsection{DL-based CSP}
AlphaCrystal-II \cite{song2024alphacrystal} is a deep learning based crystal structure prediction algorithm based on the prediction of atomic pairwise distances and distance matrix based coordinate reconstruction \cite{yang2021crystal}. This data-driven CSP algorithm exploits the implicit chemical and geometric rules embedded in existing crystal structures as deposited in material databases such as Materials Project or ICSD: for example, most cations are surrounded by anions. A deep neural network is trained to predict the distance matrix given only the composition, which is then used as the objective target for a gradient free optimization (Nevergrad\cite{rapin2018nevergrad})-based crystal structure reconstruction algorithm to search for the atomic coordinates of the structures. The resulting candidate structures are then fed to the M3GNet-based fast structure relaxer to fine-tune the structures.

\subsection{CSPBenchmark test set design}
To construct a balanced and effective benchmark dataset for crystal structure prediction, we meticulously considered several key factors contributing to the complexity of this challenge. These factors include the total number of atoms and distinct elements within the compositions, the degree of symmetry as indicated by space groups, the prototype characterized by specific atomic ratios, and the shape and dimensions of the unit cell. Additionally, we accounted for the prevalence of similar compositions within established crystal structure databases to ensure comprehensive representation. We selected a total of 180 crystal structures, named CSP180, from the Materials Project database \cite{jain2013commentary}. These structures are evenly distributed among binary, ternary, and quaternary compounds, ensuring a diverse and representative sample. The selected structures exhibit a wide variety of space groups, with the most prevalent being space group 225, which appears 27 times. Other common space groups include 139, 216, 221, and 194. Regarding the crystal system distribution, most structures belong to the cubic system, followed by the tetragonal and hexagonal systems. There are fewer occurrences of orthorhombic, trigonal, monoclinic systems, and a single instance of the triclinic system. Our selection process aimed to include structures with varying levels of prediction difficulty. Table~\ref{table:dataset} presents detailed information on the 36 selected test crystals of binary structures, categorized into three difficulty levels: binary\_easy, binary\_medium, and binary\_hard, with each category containing 12 structures. The criteria for difficulty classification include factors such as space group classification, template-based categorization, and the prototype ratios defining the crystal structures. The 180 crystal structures were chosen to cover a broad range of complexities and to provide a comprehensive benchmark for testing. For example, binary compounds like DyCu and GaCo, which belong to space group 221 and exhibit cubic symmetry, were categorized as binary\_easy due to their simpler and more predictable structures. In contrast, more complex structures, such as those with trigonal symmetry or multiple elements with varying oxidation states, were placed in higher difficulty categories. This careful selection ensures that the dataset not only includes easily predictable structures but also those that present significant challenges, thereby testing the robustness and accuracy of prediction algorithms.
Additional test structures and their corresponding data are provided in the supplementary file for further reference.

\renewcommand{\arraystretch}{1.4}
\begin{longtable}{l l l l l l}
\caption{\textbf{Details of the binary\_easy, binary\_medium and binary\_hard data for benchmark crystals used in this work. See Table S1 for the whole set.}}
\label{table:dataset}\\
\hline
\hline
\textbf{Material id} & \multicolumn{1}{l}{\begin{tabular}[c]{@{}c@{}}\textbf{Pretty formula}\end{tabular}} & \textbf{Space group} & \multicolumn{1}{l}{\begin{tabular}[c]{@{}c@{}}\textbf{Crystal system}\end{tabular}} & \multicolumn{1}{l}{\begin{tabular}[c]{@{}c@{}}\textbf{Category}\end{tabular}}\\
\hline
\hline
mp-2334      & DyCu               & 221        & Cubic         & binary\_easy   \\
mp-2226      & DyPd               & 221        & Cubic         & binary\_easy   \\
mp-1121      & GaCo               & 221        & Cubic         & binary\_easy   \\
mp-2735      & PaO                & 225        & Cubic         & binary\_easy   \\
mp-1169      & ScCu               & 221        & Cubic         & binary\_easy   \\
mp-30746     & YIr                & 221        & Cubic         & binary\_easy   \\
mp-24658     & SmH$_2$               & 225        & Cubic         & binary\_easy   \\
mp-20225     & CePb$_3$              & 221        & Cubic         & binary\_easy   \\
mp-788       & Co$_2$Te$_2$             & 194        & Hexagonal     & binary\_easy   \\
mp-20176     & DyPb$_3$              & 221        & Cubic         & binary\_easy   \\
mp-1231      & Cr6Ga$_2$             & 223        & Cubic         & binary\_easy   \\
mp-12570     & ThB$_{12}$             & 225        & Cubic         & binary\_easy   \\
mp-20132     & InHg               & 166        & Trigonal      & binary\_medium \\
mp-2209      & CeGa$_2$              & 191        & Hexagonal     & binary\_medium \\
mp-30497     & TbCd$_2$              & 191        & Hexagonal     & binary\_medium \\
mp-30725     & YHg$_2$               & 191        & Hexagonal     & binary\_medium \\
mp-2731      & TiGa$_3$              & 139        & Tetragonal    & binary\_medium \\
mp-2510      & ZrHg               & 123        & Tetragonal    & binary\_medium \\
mp-2740      & ErCo$_5$               & 191        & Hexagonal     & binary\_medium \\
mp-570875    & Ga$_4$Os$_2$              & 70         & Orthorhombic  & binary\_medium \\
mp-861       & Hf4Ni$_2$             & 140        & Tetragonal    & binary\_medium \\
mp-1566      & SmFe$_5$               & 191        & Hexagonal     & binary\_medium \\
mp-2387      & Th$_4$Zn$_2$             & 140        & Tetragonal    & binary\_medium \\
mp-1607      & YbCu$_5$             & 191        & Hexagonal     & binary\_medium \\
mp-13452     & BePd$_2$              & 139        & Tetragonal    & binary\_hard   \\
mp-11359     & Ga$_2$Cu              & 123        & Tetragonal    & binary\_hard   \\
mp-1995      & PrC$_2$               & 139        & Tetragonal    & binary\_hard   \\
mp-30501     & Ti$_2$Cd              & 139        & Tetragonal    & binary\_hard   \\
mp-30789     & U$_2$Mo               & 139        & Tetragonal    & binary\_hard   \\
mp-454       & NaGa$_4$              & 139        & Tetragonal    & binary\_hard   \\
mp-1827      & SrGa$_4$              & 139        & Tetragonal    & binary\_hard   \\
mp-2129      & Nd$_2$Ge$_4$             & 141        & Tetragonal    & binary\_hard   \\
mp-30682     & ZrGa               & 141        & Tetragonal    & binary\_hard   \\
mp-2128      & Sn$_8$Pd$_2$             & 68         & Orthorhombic  & binary\_hard   \\
mp-1208467   & Tb$_8$Al$_2$             & 227        & Cubic         & binary\_hard   \\
mp-640079    & Mn$_9$Au$_3$             & 123        & Tetragonal    & binary\_hard  \\
\hline
\end{longtable}

\subsection{Evaluation procedure and running parameters for different algorithms}

We substituted DFT calculations with the M3GNet potential, a graph neural network-based surrogate potential model \cite{chen2022universal}, to compute the energy distance for relaxing both the ground truth structure and the predicted structure. Subsequently, we utilized this model to determine the energy distance between these structures. The running parameters and configuration for all CSP algorithms are shown in Table S6 in the supplementary file.

\subsection{Evaluation metrics}
Evaluation metrics are essential in materials science research as they quantitatively assess the performance and effectiveness of different materials. Currently, numerous evaluation metrics exist in molecular research, such as RDKit \cite{landrum2013rdkit} and MOSES \cite{polykovskiy2020molecular}. However, in the field of materials informatics, there is no unified standard for evaluating new structures. Recently, we introduced a set of distance metrics for CSP performance comparisons in benchmark studies \cite{wei2024towards}, including M3GNet energy distance, minimal rmse distance, minimal mae distance, rms distance, rms anonymous distance, Sinkhorn distance, Chamfer distance, Hausdorff distance, superpose rmsd distance, edit graph distance, Fingerprint distance, to standardize the training and comparison of material structure generation models. For test structures in the polymorph category, we employ a detailed evaluation approach. We compare the predicted structures with multiple ground truth structures, each representing different polymorphs. As each sample corresponds to multiple ground truth polymorphs, this results in several evaluation metrics for each sample. To identify the most accurate predictions, we select the evaluation metrics associated with the ground truth structure that has the minimum M3GNet energy distance. This method ensures that the selected metrics reflect the closest match to the predicted structure, providing a reliable measure of prediction accuracy. The distance metrics are shown below. Table \ref{tab:scores} shows selected distance scores for various test samples generated by the AGOX-pt algorithm.

\begin{itemize}
    \item Wyckoff position fraction coordinate RMSE distance
    \item Wyckoff Minimal MAE distance
    \item M3GNet Energy distance
    \item Pymatgen RMS distance
    \item Sinkhorn distance
    \item Chamfer distance
    \item Hausdorff distance
    \item Superpose RMS distance
    \item CrystalNN Fingerprint distance
    \item Edit Graph distance
    \item XRD distance
    \item OFM distance
\end{itemize}

In addition to the above quantitative distance metrics, we also used Pymatgen's StructureMatcher to calculate the success rate of crystal structure prediction by identifying if similar structures exist in the MP database with the following default parameters: ltol=0.2, stol=0.3, angle\_tol=5, as used in \cite{luo2024deep}. 
It is important to note that, unlike previous research, we found that StructureMatcher can incorrectly declare structure identity when two similar structures have different space groups (see Discussion section). Therefore, we interpreted the success rate along with the space group matching rate in our results. These additional evaluations provide a more comprehensive assessment of the generated structures' performance across different algorithms.

\paragraph{Ranking scores of algorithms:} To evaluate how different performance metrics reflect the actual closeness of the predicted structures to the ground truth structure, we employed quantitative distance matrices of CSP \cite{wei2024towards} to assess the quality of all structures generated by the algorithms. We adopted a ranking scheme to evaluate candidate CSP algorithms based on the quality of their predicted structure against the ground truth structure. For each test structure, all algorithms are first ranked based on the quality of their predicted structures, i.e., their distances to the ground truth structure. Ranking scores on a 0-100 scale are assigned to the algorithms using a standardized scoring method to ensure fairness in ranking. The ranking scheme is illustrated as follows: for example, if there are five algorithms for comparison, five evenly distributed scores ranging from 100 to 0 are assigned to the five algorithms sorted by their performance from the highest to the lowest. Specifically, the algorithm in the first place receives a score of 100 (reflecting the smallest distance), and the second-placed algorithm earns a score of 75, followed by 50 for the third place, 25 for the fourth place, and 0 for the fifth place. In cases where multiple algorithms produced structures with identical quality/distances, they were assigned the same rank, and scores were averaged according to their rankings. For instance, if the first and second place algorithms tie in the quality of their predicted structures, their scores are set as the average of 100 and 75  [(100 + 75)/2]. Similarly, if all five algorithms have the same performance, their scores are set as the average of the five scores [(100 + 75 + 50 + 25 + 0)/5]. Figure \ref{fig:overall_avg_scores}
shows the ranking scores based on the overall average distances for each algorithm.

\begin{table}[]
\centering
\caption{Metrics of seven distances for partial test data of the AGOX-pt algorithm, including M3GNet Energy Distance (eV/atom); Wyckoff RMSE, Chamfer Distance, Superpose RMSD, and Fingerprint Distance (all in Å); and XRD Spectrum Distance (counts/sec).}
\centering
\begin{tabular}{|l|l|l|l|l|l|l|l|}
\hline
\textbf{\begin{tabular}[c]{@{}l@{}}Primitive \\ Formula\end{tabular}} & \multicolumn{1}{l|}{\textbf{\begin{tabular}[c]{@{}l@{}}M3GNet \\ Energy \\ Distance\end{tabular}}} & \multicolumn{1}{l|}{\textbf{\begin{tabular}[c]{@{}l@{}}Wyckoff \\ RMSE\end{tabular}}} & \multicolumn{1}{l|}{\textbf{\begin{tabular}[c]{@{}l@{}}Sinkhorn \\ Distance\end{tabular}}} & \multicolumn{1}{l|}{\textbf{\begin{tabular}[c]{@{}l@{}}Chamfer \\ Distance\end{tabular}}} & \multicolumn{1}{l|}{\textbf{\begin{tabular}[c]{@{}l@{}}Superpose\\ RMSD\end{tabular}}} & \multicolumn{1}{l|}{\textbf{\begin{tabular}[c]{@{}l@{}}Fingerprint\\ Distance\end{tabular}}} & \multicolumn{1}{l|}{\textbf{\begin{tabular}[c]{@{}l@{}}XRD \\ Distance\end{tabular}}} \\ \hline
YHg$_2$                                                               & 1.02                                                                                               & 0.27                                                                                  & 38.60                                                                                      & 23.01                                                                                     & 13.20                                                                                  & 1.87                                                                                         & 1.44                                                                                  \\ \hline
ScCu                                                                  & 2.69                                                                                               & 0.43                                                                                  & 22.04                                                                                      & 19.80                                                                                     & 10.94                                                                                  & 2.35                                                                                         & 2.85                                                                                  \\ \hline
K$_4$Na$_2$Ga$_2$P$_4$                                                & 0.22                                                                                               & 0.31                                                                                  & 92.57                                                                                      & 11.88                                                                                     & 1.98                                                                                   & 1.75                                                                                         & 1.13                                                                                  \\ \hline
Re$_2$O$_6$                                                           & 0.21                                                                                               & 0.31                                                                                  & 71.15                                                                                      & 13.76                                                                                     & 1.00                                                                                   & 2.89                                                                                         & 1.73                                                                                  \\ \hline
CrFeCoSi                                                              & 2.25                                                                                               & 0.32                                                                                  & 59.11                                                                                      & 23.89                                                                                     & 14.55                                                                                  & N/A                                                                                      & 2.92                                                                                  \\ \hline
Ba$_2$YRuO$_6$                                                        & 1.07                                                                                               & 0.31                                                                                  & 154.98                                                                                     & 23.12                                                                                     & 16.07                                                                                  & 2.54                                                                                         & 1.95                                                                                  \\ \hline
Li$_2$NiO$_2$                                                         & 0.91                                                                                               & 0.31                                                                                  & 51.95                                                                                      & 18.29                                                                                     & 10.82                                                                                  & N/A                                                                                      & 1.42                                                                                  \\ \hline
PrC$_2$                                                               & 1.76                                                                                               & 0.35                                                                                  & 23.67                                                                                      & 14.92                                                                                     & 8.06                                                                                   & 1.42                                                                                         & 1.43                                                                                  \\ \hline
Ge$_{12}$Rh$_3$                                                       & 0.03                                                                                               & 0.35                                                                                  & 207.02                                                                                     & 24.41                                                                                     & 15.53                                                                                  & 1.84                                                                                         & 1.61                                                                                  \\ \hline
MgV$_4$SnO$_{12}$                                                     & 0.39                                                                                               & 0.30                                                                                  & 181.78                                                                                     & 15.46                                                                                     & 1.28                                                                                   & 1.93                                                                                         & 1.35                                                                                  \\ \hline
DyPd                                                                  & 2.46                                                                                               & 0.35                                                                                  & 16.97                                                                                      & 15.52                                                                                     & 8.65                                                                                   & 2.33                                                                                         & 2.37                                                                                  \\ \hline
CeCr$_2$Si$_2$C                                                       & 1.29                                                                                               & 0.32                                                                                  & 48.70                                                                                      & 12.08                                                                                     & 1.44                                                                                   & 1.69                                                                                         & 1.39                                                                                  \\ \hline
Nb$_2$P$_2$Se$_2$                                                     & 0.64                                                                                               & 0.26                                                                                  & 38.29                                                                                      & 9.25                                                                                      & 1.49                                                                                   & 1.80                                                                                         & 1.34                                                                                  \\ \hline
BePd$_2$                                                              & 2.23                                                                                               & 0.29                                                                                  & 24.23                                                                                      & 15.05                                                                                     & 8.42                                                                                   & 2.47                                                                                         & 1.61                                                                                  \\ \hline
SrGa$_4$                                                              & 0.56                                                                                               & 0.27                                                                                  & 44.16                                                                                      & 16.13                                                                                     & 9.16                                                                                   & N/A                                                                                      & 1.38                                                                                  \\ \hline
KLi$_6$IrO$_6$                                                        & 0.55                                                                                               & 0.30                                                                                  & 162.23                                                                                     & 15.92                                                                                     & 1.40                                                                                   & 2.64                                                                                         & 1.23                                                                                  \\ \hline
Hf$_4$Mn$_8$                                                          & 1.66                                                                                               & 0.33                                                                                  & 185.07                                                                                     & 23.21                                                                                     & 15.24                                                                                  & 2.32                                                                                         & 1.58                                                                                  \\ \hline
Fe$_2$Cu$_6$SnS$_8$                                                   & 0.28                                                                                               & 0.34                                                                                  & 81.88                                                                                      & 7.01                                                                                      & 1.82                                                                                   & 2.20                                                                                         & 2.06                                                                                  \\ \hline
KAs$_4$IO$_6$                                                         & 0.27                                                                                               & 0.35                                                                                  & 145.19                                                                                     & 20.16                                                                                     & 13.39                                                                                  & 2.17                                                                                         & 1.33                                                                                  \\ \hline
Ti$_2$Cd                                                              & 2.50                                                                                               & 0.39                                                                                  & 26.25                                                                                      & 14.15                                                                                     & 8.65                                                                                   & 2.36                                                                                         & 1.71                                                                                  \\ \hline
\end{tabular}
\label{tab:scores}
\end{table}

\section{Results}
\label{sec:others}

\subsection{Performance comparison of CSP algorithms over all test structures}

\begin{figure}[tbh!]
  \centering
  \includegraphics[width=0.9\linewidth]{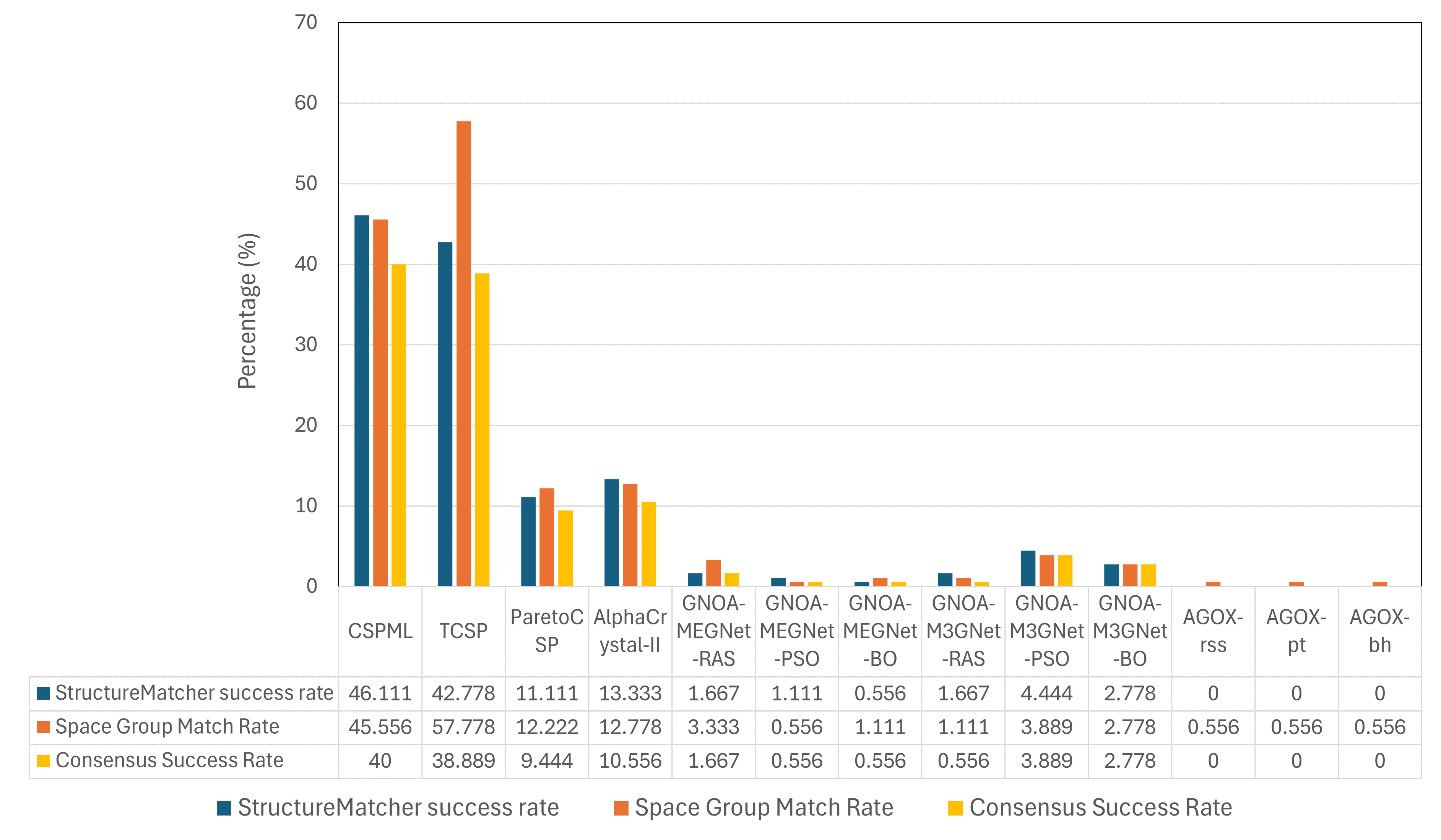}
  \caption{Comparison of structure prediction performance in terms of the success rate (\%) measured as the percentage of the predicted structures (out of 180 test samples) that match the ground truth structures according to Pymatgen StructureMatcher and have identical space groups.}
  \label{fig:succ-180}
\end{figure}

We evaluated the performance of 13 CSP algorithms in predicting the structures of 180 test samples: TCSP, CSPML, ParetoCSP, AlphaCrystal-II, GNOA-M3GNet-RAS, GNOA-M3GNet-PSO, GNOA-M3GNet-BO, GNOA-MEGNet-RAS, GNOA-MEGNet-PSO, GNOA-MEGNet-BO, AGOX-rss, AGOX-pt, AGOX-bh. 
The success rates varied across the different algorithms. TCSP achieved the highest success rate, efficiently predicting all 180 structures, while each of the AGOX algorithms (AGOX-rss, AGOX-pt, and AGOX-bh) successfully predicted structures for 175 out of the 180 samples. ParetoCSP and CSPML exhibited impressive performance, predicting 173 and 158 structures, respectively. However, AlphaCrystal-II was limited to predicting only 121 structures due to its inability to handle structures with more than 12 atoms. The GNOA algorithms showed weaknesses in predicting complex binary, ternary, and quaternary structures, with successful predictions ranging from 30 to 39 out of 180 samples.
To analyze and compare the performance of all CSP algorithms across the 180 test structures, we first calculated the StructureMatcher success rate by utilizing StructureMatcher from Pymatgen, with the following default parameters: ltol=0.2, stol=0.3, angle\_tol=5 to find out if similar materials already existed in Materials Project database. As shown in Figure \ref{fig:succ-180}, we find that two template-based CSP algorithms, TCSP and CSPML, stand out for their high performance in generating structures with similar space groups to the ground truth structures. CSPML and TCSP achieve the best performance with success rates of 46.111\% and 42.778\%, respectively. In contrast, AlphaCrystal-II and ParetoCSP showed significantly lower performance, with StructureMatcher success rates of 13.333\% and 11.111\%, respectively. The GNOA algorithms generally underperformed, with success rates ranging from 0.556\% to 4.444\%. Among the three M3GNet-based GNOA CSP algorithms, it can be found that the GNOA-M3GNet-PSO (particle swarm optimization) is better than BO (bayesian optimization) and  RAS (random search) based algorithms, reflecting the importance of the search capability used in the CSP algorithms. We can also find that three MEGNet based GNOA algorithms all perform poorly here due to their MEGNet energy potential with lower accuracy. We also evaluated the performance of three AGOX algorithms using different optimization strategies: Basin Hopping (BH), parallel tempering (PT), and random search (RSS). The AGOX algorithms failed to predict any structures matching those in the MP database.
Symmetry prediction performance also plays an important role. We computed the space group match rate for each algorithm, which indicates whether the predicted structure has the same space group number as the ground truth structure. TCSP achieves the best performance with 57.778\%, followed by CSPML with a space group match rate of 45.556\%. The proficiency of template-based CSP algorithms in predicting crystal structures with identical symmetries may stem from two factors. The first is their adeptness at recognizing highly similar structure templates by using oxidation state and composition-based fingerprint matching. The second reason for their success is the widespread existence of similar crystal structures with identical space groups and crystal systems, making it easier to find a template and use simple elemental substitution to determine their structures. This structural distribution pattern has been exploited by the DeepMind team to help discover more than 380,000 new hypothetical stable materials in their Nature report \cite{merchant2023scaling} in 2023. However, it is also observed that both template-based algorithms, CSPML and TCSP, fail to find structures with correct symmetry for at least 98 materials (54\%) and 76 materials (42\%) respectively, reflecting the dire demand for developing de novo CSP prediction algorithms.
Next, we found that the space group match rate and StructureMatcher success rate are consistent over the remaining algorithms. Out of the 11 de novo CSP algorithms, ParetoCSP and AlphaCrystal-II outperform all other algorithms, the space group match rates are 12.222\% and 12.778\% for ParetoCSP and AlphaCrystal-II, respectively. These are 214.27\% and 228.57\% better than the GNOA-M3GNet-PSO algorithm, the best of the remaining 9 de novo algorithms. These successes can be attributed to ParetoCSP's strong global search capability based on the age-fitness multi-objective genetic algorithm along with its usage of the M3GNet deep learning potential model \cite{chen2022universal}, and contact-map based deep learning CSP algorithm AlphaCrystal-II utilizes inter-atomic interaction patterns found in existing known crystal structures. 
GNOA algorithms exhibited low space group match rates, ranging from 0.556\% to 12.778\%. All AGOX algorithms had a space group match rate of 0.556\%, indicating their inability to find structures with the same space group number as the ground truth structures.
Overall, we find that current de novo algorithms based on machine learning potentials can only achieve moderate crystal structure performance in terms of their success rate and space group prediction accuracy, indicating the significant potential for further development in this research area. 
More details on how many of the predicted structures by each algorithm have the same symmetry with the ground truth structures in terms of their space group and crystal system are shown in Table S1 of the supplementary file.

To conduct a comprehensive analysis and comparison of each algorithm's performance, we further evaluated the algorithms using a set of quantitative metrics proposed in our work \cite{wei2024towards}. First, we used the formation energy distances of the predicted structures compared to the ground truth as the performance metric for algorithm comparison, a method widely used in previous CSP work \cite{glass2006uspex, bao2009structure, tong2018accelerating}. Using formation energy as a performance metric has a unique value as it can serve as a critical indicator of a crystal's stability in nature. To efficiently analyze and validate the performance of structures generated by various algorithms, we computed the ranking score for each algorithm based on the M3GNet formation energy distance of each structure.

\begin{figure}[!htb]
  \centering
  \includegraphics[width=0.8\linewidth]{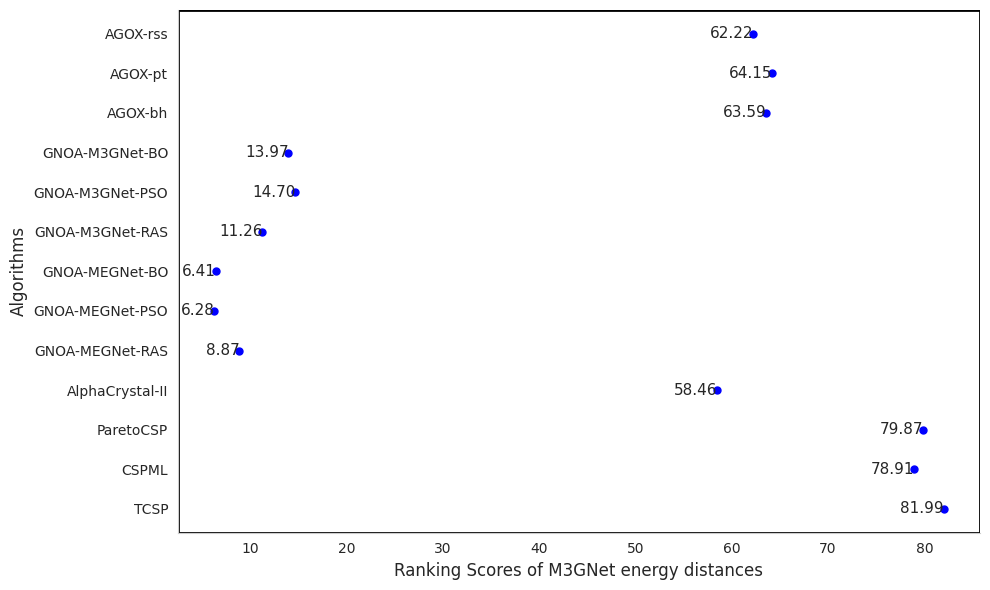}
  \caption{Comparison of CSP algorithms by their ranking scores based on the average M3GNet energy distances of the predicted structures against the ground truth structures.}
  \label{fig:m3gnet_score}
\end{figure}

As shown in Figure \ref{fig:m3gnet_score}, TCSP achieves the best performance in terms of the average M3GNet energy distance, with a score of 81.99. Close behind, ParetoCSP and CSPML achieve ranking scores of 79.87 and 78.91, respectively. The high ranking scores for TCSP and CSPML based on M3GNet energy distance reflect these two template-based CSP algorithms' consistency in symmetry prediction performance. ParetoCSP's ranking score of 79.87 is slightly below TCSP yet considerably higher than the scores of all other 11 de novo CSP algorithms. The three AGOX algorithms achieve ranking scores of 62.22, 64.15 and 63.59, respectively. AlphaCrystal-II achieves a ranking score of 58.46, slightly lower than those of the AGOX algorithms. Despite the AGOX algorithms having the lowest StructureMatcher success rate and space group match rate, their higher M3GNet energy distance ranking scores compared to AlphaCrystal-II can be attributed to the larger number of successfully generated structures by the AGOX algorithms. Furthermore, AlphaCrystal-II demonstrates its capability by generating a larger total number of predicted structures compared to all GNOA algorithms, it surpasses the performance of GNOA-MEGNet-PSO by an astounding 830.89\% and GNOA-M3GNet-PSO by at least 297.69\%. The significantly lower ranking scores for all GNOA algorithms are primarily due to their limited ability to generate structures and their weakness in predicting structures similar to the ground truth, as reflected by the small portion of successfully generated structures.

\begin{figure}[!htb]
  \centering
  \includegraphics[width=0.8\linewidth]{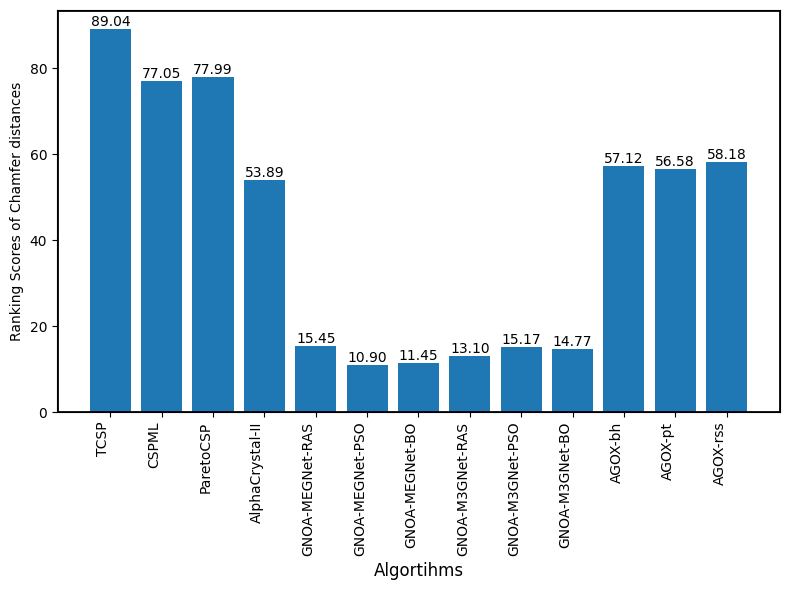}
  \caption{Comparison of CSP algorithms by their ranking scores based on the average Chamfer distances of the predicted structures against the ground truth structures.}
  \label{fig:avg_chamfer}
\end{figure}

\begin{figure}[!htb]
  \centering
  \includegraphics[width=0.85\linewidth]{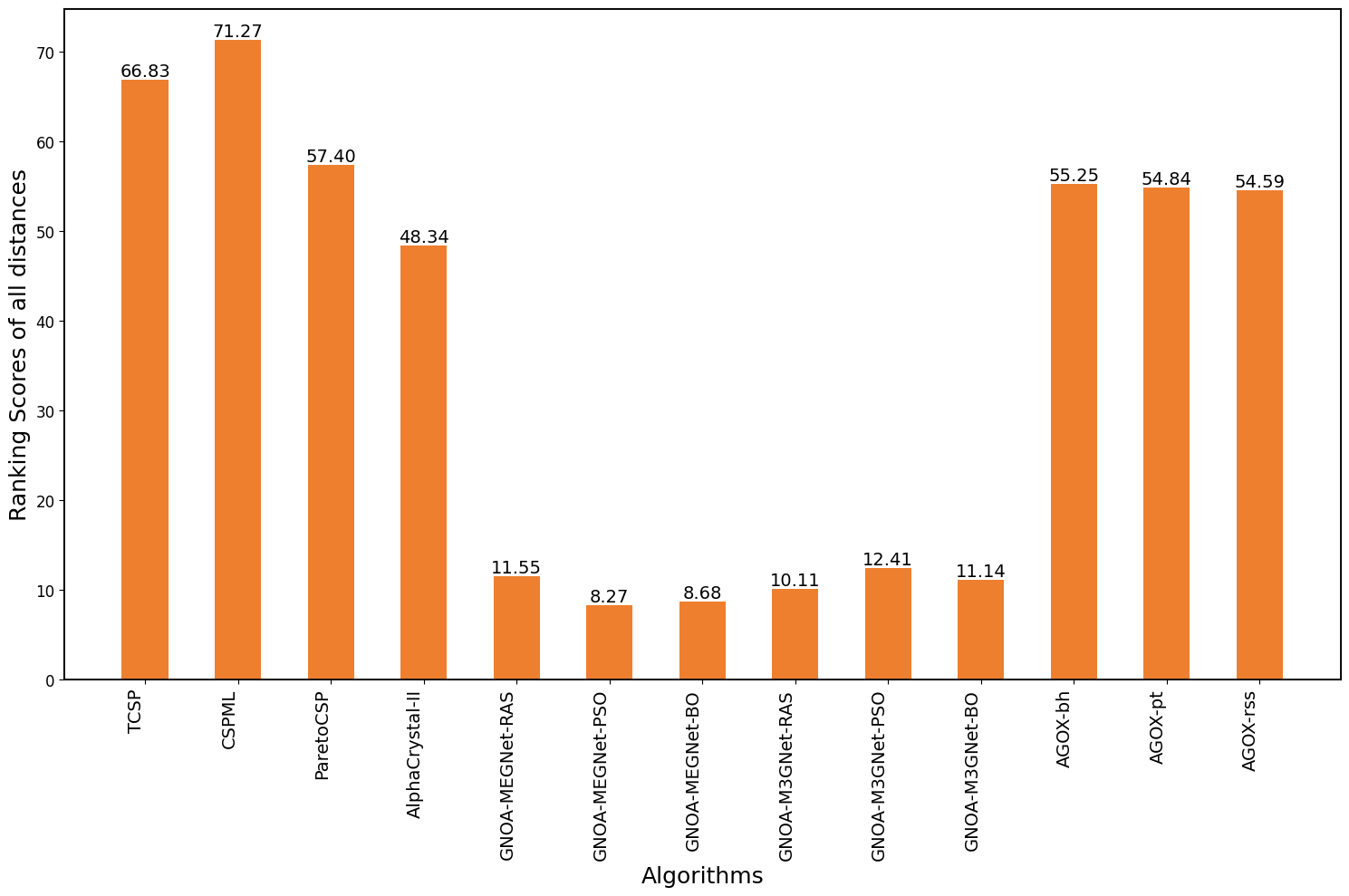}
  \caption{Comparison of CSP algorithms by their ranking scores based on the average of all 12 distance metrics of the predicted structures against the ground truth structures.}
  \label{fig:overall_avg_scores}
\end{figure}

We further utilized the average Chamfer distance as a metric, which is calculated as the mean of the squared distances between each atomic site in one structure and its nearest neighbor in another, and serves as a robust measure of structural congruence. By capturing the spatial correlation between atomic sites, the Chamfer distance offers a comprehensive and nuanced measure of similarity. This sensitivity to the intricate details of crystallographic arrangements allows for a nuanced evaluation of model performance across a diverse range of structures. The ranking scores based on the average Chamfer distances are depicted in Figure \ref{fig:avg_chamfer}. 
Among the algorithms evaluated, TCSP emerges as the top performer with the highest ranking score of 89.04, closely followed by CSPML with a score of 77.05. These template-based CSP algorithms excel in generating structures that are more congruent with the ground truth structures, demonstrating their strong predictive capabilities. ParetoCSP, with a ranking score of 77.99, also exhibits good performance by the age-fitness Pareto genetic algorithm. The AGOX family of algorithms, including AGOX-bh, AGOX-pt, and AGOX-rss, show relatively consistent performance, with scores ranging from 56.58 to 58.18. AlphaCrystal-II achieves a comparable ranking score of 53.89.
The performances of ParetoCSP, the AGOX family, and AlphaCrystal-II significantly outperform the GNOA family of algorithms. AGOX-rss outperforms GNOA-M3GNet-BO by a substantial 293.90\%. While GNOA algorithms demonstrate superior performance for simpler binary structures, their predictive accuracy diminishes for more complex structures. This is evident in their lower average ranking scores on the Chamfer distance metric, ranging from 10.90 to 15.45, highlighting the challenge of accurately predicting complex crystal structures.
Common limitations of current methods include their dependency on template-based approaches and difficulties in predicting structures with complex symmetries and compositions. These constraints necessitate the development of more advanced, de novo CSP prediction algorithms.
Overall, the ranking scores based on the average Chamfer distances provide a meaningful evaluation of the algorithms' ability to predict crystal structures that closely match the ground truth structures, with template-based approaches generally outperforming other methods for the given set of structures.
Finally, in our endeavor to thoroughly and accurately assess the performance of each algorithm, we conducted a comprehensive analysis by computing overall ranking scores based on all 12 distance metrics. As depicted in Figure \ref{fig:overall_avg_scores}, CSPML maintains its dominance with a score of 71.27, followed closely by the TCSP algorithm at 66.83. CSPML outperforms the TCSP algorithm due to its utilization of chemical composition descriptors and crystal structure descriptors. This comprehensive approach results in a higher ranking score for CSPML.
The AGOX family achieved scores of 55.25, 54.84, and 54.59, respectively, while ParetoCSP had a ranked score of 57.40. In contrast, AlphaCrystal-II received a slightly lower score of 48.34, and the GNOA family ranged from 8.27 to 12.41. These results underscore the varied capabilities of the algorithms, not only affirming the importance of integrating diverse descriptors for enhancing predictive accuracy but also highlighting the challenges and potential areas for improvement in crystal structure prediction algorithms.

\subsection{Performance comparison over binary structures}

\begin{figure}[!htb]
  \centering
  \includegraphics[width=0.85\linewidth]{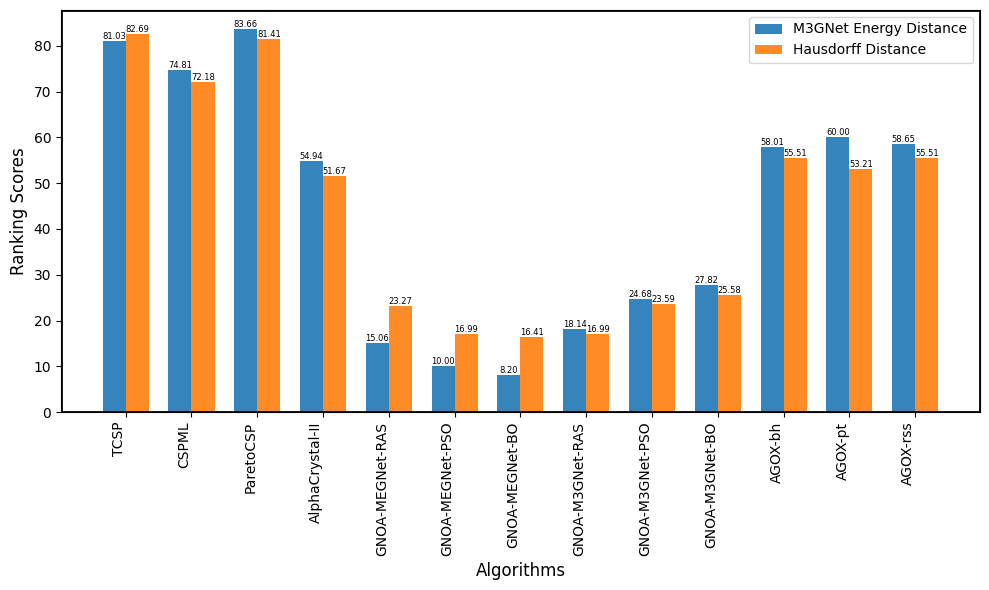}
  \caption{Binary structure prediction performance comparison of CSP algorithms by their ranking scores based on the average of M3GNet energy distance and Hausdroff distance across 60 binary predicted structures against the ground truth structures.}
\label{fig:binary}
\end{figure}

To further analyze the performance of various algorithms, we focus on their predictions across 60 binary structures, employing the M3GNet energy distance and Hausdorff distance as key metrics. Hausdorff distance is a structure similarity metric that represents the maximum deviation between two structures which has the advantage of being invariant to rigid transformations, such as translations, rotations, and reflections. The average ranking scores derived from both metrics for these binary test structures are detailed in Figure \ref{fig:binary}, showcasing the average M3GNet energy distances and Hausdorff distances in comparison to the ground truth structures for all predictions. Among the algorithms, ParetoCSP achieved the highest ranking scores, with 83.66 for the M3GNet energy distance metric and 81.41 for the Hausdorff distance. TCSP also performed well, with a ranking score of 81.03 for the M3GNet energy distance and 82.69 for the Hausdorff distance, indicating strong performance in both metrics. CSPML, while slightly lower than TCSP and ParetoCSP, still showed robust performance with ranking scores of 74.81 and 72.18 for the M3GNet energy distance and Hausdorff distance, respectively. The AlphaCrystal-II and AGOX algorithms (AGOX-bh, AGOX-pt, and AGOX-rss) demonstrated relatively good performance, with ranking scores ranging from 54.94 to 60.00 for the M3GNet energy distance and 51.67 to 55.51 for the Hausdorff distance. Additionally, the GNOA algorithms with the M3GNet potential (GNOA-M3GNet-RAS, GNOA-M3GNet-PSO, and GNOA-M3GNet-BO) generally achieved higher ranking scores based on both the M3GNet energy distance and Hausdorff distance metrics compared to the GNOA algorithms with the MEGNet potential. However, the GNOA algorithms still struggled with predicting more complex structures, even within the binary structures category, especially those that do not consist of a simple 1:1 ratio of atoms. Their ranking scores for the Hausdorff distance ranged from 16.41 to 25.28, highlighting the need for further improvements to better handle more intricate crystal structures. Overall, despite the relatively simpler nature of binary structures, some algorithms still faced challenges in accurately predicting their configurations. This underscores the importance of continuous improvement and the development of more robust predictive models.

\subsection{Performance comparison over ternary structures}

\begin{figure}[!htb]
  \centering
  \includegraphics[width=0.85\linewidth]{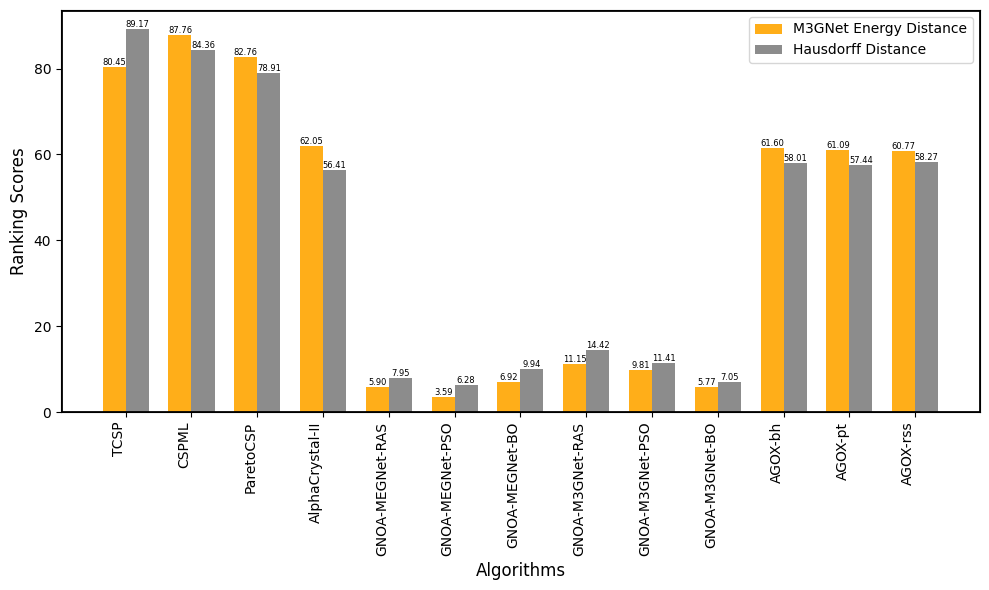}
  \caption{Ternary structure prediction performance comparison of CSP algorithms by their ranking scores based on the average of M3GNet energy distance and Hausdroff distance across 60 ternary predicted structures against the ground truth structures.}
  \label{fig:ternary}
\end{figure}

Given the relative simplicity of binary structures in our dataset, we extend our analysis to ternary and quaternary crystal structures to evaluate the strengths and weaknesses of various CSP algorithms across different structural complexities.
We compared the algorithm performances for ternary structures using the M3GNet energy distance and the Hausdorff distance based ranking scores. 
As shown in Figure \ref{fig:ternary}, CSPML and ParetoCSP achieve the highest ranking scores in terms of M3GNet energy distance of 87.76 and 82.76, respectively. Meanwhile, TCSP attains the highest ranking scores for the Hausdorff distance, recording 89.17, with a closely comparable score of 80.45 on the M3GNet energy distance. Similar to their performance on binary structures, the AGOX algorithms exhibited lower scores, ranging from 60.77 to 61.60 on the M3GNet energy distance and from 57.44 to 58.27 on the Hausdorff distance. AlphaCrystal-II achieved ranking scores of 62.05 on the M3GNet energy distance and 56.41 on the Hausdorff distance. However, the GNOA algorithms faced significant limitations in predicting more complex structures, resulting in low ranking scores on both distance metrics, ranging from 3.59 to 14.42. Comparing the ranking scores of different algorithms in Figure \ref{fig:ternary} to those for binary test structures (Figure \ref{fig:binary}), we find that the template-based algorithms and ParetoCSP maintain similar ranking scores. In contrast, the scores for the AGOX algorithms and AlphaCrystal-II increase, indicating improved performance with more complex structures. However,  these improvements come at the cost of the ranking scores of the GNOA family of algorithms. This performance gap clearly indicates the necessity for further development and refinement within the GNOA algorithms to enhance their predictive accuracy and reliability in handling complex crystal structures.

\subsection{Performance comparison over quarternary structures}

\begin{figure}[!htb]
  \centering
  \includegraphics[width=0.85\linewidth]{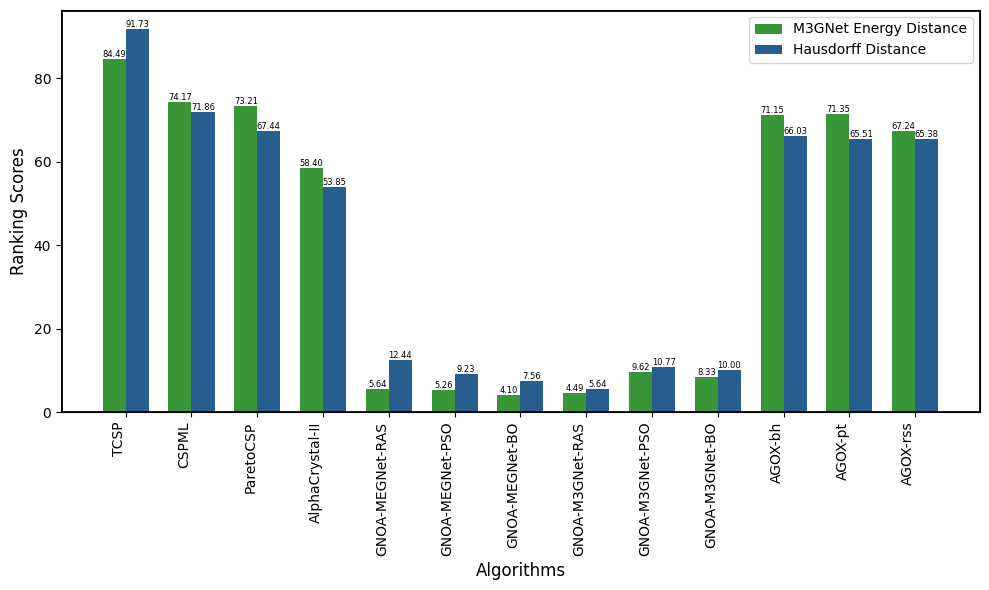}
  \caption{Quaternary structure prediction performance comparison of CSP algorithms by their ranking scores based on the average of M3GNet energy distance and Hausdroff distance across 60 quaternary predicted structures against the ground truth structures.}
  \label{fig:quaternary}
\end{figure}

We evaluated the ranking scores for quaternary structure predictions across all CSP algorithms using M3GNet energy distance and Hausdorff distances as metrics. Figure \ref{fig:quaternary} illustrates the ranking scores for each algorithm. shows the ranking scores for each algorithm. Using the Hausdorff distance as a metric, the TCSP algorithm achieved the highest ranking score of 91.73, followed by CSPML and ParetoCSP with the scores of 71.86 and 67.44. Scores for the AGOX family are competitive, with 66.03, 65.51, and 65.38, respectively. On the other hand, ParetoCSP presents a challenge in addressing quaternary structures, reflected by its slightly lower score compared to the ranking scores for binary (Figure \ref{fig:binary}) and ternary structures (Figure \ref{fig:ternary}). The GNOA algorithms, which encompass six distinct approaches, record scores ranging from 4.10 to 12.44, indicating significant difficulties in predicting quaternary compounds. When evaluated based on the M3GNet energy distance, TCSP again leads with the highest ranking score of 84.49. CSPML, ParetoCSP, and AGOX algorithms also demonstrate competitive performance with scores from 67.24 to 74.17. Among the algorithms assessed, seven others outperformed the GNOA algorithms, reflecting the inherent challenges in predicting quaternary compounds.
This comprehensive analysis highlights the strengths and limitations of different CSP algorithms across varying structural complexities, emphasizing the need for continuous improvement and refinement to handle more complex crystal structures effectively.
To provide a comprehensive analysis, additional performance comparisons utilizing the Sinkhorn distance, superpose RMSD, Wyckoff RMSE, XRD distance, and OFM distance across binary, ternary, and quaternary test structures, as detailed in the Figure S2, S3 and S4 in the supplementary file.

\FloatBarrier

\begin{table}[tbh!]
\caption{Performance comparison of CSP algorithms CALYPSO, CSPML, ParetoCSP, and AGOX-pt in terms of M3GNet Energy Distance (eV/atom) (ED) and Hausdorff Distance (HD) (Å). Values highlighted in bold represent the minimum ED or HD computed from the predicted and ground truth structures for each test sample across various algorithms.}
\centering
\begin{tabular}{|ll|ll|ll|ll|ll|}
\hline
\multicolumn{2}{|c|}{\textbf{Algorithm}}                                                          & \multicolumn{2}{c|}{\cellcolor[HTML]{FFFFFF}{\color[HTML]{1F1F1F} \textbf{CALYPSO}}} & \multicolumn{2}{c|}{\cellcolor[HTML]{FFFFFF}{\color[HTML]{1F1F1F} \textbf{CSPML}}}                   & \multicolumn{2}{c|}{\cellcolor[HTML]{FFFFFF}{\color[HTML]{1F1F1F} \textbf{ParetoCSP}}}               & \multicolumn{2}{c|}{\cellcolor[HTML]{FFFFFF}{\color[HTML]{1F1F1F} \textbf{AGOX-pt}}}                  \\ \hline
\multicolumn{1}{|l|}{primitive formula}                                        & mp-id            & \multicolumn{1}{l|}{ED}                 & \multicolumn{1}{l|}{HD}                    & \multicolumn{1}{l|}{ED}                                     & \multicolumn{1}{l|}{HD}                & \multicolumn{1}{l|}{ED}                                     & \multicolumn{1}{l|}{HD}                & \multicolumn{1}{l|}{ED}                                     & \multicolumn{1}{l|}{HD}                 \\ \hline
\multicolumn{1}{|l|}{\cellcolor[HTML]{FFFFFF}{\color[HTML]{0D0D0D} Ca$_3$SnO}}    & mp-29241         & \multicolumn{1}{r|}{0.002}     & \cellcolor[HTML]{FFFFFF}2.413     & \multicolumn{1}{r|}{\cellcolor[HTML]{FFFFFF}\textbf{0.001}} & \cellcolor[HTML]{FFFFFF}\textbf{0.021} & \multicolumn{1}{r|}{\cellcolor[HTML]{FFFFFF}\textbf{0.001}} & \cellcolor[HTML]{FFFFFF}0.023          & \multicolumn{1}{r|}{\cellcolor[HTML]{FFFFFF}1.099} & \cellcolor[HTML]{FFFFFF}9.715           \\ \hline
\multicolumn{1}{|l|}{\cellcolor[HTML]{FFFFFF}{\color[HTML]{0D0D0D} Co$_2$Ni$_2$Sn$_2$}} & mp-20237         & \multicolumn{1}{r|}{0.061}              & \cellcolor[HTML]{FFFFFF}5.489              & \multicolumn{1}{r|}{\cellcolor[HTML]{FFFFFF}\textbf{0.000}} & \cellcolor[HTML]{FFFFFF}2.557 & \multicolumn{1}{r|}{\cellcolor[HTML]{FFFFFF}0.002}          & \cellcolor[HTML]{FFFFFF}\textbf{0.056} & \multicolumn{1}{r|}{\cellcolor[HTML]{FFFFFF}1.210}          & \cellcolor[HTML]{FFFFFF}15.062          \\ \hline
\multicolumn{1}{|l|}{\cellcolor[HTML]{FFFFFF}{\color[HTML]{0D0D0D} Co$_2$Te$_2$}}    & mp-788           & \multicolumn{1}{r|}{\textbf{0.028}}     & \cellcolor[HTML]{FFFFFF}6.520     & \multicolumn{1}{r|}{\cellcolor[HTML]{FFFFFF}0.220}          & \cellcolor[HTML]{FFFFFF}\textbf{2.475} & \multicolumn{1}{r|}{\cellcolor[HTML]{FFFFFF}0.050}          & \cellcolor[HTML]{FFFFFF}4.573          & \multicolumn{1}{r|}{\cellcolor[HTML]{FFFFFF}0.879} & \cellcolor[HTML]{FFFFFF}20.100          \\ \hline
\multicolumn{1}{|l|}{\cellcolor[HTML]{FFFFFF}{\color[HTML]{0D0D0D} Cr$_6$Ga$_2$}}    & mp-1231          & \multicolumn{1}{r|}{2.016}              & \cellcolor[HTML]{FFFFFF}7.001              & \multicolumn{1}{r|}{\cellcolor[HTML]{FFFFFF}0.096} & \cellcolor[HTML]{FFFFFF}5.710 & \multicolumn{1}{r|}{\cellcolor[HTML]{FFFFFF}\textbf{0.015}} & \cellcolor[HTML]{FFFFFF}\textbf{1.622} & \multicolumn{1}{r|}{\cellcolor[HTML]{FFFFFF}1.494}          & \cellcolor[HTML]{FFFFFF}6.864           \\ \hline
\multicolumn{1}{|l|}{\cellcolor[HTML]{FFFFFF}{\color[HTML]{0D0D0D} Hf$_4$Mn$_8$}}    & mp-11449         & \multicolumn{1}{r|}{\textbf{0.002}}     & \cellcolor[HTML]{FFFFFF}\textbf{6.383}     & \multicolumn{1}{r|}{\cellcolor[HTML]{FFFFFF}0.129}          & \cellcolor[HTML]{FFFFFF}8.715 & \multicolumn{1}{r|}{\cellcolor[HTML]{FFFFFF}0.266}          & \cellcolor[HTML]{FFFFFF}6.457          & \multicolumn{1}{r|}{\cellcolor[HTML]{FFFFFF}1.660}          & \cellcolor[HTML]{FFFFFF}15.644          \\ \hline
\multicolumn{1}{|l|}{\cellcolor[HTML]{FFFFFF}{\color[HTML]{0D0D0D} Hf$_4$Ni$_2$}}    & mp-861           & \multicolumn{1}{r|}{\textbf{0.014}}     & \cellcolor[HTML]{FFFFFF}\textbf{4.064}     & \multicolumn{1}{r|}{\cellcolor[HTML]{FFFFFF}1.274}          & \cellcolor[HTML]{FFFFFF}11.162         & \multicolumn{1}{r|}{\cellcolor[HTML]{FFFFFF}0.039}          & \cellcolor[HTML]{FFFFFF}7.752          & \multicolumn{1}{r|}{\cellcolor[HTML]{FFFFFF}1.823}          & \cellcolor[HTML]{FFFFFF}11.395          \\ \hline
\multicolumn{1}{|l|}{\cellcolor[HTML]{FFFFFF}{\color[HTML]{0D0D0D} HfCo$_2$Sn}}   & mp-20730         & \multicolumn{1}{r|}{0.054}     & \cellcolor[HTML]{FFFFFF}3.928     & \multicolumn{1}{r|}{\cellcolor[HTML]{FFFFFF}\textbf{0.002}} & \cellcolor[HTML]{FFFFFF}\textbf{0.046} & \multicolumn{1}{r|}{\cellcolor[HTML]{FFFFFF}0.038}          & \cellcolor[HTML]{FFFFFF}9.083          & \multicolumn{1}{r|}{\cellcolor[HTML]{FFFFFF}2.175}          & \cellcolor[HTML]{FFFFFF}16.670          \\ \hline
\multicolumn{1}{|l|}{\cellcolor[HTML]{FFFFFF}{\color[HTML]{0D0D0D} InHg}}      & mp-20132         & \multicolumn{1}{r|}{\textbf{0.012}}     & \cellcolor[HTML]{FFFFFF}10.296             & \multicolumn{1}{r|}{\cellcolor[HTML]{FFFFFF}0.015}          & \cellcolor[HTML]{FFFFFF}7.968          & \multicolumn{1}{r|}{\cellcolor[HTML]{FFFFFF}0.069}          & \cellcolor[HTML]{FFFFFF}\textbf{7.379} & \multicolumn{1}{r|}{\cellcolor[HTML]{FFFFFF}0.191} & \cellcolor[HTML]{FFFFFF}12.479 \\ \hline
\multicolumn{1}{|l|}{\cellcolor[HTML]{FFFFFF}{\color[HTML]{0D0D0D} Li$_2$CuSn}}   & mp-30591         & \multicolumn{1}{r|}{\textbf{0.004}}     & \cellcolor[HTML]{FFFFFF}3.933     & \multicolumn{1}{r|}{\cellcolor[HTML]{FFFFFF}0.111} & \cellcolor[HTML]{FFFFFF}\textbf{0.129} & \multicolumn{1}{r|}{\cellcolor[HTML]{FFFFFF}0.012}          & \cellcolor[HTML]{FFFFFF}8.105          & \multicolumn{1}{r|}{\cellcolor[HTML]{FFFFFF}0.818}          & \cellcolor[HTML]{FFFFFF}16.551          \\ \hline
\multicolumn{1}{|l|}{\cellcolor[HTML]{FFFFFF}{\color[HTML]{0D0D0D} LiMg$_2$Ga}}   & mp-30648         & \multicolumn{1}{r|}{0.031}     & \cellcolor[HTML]{FFFFFF}7.062     & \multicolumn{1}{r|}{\cellcolor[HTML]{FFFFFF}\textbf{0.000}} & \cellcolor[HTML]{FFFFFF}\textbf{2.892} & \multicolumn{1}{r|}{\cellcolor[HTML]{FFFFFF}0.032}          & \cellcolor[HTML]{FFFFFF}9.908          & \multicolumn{1}{r|}{\cellcolor[HTML]{FFFFFF}0.773}          & \cellcolor[HTML]{FFFFFF}19.834          \\ \hline
\multicolumn{1}{|l|}{\cellcolor[HTML]{FFFFFF}{\color[HTML]{0D0D0D} MgCu$_4$Sn}}   & mp-3676          & \multicolumn{1}{r|}{\textbf{0.006}}     & \cellcolor[HTML]{FFFFFF}\textbf{3.194}     & \multicolumn{1}{r|}{\cellcolor[HTML]{FFFFFF}0.167}          & \cellcolor[HTML]{FFFFFF}5.256          & \multicolumn{1}{r|}{\cellcolor[HTML]{FFFFFF}0.085}          & \cellcolor[HTML]{FFFFFF}3.881          & \multicolumn{1}{r|}{\cellcolor[HTML]{FFFFFF}0.942}          & \cellcolor[HTML]{FFFFFF}16.160          \\ \hline
\multicolumn{1}{|l|}{\cellcolor[HTML]{FFFFFF}{\color[HTML]{0D0D0D} MgInCu$_4$}}   & mp-30587         & \multicolumn{1}{r|}{0.070}     & \cellcolor[HTML]{FFFFFF}4.861     & \multicolumn{1}{r|}{\cellcolor[HTML]{FFFFFF}\textbf{0.010}} & \cellcolor[HTML]{FFFFFF}\textbf{1.704} & \multicolumn{1}{r|}{\cellcolor[HTML]{FFFFFF}0.079}          & \cellcolor[HTML]{FFFFFF}5.029          & \multicolumn{1}{r|}{\cellcolor[HTML]{FFFFFF}0.986}          & \cellcolor[HTML]{FFFFFF}20.072          \\ \hline
\multicolumn{1}{|l|}{\cellcolor[HTML]{FFFFFF}{\color[HTML]{0D0D0D} NaGa$_4$}}     & mp-454           & \multicolumn{1}{r|}{\textbf{0.021}}     & \cellcolor[HTML]{FFFFFF}2.473              & \multicolumn{1}{r|}{\cellcolor[HTML]{FFFFFF}0.388} & \cellcolor[HTML]{FFFFFF}5.206 & \multicolumn{1}{r|}{\cellcolor[HTML]{FFFFFF}\textbf{0.009}} & \cellcolor[HTML]{FFFFFF}\textbf{1.236} & \multicolumn{1}{r|}{\cellcolor[HTML]{FFFFFF}0.297}          & \cellcolor[HTML]{FFFFFF}9.522           \\ \hline
\multicolumn{1}{|l|}{\cellcolor[HTML]{FFFFFF}{\color[HTML]{0D0D0D} ScCu}}      & mp-1169          & \multicolumn{1}{r|}{0.004}     & \cellcolor[HTML]{FFFFFF}1.701     & \multicolumn{1}{r|}{\cellcolor[HTML]{FFFFFF}0.108} & \cellcolor[HTML]{FFFFFF}3.681 & \multicolumn{1}{r|}{\cellcolor[HTML]{FFFFFF}\textbf{0.000}} & \cellcolor[HTML]{FFFFFF}\textbf{0.006} & \multicolumn{1}{r|}{\cellcolor[HTML]{FFFFFF}2.694}          & \cellcolor[HTML]{FFFFFF}11.775          \\ \hline
\multicolumn{1}{|l|}{\cellcolor[HTML]{FFFFFF}{\color[HTML]{0D0D0D} SrGa$_4$}}     & mp-1827          & \multicolumn{1}{r|}{\textbf{0.003}}     & \cellcolor[HTML]{FFFFFF}2.685              & \multicolumn{1}{r|}{\cellcolor[HTML]{FFFFFF}0.722} & \cellcolor[HTML]{FFFFFF}6.777          & \multicolumn{1}{r|}{\cellcolor[HTML]{FFFFFF}0.009}          & \cellcolor[HTML]{FFFFFF}\textbf{2.229} & \multicolumn{1}{r|}{\cellcolor[HTML]{FFFFFF}0.565}          & \cellcolor[HTML]{FFFFFF}10.163          \\ \hline
\multicolumn{1}{|l|}{\cellcolor[HTML]{FFFFFF}{\color[HTML]{0D0D0D} SrGaCu$_2$}}   & mp-30580         & \multicolumn{1}{r|}{\textbf{0.000}}     & \cellcolor[HTML]{FFFFFF}8.402     & \multicolumn{1}{r|}{\cellcolor[HTML]{FFFFFF}0.196}          & \cellcolor[HTML]{FFFFFF}\textbf{4.749} & \multicolumn{1}{r|}{\cellcolor[HTML]{FFFFFF}0.075}          & \cellcolor[HTML]{FFFFFF}4.853          & \multicolumn{1}{r|}{\cellcolor[HTML]{FFFFFF}1.026}          & \cellcolor[HTML]{FFFFFF}15.771          \\ \hline
\multicolumn{1}{|l|}{\cellcolor[HTML]{FFFFFF}{\color[HTML]{0D0D0D} Ti$_2$Cd}}     & mp-30501         & \multicolumn{1}{r|}{0.041}              & \cellcolor[HTML]{FFFFFF}3.755              & \multicolumn{1}{r|}{\cellcolor[HTML]{FFFFFF}0.061}          & \cellcolor[HTML]{FFFFFF}\textbf{1.064} & \multicolumn{1}{r|}{\cellcolor[HTML]{FFFFFF}\textbf{0.010}} & \cellcolor[HTML]{FFFFFF}5.197 & \multicolumn{1}{r|}{\cellcolor[HTML]{FFFFFF}2.497}          & \cellcolor[HTML]{FFFFFF}8.648           \\ \hline
\multicolumn{1}{|l|}{\cellcolor[HTML]{FFFFFF}{\color[HTML]{0D0D0D} TiGa3}}     & mp-2731          & \multicolumn{1}{r|}{0.023}              & \cellcolor[HTML]{FFFFFF}2.348              & \multicolumn{1}{r|}{\cellcolor[HTML]{FFFFFF}0.006} & \cellcolor[HTML]{FFFFFF}8.246 & \multicolumn{1}{r|}{\cellcolor[HTML]{FFFFFF}\textbf{0.002}} & \cellcolor[HTML]{FFFFFF}\textbf{0.221} & \multicolumn{1}{r|}{\cellcolor[HTML]{FFFFFF}1.296}          & \cellcolor[HTML]{FFFFFF}11.730          \\ \hline
\multicolumn{1}{|l|}{\cellcolor[HTML]{FFFFFF}{\color[HTML]{0D0D0D} Y$_3$Al$_9$}}     & \begin{tabular}[c]{@{}l@{}}mp-2451\\ mp-11231\end{tabular} & \multicolumn{1}{r|}{\textbf{0.001}}     & \cellcolor[HTML]{FFFFFF}\textbf{0.011}              & \multicolumn{1}{r|}{\cellcolor[HTML]{FFFFFF}0.002} & \cellcolor[HTML]{FFFFFF}3.022          & \multicolumn{1}{r|}{\cellcolor[HTML]{FFFFFF}\textbf{0.001}} & \cellcolor[HTML]{FFFFFF}3.723 & \multicolumn{1}{r|}{\cellcolor[HTML]{FFFFFF}0.893}          & \cellcolor[HTML]{FFFFFF}21.138          \\ \hline
\multicolumn{1}{|l|}{\cellcolor[HTML]{FFFFFF}{\color[HTML]{0D0D0D} YHg$_2$}}      & mp-30725         & \multicolumn{1}{r|}{\textbf{0.001}}     & \cellcolor[HTML]{FFFFFF}1.747              & \multicolumn{1}{r|}{\cellcolor[HTML]{FFFFFF}0.006}          & \cellcolor[HTML]{FFFFFF}\textbf{0.044} & \multicolumn{1}{r|}{\cellcolor[HTML]{FFFFFF}0.008}          & \cellcolor[HTML]{FFFFFF}1.741          & \multicolumn{1}{r|}{\cellcolor[HTML]{FFFFFF}1.025} & \cellcolor[HTML]{FFFFFF}13.130 \\ \hline
\multicolumn{1}{|l|}{\cellcolor[HTML]{FFFFFF}{\color[HTML]{0D0D0D} Zn$_2$C$_2$O$_6$}}   & mp-9812          & \multicolumn{1}{r|}{0.054}              & \cellcolor[HTML]{FFFFFF}10.398             & \multicolumn{1}{r|}{\cellcolor[HTML]{FFFFFF}\textbf{0.008}} & \cellcolor[HTML]{FFFFFF}\textbf{3.995} & \multicolumn{1}{r|}{\cellcolor[HTML]{FFFFFF}0.888}          & \cellcolor[HTML]{FFFFFF}11.537         & \multicolumn{1}{r|}{\cellcolor[HTML]{FFFFFF}0.679} & \cellcolor[HTML]{FFFFFF}13.052 \\ \hline
\multicolumn{1}{|l|}{\cellcolor[HTML]{FFFFFF}{\color[HTML]{0D0D0D} ZnCdPt$_2$}}   & mp-30493         & \multicolumn{1}{r|}{\textbf{0.008}}     & \cellcolor[HTML]{FFFFFF}\textbf{0.134}     & \multicolumn{1}{r|}{\cellcolor[HTML]{FFFFFF}0.086}          & \cellcolor[HTML]{FFFFFF}8.328          & \multicolumn{1}{r|}{\cellcolor[HTML]{FFFFFF}0.010}          & \cellcolor[HTML]{FFFFFF}2.038          & \multicolumn{1}{r|}{\cellcolor[HTML]{FFFFFF}1.212}          & \cellcolor[HTML]{FFFFFF}10.743          \\ \hline
\multicolumn{1}{|l|}{\cellcolor[HTML]{FFFFFF}{\color[HTML]{0D0D0D} ZrHg}}      & mp-2510          & \multicolumn{1}{r|}{0.010}              & \cellcolor[HTML]{FFFFFF}4.172              & \multicolumn{1}{r|}{\cellcolor[HTML]{FFFFFF}\textbf{0.004}} & \cellcolor[HTML]{FFFFFF}\textbf{0.463} & \multicolumn{1}{r|}{\cellcolor[HTML]{FFFFFF}0.016}          & \cellcolor[HTML]{FFFFFF}4.428          & \multicolumn{1}{r|}{\cellcolor[HTML]{FFFFFF}1.719}          & \cellcolor[HTML]{FFFFFF}7.787           \\ \hline
\multicolumn{2}{|c|}{\textbf{\# of the best}}                                                         & \multicolumn{1}{l|}{\textbf{12}}        & \multicolumn{1}{l|}{5}                     & \multicolumn{1}{l|}{7}                                      & \multicolumn{1}{l|}{\textbf{11}}       & \multicolumn{1}{l|}{7}                                      & \multicolumn{1}{l|}{7}                 & \multicolumn{1}{l|}{0}                                      & \multicolumn{1}{l|}{0}                  \\ \hline
\end{tabular}
\label{tab:dftresults}
\end{table}

\begin{figure}[tbh!]
  \centering
  \includegraphics[width=0.65\linewidth]{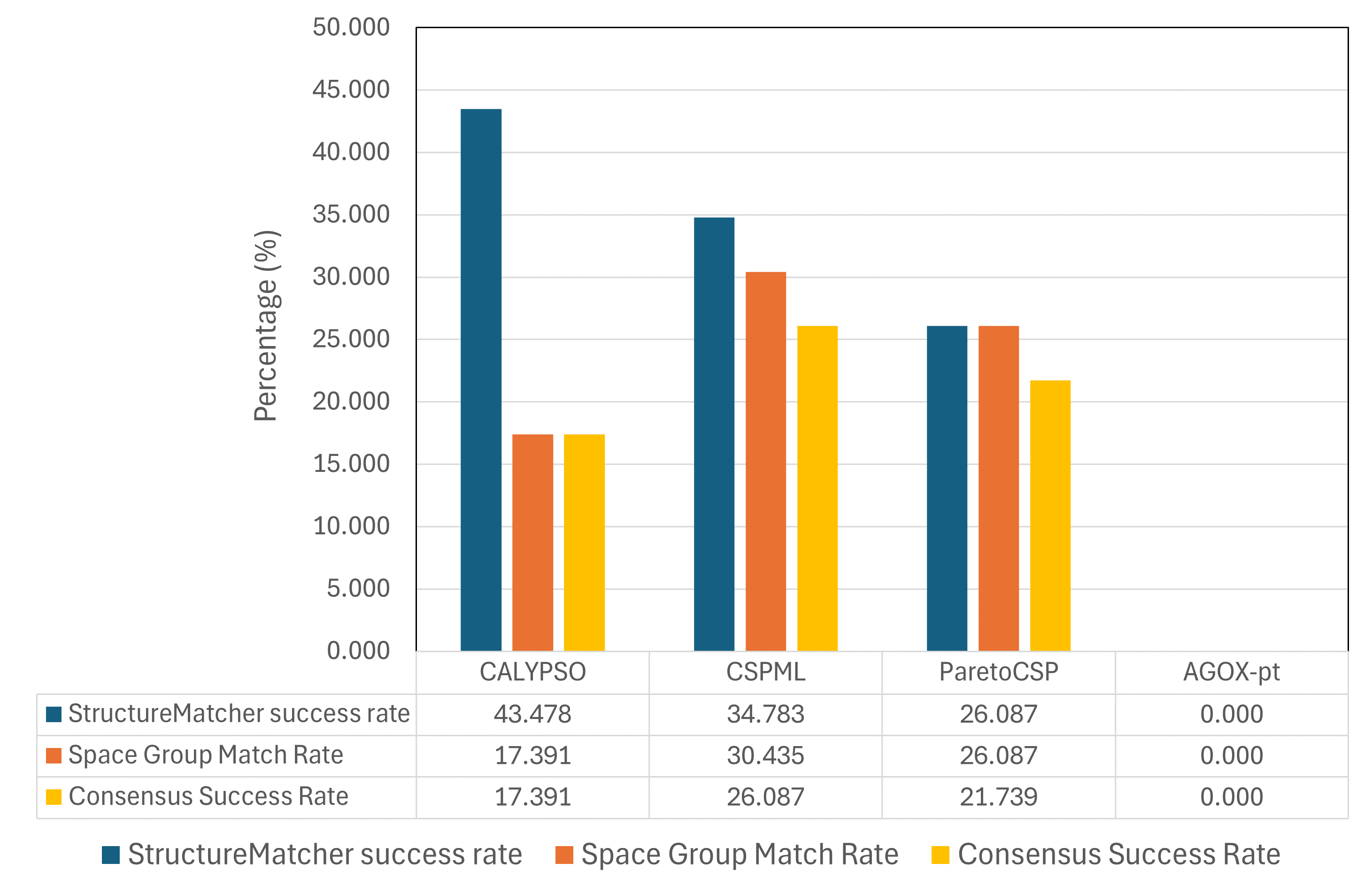}
  \caption{Comparison of structure prediction performance in terms of the success rate measured as the percentage of the predicted structures (out of 23 test samples) that match the ground truth structures and have identical space groups.}
  \label{fig:succ-23}
\end{figure}

\subsection{Performance comparison of non-DFT based CSP algorithms against DFT-based CALYPSO}

Due to the extremely demanding computational resources, it is not feasible to evaluate the DFT-based CALYPSO algorithm over all 180 test samples. Most test samples are too complex for CALYPSO to predict their structures accurately. Therefore, we selected a subset that includes 13 binary structures and 10 ternary structures for evaluating the DFT-based algorithm and compared its performance with those of non-DFT based CSP algorithms. The test set includes NaGa$_4$, Ti$_2$Cd, Y$_3$Al$_9$, ZrHg, YHg$_2$, TiGa$_3$, SrGa$_4$, ScCu, InHg, Hf$_4$Mn$_8$, Hf$_4$Ni$_2$, Cr$_6$Ga$_2$, Co$_2$Te$_2$, Zn$_2$C$_2$O$_6$, Ca$_3$SnO, ZnCdPt$_2$, SrGaCu$_2$, MgInCu$_4$, MgCu$_4$Sn, LiMg$_2$Ga, Li$_2$CuSn, HfCo$_2$Sn, Co$_2$Ni$_2$Sn$_2$. We chose M3GNet energy distance (ED) and Hausdorff distance (HD) as the evaluation metrics to compare the performances of CALYOSO, CSPML, ParetoCSP, and AGOX-pt. Notably, two structures Hf$_4$Mn$_8$ and Y$_3$Al$_9$ are categorized by polymorphy; therefore, there are two ground truth structures (mp-2451 and mp-11231) for Y$_3$Al$_9$. 
The comparison results are shown in Table \ref{tab:dftresults}. We find that CALYPSO achieves the lowest M3GNet energy distances for 12 out of 23 test samples, ranging from 0.000 to 0.070 (eV/atom). This includes 8 binary structures and 4 ternary structures. It also records the smallest Hausdorff distance for 5 out of 23 test samples. This demonstrates the superiority of the de novo CSP algorithm with DFT energy calculation for this small scale test set and its ability to find lower energy structures using DFT energy calculation. However, non-DFT based algorithms like CSPML and ParetoCSP show competitive performance as well, both achieving the lowest energy distance for 7 test samples. CSPML achieves the best performance with the lowest HD distance for 11 out of 23 samples, reflecting the effectiveness of the template-based CSP algorithm in identifying ground truth structures by finding similar template structures. Although CALYPSO has better performance on ED by utilizing DFT calculations to find structures with lower energies, CSPML achieves high performance in both ED and HD for many ternary structures, such as Ca$_3$SnO, Co$_2$Ni$_2$Sn$_2$, HfCo$_2$Sn, LiMg$_2$Ga, MgInCu$_4$, Zn$_2$C$_2$O$_6$.
Among the remaining two de novo CSP algorithms, ParetoCSP significantly outperformed AGOX-pt, achieving the best ED and HD for 7 out of 23 test samples each. In contrast, AGOX-pt did not achieve the best score in any of the test samples for ED and HD. All the HD scores of AGOX-pt are large, ranging from 6.864 Å to 21.138 Å, indicating its weakness in predicting structures with similar geometry to the ground truth.

The CSP performance comparison results need to be interpreted holistically. Good performance with a single metric can be misleading as shown in Figure \ref{fig:succ-23}, we calculate three types of success rates for CSP prediction including StructureMatcher success rate, Space group match rate, and consensus match rate (both StructureMatcher and space groups need to be matched between predicted structures and the ground truths).
CALYPSO achieves the best performance of 43.478\% on StructureMatcher success rate. However, its space group match rate is lower than those of CSPML and ParetoCSP, leading to its relatively low consensus success rate of 17.391\%. In contrast, the consensus success rates of CSPML and ParetoCSP are 26.087\% and 21.739\%, respectively, which can be attributed to their higher space group match rates. Overall, the template-based CSP algorithm CSPML shows the best performance while the de novo CSP algorithm ParetoCSP achieves the competitive performance in terms of the consensus success rate. The poor performance for AGOX-pt shows its inability to accurately predict the structures for given compositions. It should be noted that space group determination is dependent on the parameter setting adopted, which may change the space group success rate results in Figure \ref{fig:succ-23}. Here the default parameters of the space group analyzer of Pymatgen are used. We also noted that the predicted structures with incorrect space groups may be fine-tuned into ones with correct space groups using DFT-based relaxation procedures.

\newpage
\subsection{Case studies}

Our benchmark results have demonstrated the limited prediction capability of current computational CSP algorithms, including both template-based and de novo algorithms. Especially, most de novo CSP algorithms cannot accurately predict the space groups for the majority of test samples. To further understand the success and failure cases of different CSP algorithms and how the performance metric scores correlate with the predicted structures, we present three case studies for ErCo$_5$, Ca$_3$SnO, KAsIO$_6$. Additionally, two more case studies, ZnCdPt$_2$ and Ga$_2$Cu, are provided in the supplementary file Figure S5, S6 and Table S8, S9 for further reference.

\begin{figure}[!htb]
  \includegraphics[width=1\linewidth]{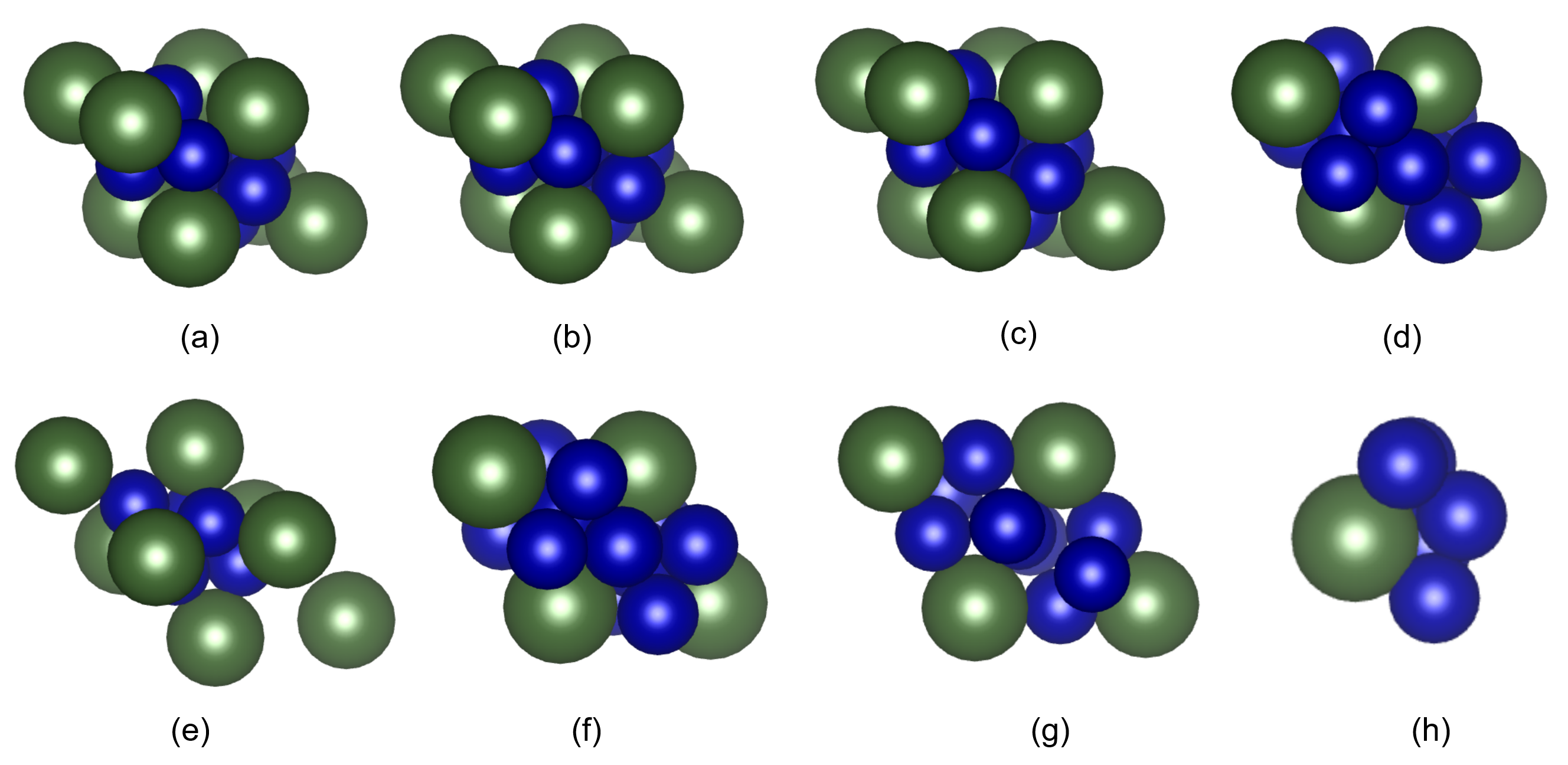}
  \caption{Comparison of the ground truth and predicted crystal structures of ErCo$_5$. (a). Ground truth structure (b). Predicted by ParetoCSP (c). Predicted by TCSP (d). Predicted by AlphaCrystal-II (e). Predicted by CSPML (f). Predicted by GNOA-M3GNet-PSO (g). Predicted by GNOA-M3GNet-RAS (h). Predicted by AGOX-rss}
  \label{fig:ErCo5}
\end{figure}

First, we compared the prediction structures of seven algorithms for ErCo$_5$ as shown in Figure \ref{fig:ErCo5} and Table \ref{tab:ErO5}. Out of the seven algorithms, two achieved successful predictions: ParetoCSP (Figure \ref{fig:ErCo5}(b)) and TCSP (Figure \ref{fig:ErCo5}(c)). These predicted structures closely match the ground truth structure of ErCo$_5$ obtained from the Material Project database, as reflected by the small distance scores in Table \ref{tab:ErO5} (first two rows). The structures predicted by AlphaCrystal-II (Figure \ref{fig:ErCo5}(d)) and GNOA-M3GNet-PSO (Figure \ref{fig:ErCo5}(f))  are close to the target structure and have a formation energy distance of 0.000 eV/atom. However, their Sinkhorn distance, Chamfer distance and superpose rmsd are much higher than those structures predicted by ParentoCSP and TCSP, highlighting the importance to interpret the structure similarity using a comprehensive set of criteria. On the other hand, the structures predicted by CSPML (Figure \ref{fig:ErCo5}(e)), GNOA-MEGNet-RAS (Figure \ref{fig:ErCo5}(g)) and AGOX-rss (Figure \ref{fig:ErCo5}(h)) have high energy distance score as well as other distance metrics values, showing that a large energy distance score can be used as an indicator of low quality predictions. Notably, AGOX-rss displays much higher distance scores across all metrics compared to the other algorithms, indicating its poor performance in predicting the structure of ErCo$_5$. This case, along with the benchmark studies,  underscores the importance of using a set of quantitative criteria to comprehensively evaluate the performance of CSP algorithms.

\begin{table}[tbh]
\caption{The seven distance metrics between the ground truth structure of ErCo$_5$ and the predicted structures by ParetoCSP, TCSP, AlphaCrystal-II, CSPML, GNOA-M3GNet-PSO, GNOA-M3GNet-RAS, and AGOX-rss algorithms. These include M3GNet Energy Distance (eV/atom); Wyckoff RMSE, Chamfer Distance, Hausdorff Distance, Superpose RMSD, and Fingerprint Distance (all in Å); XRD Spectrum Distance (counts/sec); and OFM Distance (measured in valence electrons).} 
\centering
\begin{tabular}{|l|l|l|l|l|l|l|l|}
\hline
\textbf{Algorithm}       & \textbf{\begin{tabular}[c]{@{}l@{}}M3gnet \\ Energy \\ Distance\end{tabular}} & \textbf{\begin{tabular}[c]{@{}l@{}}Sinkhorn \\ Distance\end{tabular}} & \textbf{\begin{tabular}[c]{@{}l@{}}Chamfer \\ Distance\end{tabular}} & \textbf{\begin{tabular}[c]{@{}l@{}}Superpose \\ RMSD\end{tabular}} & \textbf{\begin{tabular}[c]{@{}l@{}}Fingerprint \\ Distance\end{tabular}} & \textbf{\begin{tabular}[c]{@{}l@{}}XRD \\ Distance\end{tabular}} & \textbf{\begin{tabular}[c]{@{}l@{}}OFM \\ Distance\end{tabular}} \\ \hline
\textbf{ParetoCSP}       & \textbf{0.000}                                                                & \textbf{4.713}                                                        & 0.814                                                       & \textbf{0.838}                                                     & 0.011                                                           & 0.163                                                   & 0.007                                                   \\ \hline
\textbf{TCSP}            & \textbf{0.000}                                                                & 4.744                                                                 & \textbf{0.813}                                                       & \textbf{0.838}                                                     & \textbf{0.007}                                                           & 0.170                                                            & \textbf{0.006}                                                   \\ \hline
\textbf{AlphaCrystal-II} & \textbf{0.000}                                                                & 9.036                                                                 & 3.012                                                                & 1.178                                                     & 0.016                                                                    & 0.240                                                            & 0.011                                                            \\ \hline
\textbf{CSPML}           & 0.097                                                                & 24.634                                                                & 5.270                                                                & 1.791                                                              & 1.492                                                                    & 2.060                                                   & 0.350                                                            \\ \hline
\textbf{GNOA-M3GNet-PSO} & \textbf{0.000}                                                                         & 9.037                                                                 & 3.012                                                                & 1.178                                                              & 0.008                                                                    & \textbf{0.138}                                                            & \textbf{0.006}                                                   \\ \hline
\textbf{GNOA-MEGNet-RAS} & 2.215                                                                         & 15.440                                                                & 4.189                                                                & 1.694                                                              & 2.148                                                                    & 1.686                                                            & 1.831                                                            \\ \hline
\textbf{AGOX-rss}        & 1.880                                                                         & 85.596                                                                & 26.181                                                               & 14.652                                                             & 1.950                                                                    & 1.650                                                            & 2.282                                                            \\ \hline
\end{tabular}
\label{tab:ErO5}
\end{table}

\begin{figure}[!htb]
  \includegraphics[width=1\linewidth]{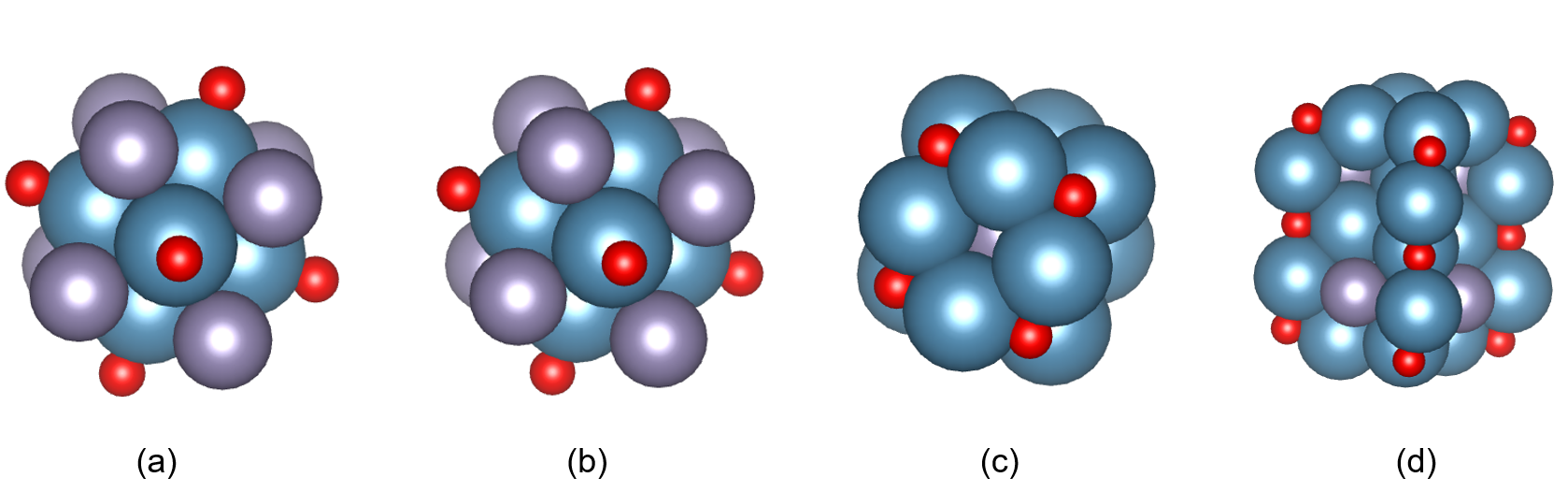}
  \caption{Comparison of the ground truth and predicted crystal structures of Ca$_3$SnO. (a). Ground truth structure (b). Predicted by CSPML(c). Predicted by CALYPSO (d). Predicted by TCSP}
  \label{fig:Ca3SnO}
\end{figure}

Next, we choose the ternary structure to compare the performance for DFT-based CSP algorithm CALYPSO with two template-based CSP algorithms, CSPML and TCSP. Figure \ref{fig:Ca3SnO} shows that the structure of Ca$_3$SnO predicted by the CSPML is more similar than those predicted by CALYPSO and TCSP. We find that the formation energy distances are small for CSPML, CALYPSO and TCSP. However, CSPML has much lower geometric distances, including a Sinkhorn distance of 0.071 Å, a Chamfer distance of 0.029 Å and a superpose RMSD of 0.010 Å. Additionally, its fingerprint score of 0.000 is much better than those predicted by CALYPSO and TCSP, which have scores of 0.116 and 1.172, respectively, indicating the consistency of the different types of CSP metrics used in this study. This consistency also applies to the OFM performance metric. However, it is recognized that the XRD distance of the prediction by CSPML is worse than the XRD score of the prediction by CALYPSO, indicating that the XRD distance alone is not a reliable metric for CSP prediction evaluation.

\begin{table}[htb!]
\caption{The metrics for seven distances of Ca$_3$SnO as evaluated by the CSPML, CALYPSO, and TCSP. This includes M3GNet Energy Distance (eV); Wyckoff RMSE, Chamfer Distance, Superpose RMSD, and FingerPrint Distance (all in Å); XRD Spectrum Distance (counts/sec); and OFM Distance (measured in valence electrons).}
\centering
\begin{tabular}{|l|l|l|l|l|l|l|l|}
\hline
\textbf{Algorithm} & \textbf{\begin{tabular}[c]{@{}l@{}}M3gnet \\ Energy \\ Distance\end{tabular}} & \textbf{\begin{tabular}[c]{@{}l@{}}Sinkhorn \\ Distance\end{tabular}} & \textbf{\begin{tabular}[c]{@{}l@{}}Chamfer \\ Distance\end{tabular}} & \textbf{\begin{tabular}[c]{@{}l@{}}Superpose \\ RMSD\end{tabular}} & \textbf{\begin{tabular}[c]{@{}l@{}}Fingerprint \\ Distance\end{tabular}} & \textbf{\begin{tabular}[c]{@{}l@{}}XRD \\ Distance\end{tabular}} & \textbf{\begin{tabular}[c]{@{}l@{}}OFM \\ Distance\end{tabular}} \\ \hline
\textbf{CSPML}     & \textbf{0.001}                                                                & \textbf{0.071}                                                        & \textbf{0.029}                                                       & \textbf{0.010}                                                     & \textbf{0.000}                                                           & 0.992                                                            & \textbf{0.007}                                                   \\ \hline
\textbf{CALPSO}    & 0.002                                                                         & 7.946                                                                 & 2.899                                                                & 1.614                                                              & 0.116                                                                    & \textbf{0.943}                                                   & 0.025                                                            \\ \hline
\textbf{TCSP}      & 0.007                                                                         & 80.852                                                                & 5.119                                                                & 1.125                                                              & 1.172                                                                    & 1.052                                                            & 0.176                                                            \\ \hline
\end{tabular}
\label{tab:Ca3Sn0}
\end{table}

\begin{figure}[!htb]
  \includegraphics[width=1\linewidth]{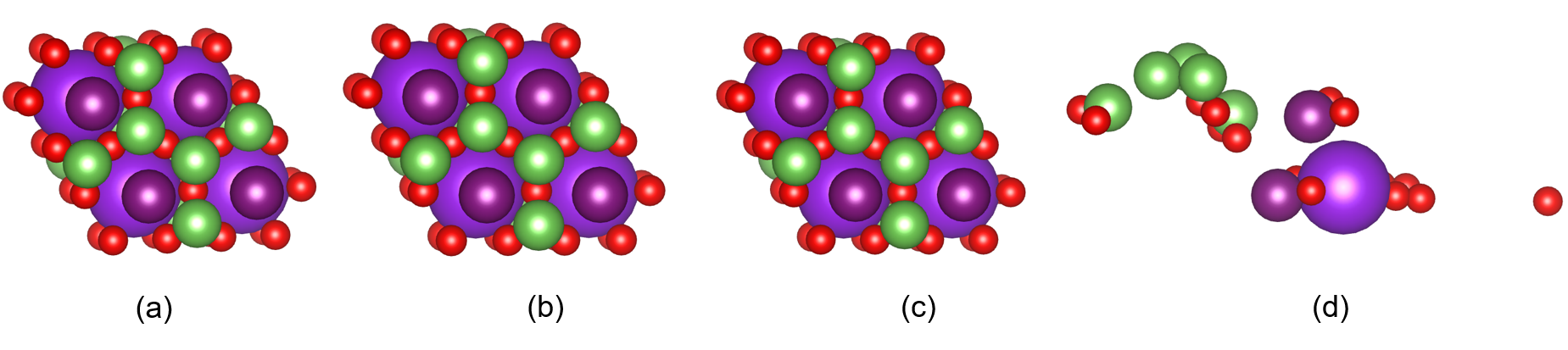}
  \caption{Comparison of the ground truth and predicted crystal structures of KAs$_4$IO$_6$Cu. (a). Ground truth structure (b). Predicted by CSPML (c). Predicted by TCSP (d). Predicted by ParetoCSP}
  \label{fig:KAs4IO6}
\end{figure}

To further understand the advantages of different algorithms, we examined the case of quaternary structure prediction for KAs$_4$IO$_6$Cu, in which the template-based methods work well while the de novo methods fail (Figure \ref{fig:KAs4IO6}. Both CSPML and TCSP produced reasonable structures but CSPML's prediction overall is better despite TCSP's result having a slightly lower XRD distance value (1.039 compared to 1.040) (See Table \ref{tab:KAs4IO6}). In contrast, the structure predicted by ParetoCSP is significantly worse across all performance metrics. The much higher Sinkhorn distance of 207.028 Å, a Chamfer distance of 23.239 Å, and a superpose RMSD of 20.014 Å indicate poor geometric similarity of the predicted structure by ParetoCSP to the ground truth structure (Figure \ref{fig}(d)).

\begin{table}[htb!]
\caption{The metrics for seven distances of KAs$_4$IO$_6$ as evaluated by the TCSP, CSPML and ParetoCSP. This includes M3GNet Energy Distance (eV/atom); Sinkhorn Distance, Chamfer Distance, Superpose RMSD, and Fingerprint Distance (all in Å); XRD Spectrum Distance (counts/sec); and OFM Distance (measured in valence electrons).}
\centering
\begin{tabular}{|l|l|l|l|l|l|l|l|}
\hline
\textbf{Algorithm} & \textbf{\begin{tabular}[c]{@{}l@{}}M3GNet \\ Energy \\ Distance\end{tabular}} & \textbf{\begin{tabular}[c]{@{}l@{}}Sinkhorn \\ Distance\end{tabular}} & \textbf{\begin{tabular}[c]{@{}l@{}}Chamfer \\ Distance\end{tabular}} & \textbf{\begin{tabular}[c]{@{}l@{}}Superpose \\ RMSD\end{tabular}} & \textbf{\begin{tabular}[c]{@{}l@{}}fingerPrint \\ Distance\end{tabular}} & \textbf{\begin{tabular}[c]{@{}l@{}}XRD \\ Distance\end{tabular}} & \textbf{\begin{tabular}[c]{@{}l@{}}OFM \\ Distance\end{tabular}} \\ \hline
\textbf{TCSP}      & \textbf{0.000}                                                                & \textbf{1.403}                                                        & \textbf{0.234}                                                       & \textbf{0.037}                                                     & \textbf{0.143}                                                           & \textbf{1.039}                                                   & \textbf{0.025}                                                   \\ \hline
\textbf{CSPML}     & \textbf{0.000}                                                                & \textbf{1.403}                                                        & \textbf{0.234}                                                       & \textbf{0.037}                                                     & \textbf{0.143}                                                           & 1.040                                                   & \textbf{0.025}                                                   \\ \hline
\textbf{ParetoCSP} & 0.705                                                                & 207.028                                                               & 23.239                                                               & 20.014                                                             & 2.254                                                                    & 2.529                                                            & 0.521                                                            \\ \hline
\end{tabular}
\label{tab:KAs4IO6}
\end{table}

\section{Discussion}

Objective and accurate evaluation and comparison of different CSP algorithms are nontrivial due to the complexity of structure comparison and the inherent symmetry of crystal structures plus the possible polymorphism of a given test structure. Here we show several aspects that need special attention to accurately interpret the evaluation results and issues that may arise during CSP algorithm performance evaluation. 

\paragraph{Polymorphism test samples:}
In our benchmark set, there are several test structures that have alternative structures with the same composition due to structural polymorphism. For these test structures, the predicted structure of a CSP algorithm is compared to each of the polymorphism ground truth structures and the one with the smallest distance is selected to calculate the distance error. Note that in this benchmark study, we only consider the top-1 prediction performance.

\paragraph{Ranking score bias:}

We would like to point out that we need to cautiously interpret the rankings of different CSP algorithms sorted by our ranking scores as shown in Figure \ref{fig:m3gnet_score}. For example, the AGOX series algorithms have shown better rankings than the GNOA series of algorithms in Figure 3 to Figure 8 while in Figure \ref{fig:succ-180}, the AGOX algorithms have zero success rate according to the StructureMatcher criterion while GNOA algorithms have successfully predicted several test structures. This discrepancy is due to the fact that our ranking score penalizes those algorithms that cannot predict any valid structures. In our case here, the GNOA cannot find any valid structures for quite many test structures, leading to their low ranks. 

\paragraph{Cautious interpretation of StructureMatcher results:}

Several studies have used the Pymatgen's StructureMatcher to check if two structures are identical \cite{xie2021crystal} and \cite{luo2024deep}. In the CDVAE experiments, two structures are deemed as identical if StructureMatcher returns true with the following parameters ltol=0.3, stol=0.5, angle\_tol=10. In a recent study \cite{luo2024deep}, a more stringen parameter settings of ltol=0.2, stol=0.3, angle\_tol=5 are used to check if two structures are identical. However, we find that there are quite many cases StructureMatcher reports two structures to be identical while their space group numbers are not even equal (See Supplementary Table S10 and S11 for examples). It is thus critical to check the space groups even the StructureMatcher reports as identical despite that the space group determination is also based on a given set of parameters (usually using the default values).

\paragraph{CrystalNN Fingerprint dependent structure similarity:}

Another structure fingerprint distance (CrystalNNFingerprint) based structure identity checking method was also used in a CSP study \cite{liu2023shotgun}. To calculate the similarity between two structures i and j, the given structures are first encoded into a vector-type structural descriptors with their local coordination information (site fingerprint) from all sites. Then, the structure similarity $\tau$ was calculated as the Euclidean distance between the crystal structure descriptors. Structures with dissimilarity $\tau \leq 0.2$ were treated as similar structures in their study \cite{liu2023shotgun}. However we find this measurement has limitations, as structure pairs with fingerprint distance less than the threshold 0.2 can also have different space groups (See Supplementary Figure S8 and Table S12) and a space group check with a set of specific parameters or default parameters is needed, similar to the case of using Pymatgen StructureMatcher. 

\section{Conclusion}

Crystal structure prediction plays a crucial role in discovering novel function materials. However, conventional first principle based CSP algorithms currently have limited capability to predict complex crystal structures. Here we conduct a comprehensive benchmark study of 13 CSP algorithms covering template-based, ML potential based, contact map based, and DFT-based CSP algorithms, aiming to illustrate the potential and performance gaps of modern ML potential and deep learning based CSP as well as the widely used template-based CSPs in terms of scalability and accuracy. The algorithms are evaluated over 180 well-selected test set comprising of binary, ternary, and quarternary crystal structures with diverse symmetries and the numbers of atoms in unit cells. All the algorithm performances are calculated using a set of quantitative metrics along with their relative ranking scores over 180 test structures, making it possible to achieve relatively objective performance comparisons. 

Our extensive benchmark experiments have uncovered several performance trends and factors that contribute to better CSP performances. First, we find that template-based CSP algorithms can achieve strong performance when a suitable template structure can be found, which is due to the ubiquitous existence of typical structural prototypes \cite{griesemer2021high,mehl2017aflow} and the wide applications of such elemental substitution CSP algorithms in discovering a large number of hypothetical materials \cite{chen2022universal,merchant2023scaling}. However, such template-based CSP algorithms cannot be used to discover materials with novel structural prototypes. Next, it is observed that the machine learning potentials based CSP algorithms have made significant progress in the past few years, leading to competitive algorithms for CSP, especially for those without good templates. For these algorithms, their performances strongly depend on the global search capability of the search algorithms and the quality of the ML potentials. For example, with the same M3GNet potential, the ParetoCSP is better than the AGOX algorithms and also outperforms GNOA algorithms in terms of quantitative metrics due to its enhanced search capability. A further comparison of the ML potential based CSP algorithms with the DFT-based CALYPSO shows that the former class of modern CSP algorithms have demonstrated strong performance and outperform the later one for most of the test samples. Even for the relatively simple crystal structure test sets, the template-based algorithms and ML-potential based CSP algorithms both showed better performance than DFT-based CALYPSO in terms of the formation energy distance and Hausdorff distance. However, our benchmark results also showed all current de novo CSP algorithms (non-template methods) are still in an early stage of development: most of them cannot even accurately predict the space group or crystal systems for a majority of the 180 test samples. Our evaluation of DFT-based CALYPSO also showed the lack of scalability for such DFT-based de novo CSP algorithms as it is almost infeasible to complete predictions for all 180 test samples (It should noted that modern CALYPSO has also incorporated the ML potential for more scable CSP). However, DFT-based de novo CSP algorithms have their unique advantage in predicting crystal structures within special conditions such as high-pressure for which there is currently no ML potential for such condition. It should be noted that due to the subjectivity of selecting the 180 test samples, our evaluation has inherent bias despite our effort in trying to cover diverse structures, symmetries, and structural prototypes. So the rankings of different algorithms should not be used to judge the superiority of any algorithm but just should be used to guide the application of appropriate algorithms based on the application scenario. For example, for a given composition, it is reasonable to predict its structure first by using the template-based algorithms such as CSPML and ML potential based de novo algorithms such as ParetoCSP or GNOA and check their formation energy, E-above-hull energy, and mechanical stability. If still not satisfactory, one can then try the DFT-based de novo methods such as CALYPSO if the composition is not too complex.

Overall, our benchmark has demonstrated the significant progress of machine learning potential based CSP algorithms and its promising prospect to achieve scalable CSP due to the emergence of better search algorithms and modern machine learning potentials. Our benchmark data and quantitative performance evaluation metrics, the open-sourced codes of such CSP algorithms and their independence of DFT calculations thus paved the way to allow researchers from a wide variety of researchers from the communities of AI, data science, statistics to explore this promising and significant CSP problem.

\section{Data and Code Availability}
The 180 test structures are obtained from the Materials Project database. Their mp-ids are available from our Github repository \url{https://github.com/usccolumbia/cspbenchmark}. 
The code for calculating ranking scores can be also downloaded from the Github repository. The performance metrics calculation is done using the code from the CSPBenchMetrics repository \url{https://github.com/usccolumbia/CSPBenchMetrics}. The open source CSP codes are available from their corresponding websites as shown in Table 1. in the main text. We have modified AGOX and GNOA to integrate them with the neural network potential model M3GNet. 

\section{Contribution}
Conceptualization, J.H.; methodology,J.H. L. W., S. O., R.D., N.F., Y.S.,E.S., M.X. ; software, L.W., S.O.,Y.S.; resources, J.H.; writing--original draft preparation, J.H., L.W., S.O., N.F., R.D.,E.S.,M.X.; writing--review and editing, J.H., R.D., S.O., N.F.; visualization, L.W.,S.O.; supervision, J.H.;  funding acquisition, J.H.

\section*{Acknowledgement}
We would like to thank the helpful discussion and suggestions of Prof. Yanchao Wang of Jilin University. The research reported in this work was supported in part by National Science Foundation under the grant and 2110033, OAC-2311203, and 2320292. The views, perspectives, and content do not necessarily represent the official views of the NSF.

\bibliographystyle{unsrt}  
\bibliography{references}

\end{document}


\section*{\Large Supplementary information for}
\section *{CSPBench: a benchmark and critical evaluation of Crystal Structure Prediction}

Lai Wei, Sadman Sadeed Omee, Rongzhi Dong, Nihang Fu, Yuqi Song, Edirisuriya M. D. Siriwardane, Meiling Xu, Chirs Wolverton, Jianjun Hu *

\section{The CSP180 Test set} \label{subsec:testSet}
We ran all the algorithms over the $180$ test samples of our benchmark dataset of binary, ternary and quaternary structures.

\renewcommand{\arraystretch}{1.4}
\begin{longtable}{l l l l l l}
\caption{\textbf{Details of the 180 benchmark crystals except the binary\_easy, binary\_medium and binary\_hard crystals used in this work. Each category of crystals is separated by a single horizontal line.}}
\label{table:dataset}\\
\hline\hline
\textbf{Material id} & \multicolumn{1}{l}{\begin{tabular}[c]{@{}c@{}}\textbf{Primitive formula}\end{tabular}} & \textbf{Space group} & \multicolumn{1}{l}{\begin{tabular}[c]{@{}c@{}}\textbf{Crystal system}\end{tabular}} & \multicolumn{1}{l}{\begin{tabular}[c]{@{}c@{}}\textbf{Category}\end{tabular}}\\ 
\hline\hline
mp-11390     & LiGaSi             & 216        & Cubic         & ternary\_easy   \\
mp-4552      & ErInCu$_2$            & 225        & Cubic         & ternary\_easy   \\
mp-20730     & HfCo$_2$Sn            & 225        & Cubic         & ternary\_easy   \\
mp-21211     & InFeCo$_2$            & 225        & Cubic         & ternary\_easy   \\
mp-30591     & Li$_2$CuSn            & 216        & Cubic         & ternary\_easy   \\
mp-30648     & LiMg$_2$Ga            & 225        & Cubic         & ternary\_easy   \\
mp-4972      & LuInCu$_2$            & 225        & Cubic         & ternary\_easy   \\
mp-5181      & LuSnPd$_2$            & 225        & Cubic         & ternary\_easy   \\
mp-20389     & Na$_2$CdPb            & 216        & Cubic         & ternary\_easy   \\
mp-30555     & TaGaCo$_2$            & 225        & Cubic         & ternary\_easy   \\
mp-29241     & Ca$_3$SnO             & 221        & Cubic         & ternary\_easy   \\
mp-20237     & Co$_2$Ni$_2$Sn$_2$          & 194        & Hexagonal     & ternary\_easy   \\
mp-4326      & KErS$_2$              & 166        & Trigonal      & ternary\_medium \\
mp-30580     & SrGaCu$_2$            & 166        & Trigonal      & ternary\_medium \\
mp-30493     & ZnCdPt$_2$            & 123        & Tetragonal    & ternary\_medium \\
mp-9570      & CaCd$_2$P$_2$            & 164        & Trigonal      & ternary\_medium \\
mp-5452      & CeCu$_2$Si$_2$           & 139        & Tetragonal    & ternary\_medium \\
mp-3147      & ErSi$_2$Au$_2$           & 139        & Tetragonal    & ternary\_medium \\
mp-13405     & LuMn$_2$Ge$_2$           & 139        & Tetragonal    & ternary\_medium \\
mp-30805     & SrNiSn$_3$            & 107        & Tetragonal    & ternary\_medium \\
mp-5615      & Ca$_3$Ag$_3$As$_3$          & 189        & Hexagonal     & ternary\_medium \\
mp-30733     & Ho$_3$Sn$_3$Pt$_3$          & 189        & Hexagonal     & ternary\_medium \\
mp-16747     & Lu$_3$Ag$_3$Pb$_3$          & 189        & Hexagonal     & ternary\_medium \\
mp-9812      & Zn$_2$C$_2$O$_6$            & 167        & Trigonal      & ternary\_medium \\
mp-12743     & CrTe$_4$Au            & 10         & Monoclinic    & ternary\_hard   \\
mp-3676      & MgCu$_4$Sn            & 216        & Cubic         & ternary\_hard   \\
mp-30587     & MgInCu$_4$            & 216        & Cubic         & ternary\_hard   \\
mp-7524      & Nb$_2$P$_2$Se$_2$           & 71         & Orthorhombic  & ternary\_hard   \\
mp-11533     & ZrZnNi$_4$            & 216        & Cubic         & ternary\_hard   \\
mp-11435     & ErTi$_2$Ga$_4$           & 139        & Tetragonal    & ternary\_hard   \\
mp-7095      & Ce$_2$Co$_2$Si$_4$          & 63         & Orthorhombic  & ternary\_hard   \\
mp-7489      & CsUF$_6$              & 148        & Trigonal      & ternary\_hard   \\
mp-11396     & Nd$_2$Ga$_4$Ni$_2$          & 65         & Orthorhombic  & ternary\_hard   \\
mp-29225     & Al$_4$Cu$_2$O$_7$           & 216        & Cubic         & ternary\_hard   \\
mp-23520     & In$_2$Pb$_4$I$_{10}$          & 140        & Tetragonal    & ternary\_hard   \\
mp-19140     & K$_3$MnO$_4$             & 121        & Tetragonal    & ternary\_hard  \\
\hline

mp-1213726   & CrFeCoSi           & 216        & Cubic         & quaternary\_easy   \\
mp-7554      & LiMgSnAu           & 216        & Cubic         & quaternary\_easy   \\
mp-11806     & LiMgSnPt           & 216        & Cubic         & quaternary\_easy   \\
mp-552674    & ZrTaNO             & 187        & Hexagonal     & quaternary\_easy   \\
mp-1071272   & CeAl$_2$BRu$_2$          & 123        & Tetragonal    & quaternary\_easy   \\
mp-13287     & Ba$_2$Cu$_2$Te$_2$F$_2$        & 129        & Tetragonal    & quaternary\_easy   \\
mp-12444     & Sr$_2$Cu$_2$S$_2$F$_2$         & 129        & Tetragonal    & quaternary\_easy   \\
mp-19093     & Ba$_2$UNiO$_6$           & 225        & Cubic         & quaternary\_easy   \\
mp-1111671   & K$_2$LiInF$_6$           & 225        & Cubic         & quaternary\_easy   \\
mp-6686      & K$_2$NaInF$_6$           & 225        & Cubic         & quaternary\_easy   \\
mp-16307     & Sr$_2$MgIrO$_6$          & 225        & Cubic         & quaternary\_easy   \\
mp-9705      & Ba$_4$NaB$_3$N$_6$          & 229        & Cubic         & quaternary\_easy   \\
mp-6258      & CeCr$_2$Si$_2$C          & 123        & Tetragonal    & quartenary\_medium \\
mp-6794      & LaB$_2$Rh$_2$C           & 139        & Tetragonal    & quartenary\_medium \\
mp-6140      & PrNi$_2$B$_2$C           & 139        & Tetragonal    & quartenary\_medium \\
mp-546790    & La$_2$Cu$_2$Te$_2$O$_2$        & 129        & Tetragonal    & quartenary\_medium \\
mp-20807     & Sr$_2$Fe$_2$As$_2$F$_2$        & 129        & Tetragonal    & quartenary\_medium \\
mp-6231      & Ba$_2$ErSbO$_6$          & 225        & Cubic         & quartenary\_medium \\
mp-19751     & Ba$_2$YbNbO$_6$          & 225        & Cubic         & quartenary\_medium \\
mp-15888     & Ba$_2$YIrO$_6$           & 225        & Cubic         & quartenary\_medium \\
mp-6428      & Ba$_2$YRuO$_6$           & 225        & Cubic         & quartenary\_medium \\
mp-19274     & BaPrMn$_2$O$_6$          & 123        & Tetragonal    & quartenary\_medium \\
mp-6586      & K$_2$NaAlF$_6$           & 225        & Cubic         & quartenary\_medium \\
mp-24412     & K$_2$NaAlH$_6$           & 225        & Cubic         & quartenary\_medium \\
mp-545788    & Ba$_3$ZnN$_2$O           & 123        & Tetragonal    & quartenary\_hard   \\
mp-18745     & La$_2$Mn$_2$Sb$_2$O$_2$        & 129        & Tetragonal    & quartenary\_hard   \\
mp-12515     & La$_2$Zn$_2$Sb$_2$O$_2$        & 129        & Tetragonal    & quartenary\_hard   \\
mp-20349     & Sm$_2$Fe$_2$As$_2$O$_2$        & 129        & Tetragonal    & quartenary\_hard   \\
mp-19118     & BaNd$_2$CoO5          & 71         & Orthorhombic  & quartenary\_hard   \\
mp-21348     & Ba$_2$EuTaO$_6$          & 87         & Tetragonal    & quartenary\_hard   \\
mp-13356     & Ba$_2$SrTeO$_6$          & 148        & Trigonal      & quartenary\_hard   \\
mp-9666      & K$_4$Na$_2$Ga$_2$P$_4$         & 72         & Orthorhombic  & quartenary\_hard   \\
mp-23126     & KAs$_4$IO$_6$            & 191        & Hexagonal     & quartenary\_hard   \\
mp-14466     & KLi$_6$IrO$_6$           & 166        & Trigonal      & quartenary\_hard   \\
mp-651268    & Fe$_2$Cu$_6$SnS$_8$         & 115        & Tetragonal    & quartenary\_hard   \\
mp-13383     & Sc$_2$Ag$_2$P$_4$Se$_{12}$       & 163        & Trigonal      & quartenary\_hard  \\
\hline
mp-8882      & Ga$_2$P$_2$              & 186                            & Hexagonal     & polymorph\_binary     \\
mp-13181     & LaF$_3$               & 225                            & Cubic         & polymorph\_binary     \\
mp-568382    & Mn$_2$Bi$_2$             & 194                            & Hexagonal     & polymorph\_binary     \\
mp-1004      & Nb$_3$Si              & 221                            & Cubic         & polymorph\_binary     \\
mp-2067      & Th$_2$Ga$_4$             & 141                            & Tetragonal    & polymorph\_binary     \\
mp-2178      & Yb$_2$Ga$_4$             & 194                            & Hexagonal     & polymorph\_binary     \\
mp-11251     & Mg$_6$Au$_2$             & 194                            & Hexagonal     & polymorph\_binary     \\
mp-9761      & Re$_2$O$_6$              & 182                            & Hexagonal     & polymorph\_binary     \\
mp-2231      & Sn$_4$S$_4$              & 62                             & Orthorhombic  & polymorph\_binary     \\
mp-11449     & Hf$_4$Mn$_8$             & 194                            & Hexagonal     & polymorph\_binary     \\
mp-2451      & Y$_3$Al$_9$              & 166                            & Trigonal      & polymorph\_binary     \\
mp-476       & Ge$_6$N$_8$              & 227                            & Cubic         & polymorph\_binary     \\
mp-19183     & Li$_2$NiO$_2$            & 71                             & Orthorhombic  & polymorph\_ternary    \\
mp-13171     & Y$_3$Mg$_3$Cu$_3$           & 189                            & Hexagonal     & polymorph\_ternary    \\
mp-19227     & Y$_2$Mn$_2$O$_6$            & 194                            & Hexagonal     & polymorph\_ternary    \\
mp-7550      & Ce$_2$Nb$_2$O$_8$           & 15                             & Monoclinic    & polymorph\_ternary    \\
mp-11658     & Cu$_4$Sn$_2$Se$_6$          & 9                              & Monoclinic    & polymorph\_ternary    \\
mp-23550     & K$_2$Br$_2$F$_8$            & 140                            & Tetragonal    & polymorph\_ternary    \\
mp-12931     & Sm$_2$Ta$_2$O$_8$           & 13                             & Monoclinic    & polymorph\_ternary    \\
mp-19418     & V$_2$Cr$_2$O$_8$            & 63                             & Orthorhombic  & polymorph\_ternary    \\
mp-7913      & Be$_4$Si$_4$N$_8$           & 33                             & Orthorhombic  & polymorph\_ternary    \\
mp-557997    & Ca$_4$Se$_4$O$_{12}$          & 14                             & Monoclinic    & polymorph\_ternary    \\
mp-27450     & K$_4$Cu$_4$Cl$_{12}$          & 14                             & Monoclinic    & polymorph\_ternary    \\
mp-5126      & Zn$_4$S$_4$O$_{16}$           & 62                             & Orthorhombic  & polymorph\_ternary    \\
mp-23127     & Ba$_2$BiSbO$_6$          & 12                             & Monoclinic    & polymorph\_quartenary \\
mp-10731     & Ba$_2$PrSbO$_6$          & 225                            & Cubic         & polymorph\_quartenary \\
mp-9349      & KRb$_2$ScF$_6$           & 225                            & Cubic         & polymorph\_quartenary \\
mp-1080753   & NaPb$_2$IO$_6$           & 225                            & Cubic         & polymorph\_quartenary \\
mp-550722    & Ba$_2$Tl$_2$CuO$_6$         & 139                            & Tetragonal    & polymorph\_quartenary \\
mp-6628      & CsCdN$_3$O$_6$           & 146                            & Trigonal      & polymorph\_quartenary \\
mp-1095437   & Lu$_2$Se$_2$O$_6$F$_2$         & 11                             & Monoclinic    & polymorph\_quartenary \\
mp-726253    & RbLi$_3$S$_2$O$_9$          & 1                              & Triclinic     & polymorph\_quartenary \\
mp-23667     & Rb$_2$P$_2$H$_4$O$_8$          & 43                             & Orthorhombic  & polymorph\_quartenary \\
mp-2233097   & MgV$_4$SnO$_{12}$          & 5                              & Monoclinic    & polymorph\_quartenary \\
mp-1105110   & Sr$_4$Bi$_4$Se$_6$O$_4$        & 14                             & Monoclinic    & polymorph\_quartenary \\
mp-23125     & K$_6$Na$_2$Fe$_2$Cl$_{12}$       & 167                            & Trigonal      & polymorph\_quartenary \\
\hline

mp-1071835   & Dy$_2$Cu$_4$             & 74         & Orthorhombic  & template-based\_binary     \\
mp-1102067   & Pu$_4$Sn$_8$             & 141        & Tetragonal    & template-based\_binary     \\
mp-1102049   & Ta$_3$Be$_9$             & 166        & Trigonal      & template-based\_binary     \\
mp-1102936   & Ta$_8$Fe$_4$             & 193        & Hexagonal     & template-based\_binary     \\
mp-1095641   & Tb$_5$S$_7$              & 12         & Monoclinic    & template-based\_binary     \\
mp-1103888   & YbB$_{12}$              & 225        & Cubic         & template-based\_binary     \\
mp-1105001   & Tm$_6$Pt$_8$             & 148        & Trigonal      & template-based\_binary     \\
mp-1104457   & Eu$_9$Au$_6$             & 148        & Trigonal      & template-based\_binary     \\
mp-1104286   & Ge$_{12}$Rh$_3$            & 152        & Trigonal      & template-based\_binary     \\
mp-1106395   & Pr$_{12}$Ir$_4$            & 62         & Orthorhombic  & template-based\_binary     \\
mp-1105958   & Yb$_{12}$Co$_4$            & 62         & Orthorhombic  & template-based\_binary     \\
mp-1190213   & Re$_4$B$_{16}$             & 194        & Hexagonal     & template-based\_binary     \\
mp-1114630   & Rb$_3$PrCl$_6$           & 225        & Cubic         & template-based\_ternary    \\
mp-1105802   & Ca$_4$Ge$_8$Pt$_4$          & 71         & Orthorhombic  & template-based\_ternary    \\
mp-1106406   & Ce$_4$Sn$_2$S$_{10}$          & 55         & Orthorhombic  & template-based\_ternary    \\
mp-1106349   & Sm$_2$Pd$_6$S$_8$           & 223        & Cubic         & template-based\_ternary    \\
mp-1106327   & Co$_4$NiSb$_{12}$          & 204        & Cubic         & template-based\_ternary    \\
mp-1106064   & Ho$_4$Ga$_{12}$Ni          & 229        & Cubic         & template-based\_ternary    \\
mp-1106196   & Lu$_4$Ga$_{12}$Ni          & 229        & Cubic         & template-based\_ternary    \\
mp-1106117   & La$_{10}$Ag$_2$Pb$_6$         & 193        & Hexagonal     & template-based\_ternary    \\
mp-1106245   & Zr$_{10}$Al$_2$Sb$_6$         & 193        & Hexagonal     & template-based\_ternary    \\
mp-1105955   & Er$_6$Cu$_6$Sb$_8$          & 220        & Cubic         & template-based\_ternary    \\
mp-1105893   & La$_6$Cu$_6$Bi$_8$          & 220        & Cubic         & template-based\_ternary    \\
mp-1105777   & U$_6$Sb$_8$Ir$_6$           & 220        & Cubic         & template-based\_ternary    \\
mp-1111927   & K$_2$LiCrF$_6$           & 225        & Cubic         & template-based\_quartenary \\
mp-1104454   & Ta$_4$GaTe$_4$Se$_4$        & 216        & Cubic         & template-based\_quartenary \\
mp-1106310   & Cu$_4$Si$_2$Hg2S$_8$        & 31         & Orthorhombic  & template-based\_quartenary \\
mp-1105386   & Yb$_2$H$_6$C$_2$N$_6$          & 176        & Hexagonal     & template-based\_quartenary \\
mp-1106402   & Rb$_4$Ti$_2$O$_2$F$_{10}$        & 63         & Orthorhombic  & template-based\_quartenary \\
mp-1106325   & CaCu$_3$Pt$_4$O$_{12}$        & 204        & Cubic         & template-based\_quartenary \\
mp-1106150   & CeMn$_4$Cu$_3$O$_{12}$        & 204        & Cubic         & template-based\_quartenary \\
mp-1106004   & HoFe$_4$Cu$_3$O$_{12}$        & 204        & Cubic         & template-based\_quartenary \\
mp-1106068   & LaCr$_4$Cu$_3$O$_{12}$        & 204        & Cubic         & template-based\_quartenary \\
mp-1105674   & Mg$_4$Be$_2$B$_4$Ir$_{10}$       & 127        & Tetragonal    & template-based\_quartenary \\
mp-1105109   & PrCu$_3$Ru$_4$O$_{12}$        & 204        & Cubic         & template-based\_quartenary \\
mp-1105290   & Co$_3$Sb$_4$O$_6$F$_6$         & 217        & Cubic         & template-based\_quartenary 

\\
\hline\hline

\end{longtable}
\FloatBarrier

\section{The CSP4x-Binary Test set} 
\label{subsec:testSet} 
\renewcommand{\arraystretch}{1.4}
\begin{longtable}{l l l l l l}
\caption{\textbf{Details of the 60 binary benchmark crystals.}}
\label{table:dataset}\\
\hline\hline
\textbf{Material id} & \multicolumn{1}{l}{\begin{tabular}[c]{@{}c@{}}\textbf{Primitive formula}\end{tabular}} & \textbf{Space group} & \multicolumn{1}{l}{\begin{tabular}[c]{@{}c@{}}\textbf{Crystal system}\end{tabular}} & \multicolumn{1}{l}{\begin{tabular}[c]{@{}c@{}}\textbf{Category}\end{tabular}}\\ 
\hline\hline
mp-2334      & DyCu               & 221        & Cubic         & binary\_easy   \\
mp-2226      & DyPd               & 221        & Cubic         & binary\_easy   \\
mp-1121      & GaCo               & 221        & Cubic         & binary\_easy   \\
mp-2735      & PaO                & 225        & Cubic         & binary\_easy   \\
mp-1169      & ScCu               & 221        & Cubic         & binary\_easy   \\
mp-30746     & YIr                & 221        & Cubic         & binary\_easy   \\
mp-24658     & SmH$_2$               & 225        & Cubic         & binary\_easy   \\
mp-20225     & CePb$_3$              & 221        & Cubic         & binary\_easy   \\
mp-788       & Co$_2$Te$_2$             & 194        & Hexagonal     & binary\_easy   \\
mp-20176     & DyPb$_3$              & 221        & Cubic         & binary\_easy   \\
mp-1231      & Cr6Ga$_2$             & 223        & Cubic         & binary\_easy   \\
mp-12570     & ThB$_{12}$             & 225        & Cubic         & binary\_easy   \\
mp-20132     & InHg               & 166        & Trigonal      & binary\_medium \\
mp-2209      & CeGa$_2$              & 191        & Hexagonal     & binary\_medium \\
mp-30497     & TbCd$_2$              & 191        & Hexagonal     & binary\_medium \\
mp-30725     & YHg$_2$               & 191        & Hexagonal     & binary\_medium \\
mp-2731      & TiGa$_3$              & 139        & Tetragonal    & binary\_medium \\
mp-2510      & ZrHg               & 123        & Tetragonal    & binary\_medium \\
mp-2740      & ErCo$_5$               & 191        & Hexagonal     & binary\_medium \\
mp-570875    & Ga$_4$Os$_2$              & 70         & Orthorhombic  & binary\_medium \\
mp-861       & Hf4Ni$_2$             & 140        & Tetragonal    & binary\_medium \\
mp-1566      & SmFe$_5$               & 191        & Hexagonal     & binary\_medium \\
mp-2387      & Th$_4$Zn$_2$             & 140        & Tetragonal    & binary\_medium \\
mp-1607      & YbCu$_5$             & 191        & Hexagonal     & binary\_medium \\
mp-13452     & BePd$_2$              & 139        & Tetragonal    & binary\_hard   \\
mp-11359     & Ga$_2$Cu              & 123        & Tetragonal    & binary\_hard   \\
mp-1995      & PrC$_2$               & 139        & Tetragonal    & binary\_hard   \\
mp-30501     & Ti$_2$Cd              & 139        & Tetragonal    & binary\_hard   \\
mp-30789     & U$_2$Mo               & 139        & Tetragonal    & binary\_hard   \\
mp-454       & NaGa$_4$              & 139        & Tetragonal    & binary\_hard   \\
mp-1827      & SrGa$_4$              & 139        & Tetragonal    & binary\_hard   \\
mp-2129      & Nd$_2$Ge$_4$             & 141        & Tetragonal    & binary\_hard   \\
mp-30682     & ZrGa               & 141        & Tetragonal    & binary\_hard   \\
mp-2128      & Sn$_8$Pd$_2$             & 68         & Orthorhombic  & binary\_hard   \\
mp-1208467   & Tb$_8$Al$_2$             & 227        & Cubic         & binary\_hard   \\
mp-640079    & Mn$_9$Au$_3$             & 123        & Tetragonal    & binary\_hard  \\

mp-8882      & Ga$_2$P$_2$              & 186                            & Hexagonal     & polymorph\_binary     \\
mp-13181     & LaF$_3$               & 225                            & Cubic         & polymorph\_binary     \\
mp-568382    & Mn$_2$Bi$_2$             & 194                            & Hexagonal     & polymorph\_binary     \\
mp-1004      & Nb$_3$Si              & 221                            & Cubic         & polymorph\_binary     \\
mp-2067      & Th$_2$Ga$_4$             & 141                            & Tetragonal    & polymorph\_binary     \\
mp-2178      & Yb$_2$Ga$_4$             & 194                            & Hexagonal     & polymorph\_binary     \\
mp-11251     & Mg$_6$Au$_2$             & 194                            & Hexagonal     & polymorph\_binary     \\
mp-9761      & Re$_2$O$_6$              & 182                            & Hexagonal     & polymorph\_binary     \\
mp-2231      & Sn$_4$S$_4$              & 62                             & Orthorhombic  & polymorph\_binary     \\
mp-11449     & Hf$_4$Mn$_8$             & 194                            & Hexagonal     & polymorph\_binary     \\
mp-2451      & Y$_3$Al$_9$              & 166                            & Trigonal      & polymorph\_binary     \\
mp-476       & Ge$_6$N$_8$              & 227                            & Cubic         & polymorph\_binary     \\

mp-1071835   & Dy$_2$Cu$_4$             & 74         & Orthorhombic  & template-based\_binary     \\
mp-1102067   & Pu$_4$Sn$_8$             & 141        & Tetragonal    & template-based\_binary     \\
mp-1102049   & Ta$_3$Be$_9$             & 166        & Trigonal      & template-based\_binary     \\
mp-1102936   & Ta$_8$Fe$_4$             & 193        & Hexagonal     & template-based\_binary     \\
mp-1095641   & Tb$_5$S$_7$              & 12         & Monoclinic    & template-based\_binary     \\
mp-1103888   & YbB$_{12}$              & 225        & Cubic         & template-based\_binary     \\
mp-1105001   & Tm$_6$Pt$_8$             & 148        & Trigonal      & template-based\_binary     \\
mp-1104457   & Eu$_9$Au$_6$             & 148        & Trigonal      & template-based\_binary     \\
mp-1104286   & Ge$_{12}$Rh$_3$            & 152        & Trigonal      & template-based\_binary     \\
mp-1106395   & Pr$_{12}$Ir$_4$            & 62         & Orthorhombic  & template-based\_binary     \\
mp-1105958   & Yb$_{12}$Co$_4$            & 62         & Orthorhombic  & template-based\_binary     \\
mp-1190213   & Re$_4$B$_{16}$             & 194        & Hexagonal     & template-based\_binary     \\
\hline\hline

\end{longtable}
\FloatBarrier

\section{The CSP5x-Ternary Test set} \label{subsec:testSet}
\label{subsec:testSet} 

\renewcommand{\arraystretch}{1.4}
\begin{longtable}{l l l l l l}
\caption{\textbf{Details of the 60 ternary benchmark crystals.}}
\label{table:dataset}\\
\hline\hline
\textbf{Material id} & \multicolumn{1}{l}{\begin{tabular}[c]{@{}c@{}}\textbf{Primitive formula}\end{tabular}} & \textbf{Space group} & \multicolumn{1}{l}{\begin{tabular}[c]{@{}c@{}}\textbf{Crystal system}\end{tabular}} & \multicolumn{1}{l}{\begin{tabular}[c]{@{}c@{}}\textbf{Category}\end{tabular}}\\ 
\hline\hline
mp-11390     & LiGaSi             & 216        & Cubic         & ternary\_easy   \\
mp-4552      & ErInCu$_2$            & 225        & Cubic         & ternary\_easy   \\
mp-20730     & HfCo$_2$Sn            & 225        & Cubic         & ternary\_easy   \\
mp-21211     & InFeCo$_2$            & 225        & Cubic         & ternary\_easy   \\
mp-30591     & Li$_2$CuSn            & 216        & Cubic         & ternary\_easy   \\
mp-30648     & LiMg$_2$Ga            & 225        & Cubic         & ternary\_easy   \\
mp-4972      & LuInCu$_2$            & 225        & Cubic         & ternary\_easy   \\
mp-5181      & LuSnPd$_2$            & 225        & Cubic         & ternary\_easy   \\
mp-20389     & Na$_2$CdPb            & 216        & Cubic         & ternary\_easy   \\
mp-30555     & TaGaCo$_2$            & 225        & Cubic         & ternary\_easy   \\
mp-29241     & Ca$_3$SnO             & 221        & Cubic         & ternary\_easy   \\
mp-20237     & Co$_2$Ni$_2$Sn$_2$          & 194        & Hexagonal     & ternary\_easy   \\
mp-4326      & KErS$_2$              & 166        & Trigonal      & ternary\_medium \\
mp-30580     & SrGaCu$_2$            & 166        & Trigonal      & ternary\_medium \\
mp-30493     & ZnCdPt$_2$            & 123        & Tetragonal    & ternary\_medium \\
mp-9570      & CaCd$_2$P$_2$            & 164        & Trigonal      & ternary\_medium \\
mp-5452      & CeCu$_2$Si$_2$           & 139        & Tetragonal    & ternary\_medium \\
mp-3147      & ErSi$_2$Au$_2$           & 139        & Tetragonal    & ternary\_medium \\
mp-13405     & LuMn$_2$Ge$_2$           & 139        & Tetragonal    & ternary\_medium \\
mp-30805     & SrNiSn$_3$            & 107        & Tetragonal    & ternary\_medium \\
mp-5615      & Ca$_3$Ag$_3$As$_3$          & 189        & Hexagonal     & ternary\_medium \\
mp-30733     & Ho$_3$Sn$_3$Pt$_3$          & 189        & Hexagonal     & ternary\_medium \\
mp-16747     & Lu$_3$Ag$_3$Pb$_3$          & 189        & Hexagonal     & ternary\_medium \\
mp-9812      & Zn$_2$C$_2$O$_6$            & 167        & Trigonal      & ternary\_medium \\
mp-12743     & CrTe$_4$Au            & 10         & Monoclinic    & ternary\_hard   \\
mp-3676      & MgCu$_4$Sn            & 216        & Cubic         & ternary\_hard   \\
mp-30587     & MgInCu$_4$            & 216        & Cubic         & ternary\_hard   \\
mp-7524      & Nb$_2$P$_2$Se$_2$           & 71         & Orthorhombic  & ternary\_hard   \\
mp-11533     & ZrZnNi$_4$            & 216        & Cubic         & ternary\_hard   \\
mp-11435     & ErTi$_2$Ga$_4$           & 139        & Tetragonal    & ternary\_hard   \\
mp-7095      & Ce$_2$Co$_2$Si$_4$          & 63         & Orthorhombic  & ternary\_hard   \\
mp-7489      & CsUF$_6$              & 148        & Trigonal      & ternary\_hard   \\
mp-11396     & Nd$_2$Ga$_4$Ni$_2$          & 65         & Orthorhombic  & ternary\_hard   \\
mp-29225     & Al$_4$Cu$_2$O$_7$           & 216        & Cubic         & ternary\_hard   \\
mp-23520     & In$_2$Pb$_4$I$_{10}$          & 140        & Tetragonal    & ternary\_hard   \\
mp-19140     & K$_3$MnO$_4$             & 121        & Tetragonal    & ternary\_hard  \\
mp-19183     & Li$_2$NiO$_2$            & 71                             & Orthorhombic  & polymorph\_ternary    \\
mp-13171     & Y$_3$Mg$_3$Cu$_3$           & 189                            & Hexagonal     & polymorph\_ternary    \\
mp-19227     & Y$_2$Mn$_2$O$_6$            & 194                            & Hexagonal     & polymorph\_ternary    \\
mp-7550      & Ce$_2$Nb$_2$O$_8$           & 15                             & Monoclinic    & polymorph\_ternary    \\
mp-11658     & Cu$_4$Sn$_2$Se$_6$          & 9                              & Monoclinic    & polymorph\_ternary    \\
mp-23550     & K$_2$Br$_2$F$_8$            & 140                            & Tetragonal    & polymorph\_ternary    \\
mp-12931     & Sm$_2$Ta$_2$O$_8$           & 13                             & Monoclinic    & polymorph\_ternary    \\
mp-19418     & V$_2$Cr$_2$O$_8$            & 63                             & Orthorhombic  & polymorph\_ternary    \\
mp-7913      & Be$_4$Si$_4$N$_8$           & 33                             & Orthorhombic  & polymorph\_ternary    \\
mp-557997    & Ca$_4$Se$_4$O$_{12}$          & 14                             & Monoclinic    & polymorph\_ternary    \\
mp-27450     & K$_4$Cu$_4$Cl$_{12}$          & 14                             & Monoclinic    & polymorph\_ternary    \\
mp-5126      & Zn$_4$S$_4$O$_{16}$           & 62                             & Orthorhombic  & polymorph\_ternary    \\

mp-1114630   & Rb$_3$PrCl$_6$           & 225        & Cubic         & template-based\_ternary    \\
mp-1105802   & Ca$_4$Ge$_8$Pt$_4$          & 71         & Orthorhombic  & template-based\_ternary    \\
mp-1106406   & Ce$_4$Sn$_2$S$_{10}$          & 55         & Orthorhombic  & template-based\_ternary    \\
mp-1106349   & Sm$_2$Pd$_6$S$_8$           & 223        & Cubic         & template-based\_ternary    \\
mp-1106327   & Co$_4$NiSb$_{12}$          & 204        & Cubic         & template-based\_ternary    \\
mp-1106064   & Ho$_4$Ga$_{12}$Ni          & 229        & Cubic         & template-based\_ternary    \\
mp-1106196   & Lu$_4$Ga$_{12}$Ni          & 229        & Cubic         & template-based\_ternary    \\
mp-1106117   & La$_{10}$Ag$_2$Pb$_6$         & 193        & Hexagonal     & template-based\_ternary    \\
mp-1106245   & Zr$_{10}$Al$_2$Sb$_6$         & 193        & Hexagonal     & template-based\_ternary    \\
mp-1105955   & Er$_6$Cu$_6$Sb$_8$          & 220        & Cubic         & template-based\_ternary    \\
mp-1105893   & La$_6$Cu$_6$Bi$_8$          & 220        & Cubic         & template-based\_ternary    \\
mp-1105777   & U$_6$Sb$_8$Ir$_6$           & 220        & Cubic         & template-based\_ternary    \\

\hline\hline

\end{longtable}
\FloatBarrier

\section{The CSP5x-Quartenary Test set} \label{subsec:testSet}

\renewcommand{\arraystretch}{1.4}
\begin{longtable}{l l l l l l}
\caption{\textbf{Details of the 60 quaternary benchmark crystals.}}
\label{table:dataset}\\
\hline\hline
\textbf{Material id} & \multicolumn{1}{l}{\begin{tabular}[c]{@{}c@{}}\textbf{Primitive formula}\end{tabular}} & \textbf{Space group} & \multicolumn{1}{l}{\begin{tabular}[c]{@{}c@{}}\textbf{Crystal system}\end{tabular}} & \multicolumn{1}{l}{\begin{tabular}[c]{@{}c@{}}\textbf{Category}\end{tabular}}\\ 
\hline\hline
mp-1213726   & CrFeCoSi           & 216        & Cubic         & quaternary\_easy   \\
mp-7554      & LiMgSnAu           & 216        & Cubic         & quaternary\_easy   \\
mp-11806     & LiMgSnPt           & 216        & Cubic         & quaternary\_easy   \\
mp-552674    & ZrTaNO             & 187        & Hexagonal     & quaternary\_easy   \\
mp-1071272   & CeAl$_2$BRu$_2$          & 123        & Tetragonal    & quaternary\_easy   \\
mp-13287     & Ba$_2$Cu$_2$Te$_2$F$_2$        & 129        & Tetragonal    & quaternary\_easy   \\
mp-12444     & Sr$_2$Cu$_2$S$_2$F$_2$         & 129        & Tetragonal    & quaternary\_easy   \\
mp-19093     & Ba$_2$UNiO$_6$           & 225        & Cubic         & quaternary\_easy   \\
mp-1111671   & K$_2$LiInF$_6$           & 225        & Cubic         & quaternary\_easy   \\
mp-6686      & K$_2$NaInF$_6$           & 225        & Cubic         & quaternary\_easy   \\
mp-16307     & Sr$_2$MgIrO$_6$          & 225        & Cubic         & quaternary\_easy   \\
mp-9705      & Ba$_4$NaB$_3$N$_6$          & 229        & Cubic         & quaternary\_easy   \\
mp-6258      & CeCr$_2$Si$_2$C          & 123        & Tetragonal    & quartenary\_medium \\
mp-6794      & LaB$_2$Rh$_2$C           & 139        & Tetragonal    & quartenary\_medium \\
mp-6140      & PrNi$_2$B$_2$C           & 139        & Tetragonal    & quartenary\_medium \\
mp-546790    & La$_2$Cu$_2$Te$_2$O$_2$        & 129        & Tetragonal    & quartenary\_medium \\
mp-20807     & Sr$_2$Fe$_2$As$_2$F$_2$        & 129        & Tetragonal    & quartenary\_medium \\
mp-6231      & Ba$_2$ErSbO$_6$          & 225        & Cubic         & quartenary\_medium \\
mp-19751     & Ba$_2$YbNbO$_6$          & 225        & Cubic         & quartenary\_medium \\
mp-15888     & Ba$_2$YIrO$_6$           & 225        & Cubic         & quartenary\_medium \\
mp-6428      & Ba$_2$YRuO$_6$           & 225        & Cubic         & quartenary\_medium \\
mp-19274     & BaPrMn$_2$O$_6$          & 123        & Tetragonal    & quartenary\_medium \\
mp-6586      & K$_2$NaAlF$_6$           & 225        & Cubic         & quartenary\_medium \\
mp-24412     & K$_2$NaAlH$_6$           & 225        & Cubic         & quartenary\_medium \\
mp-545788    & Ba$_3$ZnN$_2$O           & 123        & Tetragonal    & quartenary\_hard   \\
mp-18745     & La$_2$Mn$_2$Sb$_2$O$_2$        & 129        & Tetragonal    & quartenary\_hard   \\
mp-12515     & La$_2$Zn$_2$Sb$_2$O$_2$        & 129        & Tetragonal    & quartenary\_hard   \\
mp-20349     & Sm$_2$Fe$_2$As$_2$O$_2$        & 129        & Tetragonal    & quartenary\_hard   \\
mp-19118     & BaNd$_2$CoO5          & 71         & Orthorhombic  & quartenary\_hard   \\
mp-21348     & Ba$_2$EuTaO$_6$          & 87         & Tetragonal    & quartenary\_hard   \\
mp-13356     & Ba$_2$SrTeO$_6$          & 148        & Trigonal      & quartenary\_hard   \\
mp-9666      & K$_4$Na$_2$Ga$_2$P$_4$         & 72         & Orthorhombic  & quartenary\_hard   \\
mp-23126     & KAs$_4$IO$_6$            & 191        & Hexagonal     & quartenary\_hard   \\
mp-14466     & KLi$_6$IrO$_6$           & 166        & Trigonal      & quartenary\_hard   \\
mp-651268    & Fe$_2$Cu$_6$SnS$_8$         & 115        & Tetragonal    & quartenary\_hard   \\
mp-13383     & Sc$_2$Ag$_2$P$_4$Se$_{12}$       & 163        & Trigonal      & quartenary\_hard  \\
mp-23127     & Ba$_2$BiSbO$_6$          & 12                             & Monoclinic    & polymorph\_quartenary \\
mp-10731     & Ba$_2$PrSbO$_6$          & 225                            & Cubic         & polymorph\_quartenary \\
mp-9349      & KRb$_2$ScF$_6$           & 225                            & Cubic         & polymorph\_quartenary \\
mp-1080753   & NaPb$_2$IO$_6$           & 225                            & Cubic         & polymorph\_quartenary \\
mp-550722    & Ba$_2$Tl$_2$CuO$_6$         & 139                            & Tetragonal    & polymorph\_quartenary \\
mp-6628      & CsCdN$_3$O$_6$           & 146                            & Trigonal      & polymorph\_quartenary \\
mp-1095437   & Lu$_2$Se$_2$O$_6$F$_2$         & 11                             & Monoclinic    & polymorph\_quartenary \\
mp-726253    & RbLi$_3$S$_2$O$_9$          & 1                              & Triclinic     & polymorph\_quartenary \\
mp-23667     & Rb$_2$P$_2$H$_4$O$_8$          & 43                             & Orthorhombic  & polymorph\_quartenary \\
mp-2233097   & MgV$_4$SnO$_{12}$          & 5                              & Monoclinic    & polymorph\_quartenary \\
mp-1105110   & Sr$_4$Bi$_4$Se$_6$O$_4$        & 14                             & Monoclinic    & polymorph\_quartenary \\
mp-23125     & K$_6$Na$_2$Fe$_2$Cl$_{12}$       & 167                            & Trigonal      & polymorph\_quartenary \\
mp-1111927   & K$_2$LiCrF$_6$           & 225        & Cubic         & template-based\_quartenary \\
mp-1104454   & Ta$_4$GaTe$_4$Se$_4$        & 216        & Cubic         & template-based\_quartenary \\
mp-1106310   & Cu$_4$Si$_2$Hg2S$_8$        & 31         & Orthorhombic  & template-based\_quartenary \\
mp-1105386   & Yb$_2$H$_6$C$_2$N$_6$          & 176        & Hexagonal     & template-based\_quartenary \\
mp-1106402   & Rb$_4$Ti$_2$O$_2$F$_{10}$        & 63         & Orthorhombic  & template-based\_quartenary \\
mp-1106325   & CaCu$_3$Pt$_4$O$_{12}$        & 204        & Cubic         & template-based\_quartenary \\
mp-1106150   & CeMn$_4$Cu$_3$O$_{12}$        & 204        & Cubic         & template-based\_quartenary \\
mp-1106004   & HoFe$_4$Cu$_3$O$_{12}$        & 204        & Cubic         & template-based\_quartenary \\
mp-1106068   & LaCr$_4$Cu$_3$O$_{12}$        & 204        & Cubic         & template-based\_quartenary \\
mp-1105674   & Mg$_4$Be$_2$B$_4$Ir$_{10}$       & 127        & Tetragonal    & template-based\_quartenary \\
mp-1105109   & PrCu$_3$Ru$_4$O$_{12}$        & 204        & Cubic         & template-based\_quartenary \\
mp-1105290   & Co$_3$Sb$_4$O$_6$F$_6$         & 217        & Cubic         & template-based\_quartenary \\
\hline\hline

\end{longtable}
\FloatBarrier

\section{The CSP5x-Template Test set} \label{subsec:testSet}

\renewcommand{\arraystretch}{1.4}
\begin{longtable}{l l l l l l}
\caption{\textbf{Details of the 36 template based test data benchmark crystals.}}
\label{table:dataset}\\
\hline\hline
\textbf{Material id} & \multicolumn{1}{l}{\begin{tabular}[c]{@{}c@{}}\textbf{Primitive formula}\end{tabular}} & \textbf{Space group} & \multicolumn{1}{l}{\begin{tabular}[c]{@{}c@{}}\textbf{Crystal system}\end{tabular}} & \multicolumn{1}{l}{\begin{tabular}[c]{@{}c@{}}\textbf{Category}\end{tabular}}\\ 
\hline\hline
mp-1071835   & Dy$_2$Cu$_4$             & 74         & Orthorhombic  & template-based\_binary     \\
mp-1102067   & Pu$_4$Sn$_8$             & 141        & Tetragonal    & template-based\_binary     \\
mp-1102049   & Ta$_3$Be$_9$             & 166        & Trigonal      & template-based\_binary     \\
mp-1102936   & Ta$_8$Fe$_4$             & 193        & Hexagonal     & template-based\_binary     \\
mp-1095641   & Tb$_5$S$_7$              & 12         & Monoclinic    & template-based\_binary     \\
mp-1103888   & YbB$_{12}$              & 225        & Cubic         & template-based\_binary     \\
mp-1105001   & Tm$_6$Pt$_8$             & 148        & Trigonal      & template-based\_binary     \\
mp-1104457   & Eu$_9$Au$_6$             & 148        & Trigonal      & template-based\_binary     \\
mp-1104286   & Ge$_{12}$Rh$_3$            & 152        & Trigonal      & template-based\_binary     \\
mp-1106395   & Pr$_{12}$Ir$_4$            & 62         & Orthorhombic  & template-based\_binary     \\
mp-1105958   & Yb$_{12}$Co$_4$            & 62         & Orthorhombic  & template-based\_binary     \\
mp-1190213   & Re$_4$B$_{16}$             & 194        & Hexagonal     & template-based\_binary     \\
mp-1114630   & Rb$_3$PrCl$_6$           & 225        & Cubic         & template-based\_ternary    \\
mp-1105802   & Ca$_4$Ge$_8$Pt$_4$          & 71         & Orthorhombic  & template-based\_ternary    \\
mp-1106406   & Ce$_4$Sn$_2$S$_{10}$          & 55         & Orthorhombic  & template-based\_ternary    \\
mp-1106349   & Sm$_2$Pd$_6$S$_8$           & 223        & Cubic         & template-based\_ternary    \\
mp-1106327   & Co$_4$NiSb$_{12}$          & 204        & Cubic         & template-based\_ternary    \\
mp-1106064   & Ho$_4$Ga$_{12}$Ni          & 229        & Cubic         & template-based\_ternary    \\
mp-1106196   & Lu$_4$Ga$_{12}$Ni          & 229        & Cubic         & template-based\_ternary    \\
mp-1106117   & La$_{10}$Ag$_2$Pb$_6$         & 193        & Hexagonal     & template-based\_ternary    \\
mp-1106245   & Zr$_{10}$Al$_2$Sb$_6$         & 193        & Hexagonal     & template-based\_ternary    \\
mp-1105955   & Er$_6$Cu$_6$Sb$_8$          & 220        & Cubic         & template-based\_ternary    \\
mp-1105893   & La$_6$Cu$_6$Bi$_8$          & 220        & Cubic         & template-based\_ternary    \\
mp-1105777   & U$_6$Sb$_8$Ir$_6$           & 220        & Cubic         & template-based\_ternary    \\
mp-1111927   & K$_2$LiCrF$_6$           & 225        & Cubic         & template-based\_quartenary \\
mp-1104454   & Ta$_4$GaTe$_4$Se$_4$        & 216        & Cubic         & template-based\_quartenary \\
mp-1106310   & Cu$_4$Si$_2$Hg2S$_8$        & 31         & Orthorhombic  & template-based\_quartenary \\
mp-1105386   & Yb$_2$H$_6$C$_2$N$_6$          & 176        & Hexagonal     & template-based\_quartenary \\
mp-1106402   & Rb$_4$Ti$_2$O$_2$F$_{10}$        & 63         & Orthorhombic  & template-based\_quartenary \\
mp-1106325   & CaCu$_3$Pt$_4$O$_{12}$        & 204        & Cubic         & template-based\_quartenary \\
mp-1106150   & CeMn$_4$Cu$_3$O$_{12}$        & 204        & Cubic         & template-based\_quartenary \\
mp-1106004   & HoFe$_4$Cu$_3$O$_{12}$        & 204        & Cubic         & template-based\_quartenary \\
mp-1106068   & LaCr$_4$Cu$_3$O$_{12}$        & 204        & Cubic         & template-based\_quartenary \\
mp-1105674   & Mg$_4$Be$_2$B$_4$Ir$_{10}$       & 127        & Tetragonal    & template-based\_quartenary \\
mp-1105109   & PrCu$_3$Ru$_4$O$_{12}$        & 204        & Cubic         & template-based\_quartenary \\
mp-1105290   & Co$_3$Sb$_4$O$_6$F$_6$         & 217        & Cubic         & template-based\_quartenary   \\
\hline\hline

\end{longtable}

\FloatBarrier
\section{Parameters and configuration for all algorithms}
\begin{table}[]
\caption{Parameters and configuration for all algorithms.}
\centering
\begin{tabular}{|l|l|}
\hline
\textbf{Algorithms} & \textbf{Parameters and configuration} 
\\ \hline
CALYPSO        &     \begin{tabular}[c]{@{}l@{}}Generation = 40 \\ Popsize = 50 \\ energy evaluation <= 20,000 
\end{tabular}      

\\ \hline
CSPML               & \begin{tabular}[c]{@{}l@{}}Topn =10, select the best one with the smallest Euclidean distance\\  of the descriptors.\end{tabular}                                                                                                                                                                               \\ \hline
TCSP                & Topn = 5, select the best one with the lowest ElMD score.                                                                                                                                                                                                                                                   \\ \hline
ParetoCSP           & \begin{tabular}[c]{@{}l@{}}Number of steps = 5,000\\ Crossover probability = 0.8\\ Crossover operator  = Simulated Binary Crossover\\ Mutation probability = 0.01\\ Population size = 100\\ Mutation operator = Polynomial Mutation\\ Tournament size (for the Pareto tournament selection) = 2\end{tabular} \\ \hline
AlphaCrystal-II        &     \begin{tabular}[c]{@{}l@{}}Atom number <= 12\\ Atomic features = 11\\ Label of distance matrix = 10\\ Relax and Formation energy Calculator = M3GNet \\ Topn = 10, select the best one with lowest formation energy.
\end{tabular}                                                                                                                                                                                                                   \\ \hline
GNOA-MEGNet-RAS     & max\_step = 20,000                                                                                                                                                                                                                                                                                           \\ \hline
GNOA-MEGNet-PSO     & max\_step = 20,000                                                                                                                                                                                                                                                                                           \\ \hline
GNOA-MEGNet-BO      & \begin{tabular}[c]{@{}l@{}}n\_init = 200\\ max\_step = 20,000\end{tabular}                                                                                                                                                                                                                                   \\ \hline
GNOA-M3GNet-RAS     & max\_step = 20,000                                                                                                                                                                                                                                                                                           \\ \hline
GNOA-M3GNet-PSO     & max\_step = 20,000                                                                                                                                                                                                                                                                                           \\ \hline
GNOA-M3GNet-BO      & \begin{tabular}[c]{@{}l@{}}n\_init = 200\\ max\_step = 20,000\end{tabular}                                                                                                                                                                                                                                   \\ \hline
AGOX-bh             & \begin{tabular}[c]{@{}l@{}}Number of iterations = 5,000\\ fmax = 0.05\\ Generator = RattleGenerator\\ Calculator = M3GNet\\ temperature = 0.25\\ sampler = MetropolisSampler\end{tabular}                                                                                                                    \\ \hline
AGOX-pt             & \begin{tabular}[c]{@{}l@{}}Number of iterations = 5,000\\ fmax = 0.05\\ Generator = RattleGenerator\\ Calculator = M3GNet\\ swap\_frequency = 10\\ swap\_order = 5\end{tabular}                                                                                                                              \\ \hline
AGOX-rss            & \begin{tabular}[c]{@{}l@{}}Number of iterations = 5,000\\ fmax = 0.05\\ Generator = RandomGenerator\\ Calculator = M3GNet

\end{tabular}                                                                                                                                                                       \\ \hline
\end{tabular}

\label{tab:parameters}
\end{table}

\begin{table}[]
\caption{The relevant parameter settings for DFT calculations of CALYPSO during the optimization process.}
\centering
\begin{tabular}{|l|l|l|l|l|l|l|l|l|}
\hline
                 & \textbf{PREC} & \textbf{\begin{tabular}[c]{@{}l@{}}Cutoff \\ energy (eV)\end{tabular}} & \textbf{\begin{tabular}[c]{@{}l@{}}Energy \\ convergence \\ criterion (eV)\end{tabular}} & \textbf{\begin{tabular}[c]{@{}l@{}}Force \\ convergence \\ criterion (eV/Å)\end{tabular}} & \textbf{K-meshes} & \textbf{\begin{tabular}[c]{@{}l@{}}Optimization\\ algorithms\end{tabular}} & \textbf{ISIF} & \textbf{POTIM} \\ \hline
\textbf{INCAR-1} & LOW           & -                                                                      & 5e-3                                                                                     & -4e-2                                                                                     & 2$\pi$x0.06       & quasi-Newton                                                               & 2             & 0.5            \\ \hline
\textbf{INCAR-2} & Normal        & -                                                                      & 1e-3                                                                                     & -3e-2                                                                                     & 2$\pi$x0.05       & CG                                                                         & 3             & 0.3            \\ \hline
\textbf{INCAR-3} & Normal        & 450                                                                    & 1e-4                                                                                     & -2e-2                                                                                     & 2$\pi$x0.04       & CG                                                                         & 3             & 0.1            \\ \hline
\end{tabular}
\end{table}

\FloatBarrier

\section{Additional performance comparisons on binary/ternary/quartenary test structures}

\begin{figure}[tbh!]
  \centering
  \includegraphics[width=0.8\linewidth]{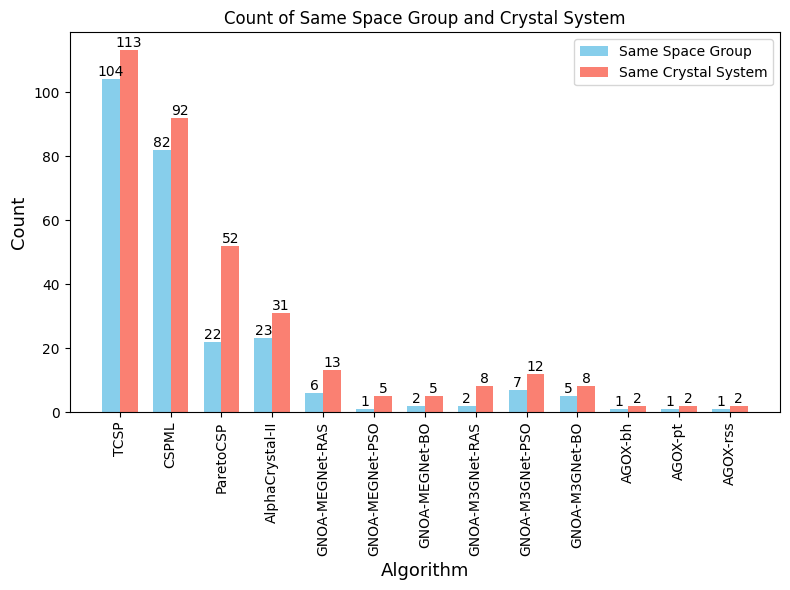}
  \caption{Comparison of structure symmetry prediction performance in terms of the number of the predicted structures (out of 180 test samples) with identical space groups and crystal systems compared to the ground truth structures.}
  \label{fig:spaceGroup_crystalSystem}
\end{figure}

\begin{figure}[!htb]
  \centering
  \includegraphics[width=0.85\linewidth]{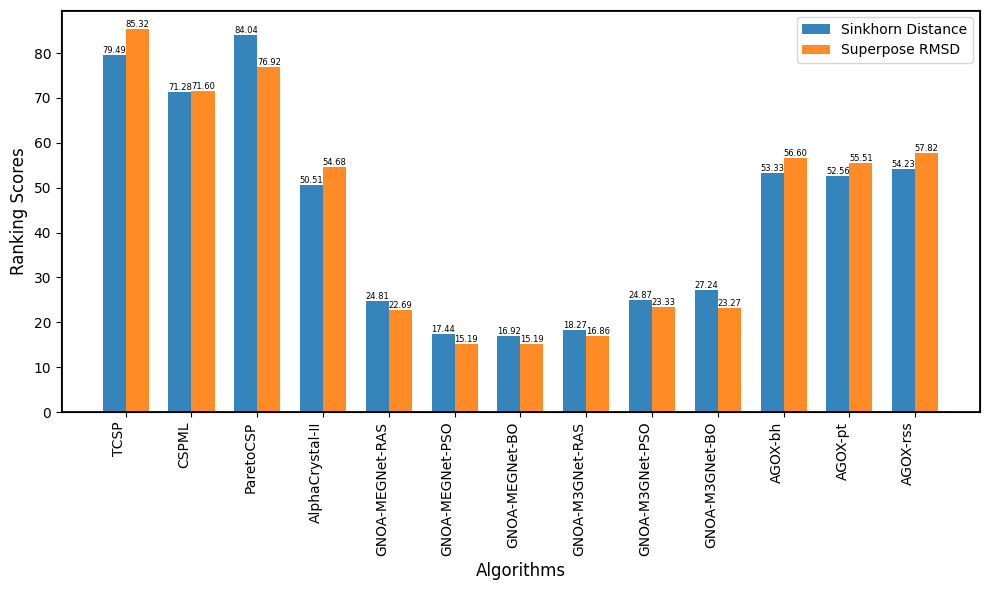}
  \caption{Comparison of CSP algorithms by their ranking scores based on the average of Sinkhorn distance and superpose RMSD across 60 binary predicted structures against the ground truth structures.}
\label{fig:binary}
\end{figure}

\begin{figure}[!htb]
  \centering
  \includegraphics[width=0.85\linewidth]{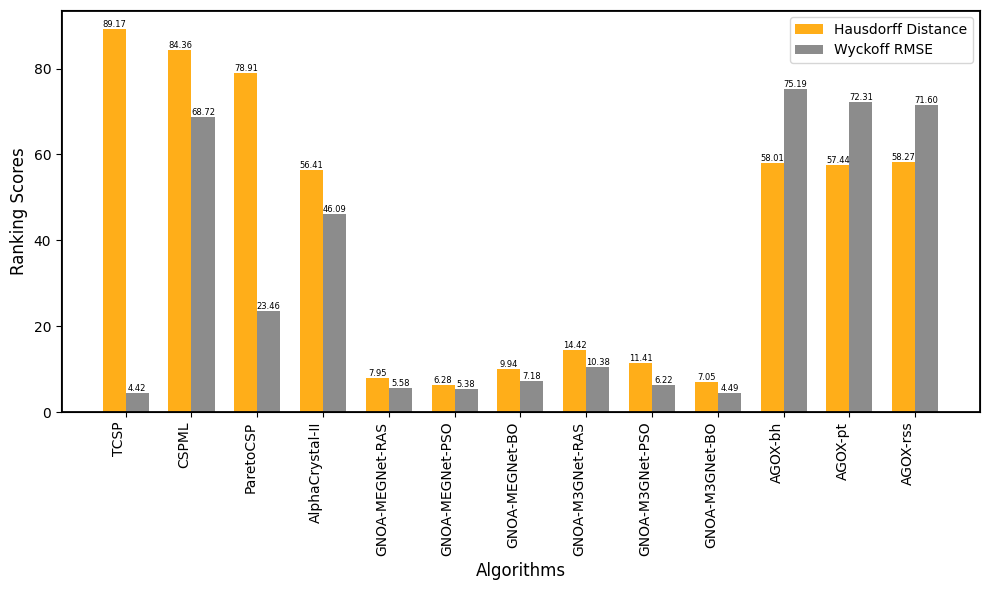}
  \caption{Comparison of CSP algorithms by their ranking scores based on the average of Hausdorff distance and Wyckoff RMSE across 60 ternary predicted structures against the ground truth structures.}
  \label{fig:ternary}
\end{figure}

\begin{figure}[!htb]
  \centering
  \includegraphics[width=0.85\linewidth]{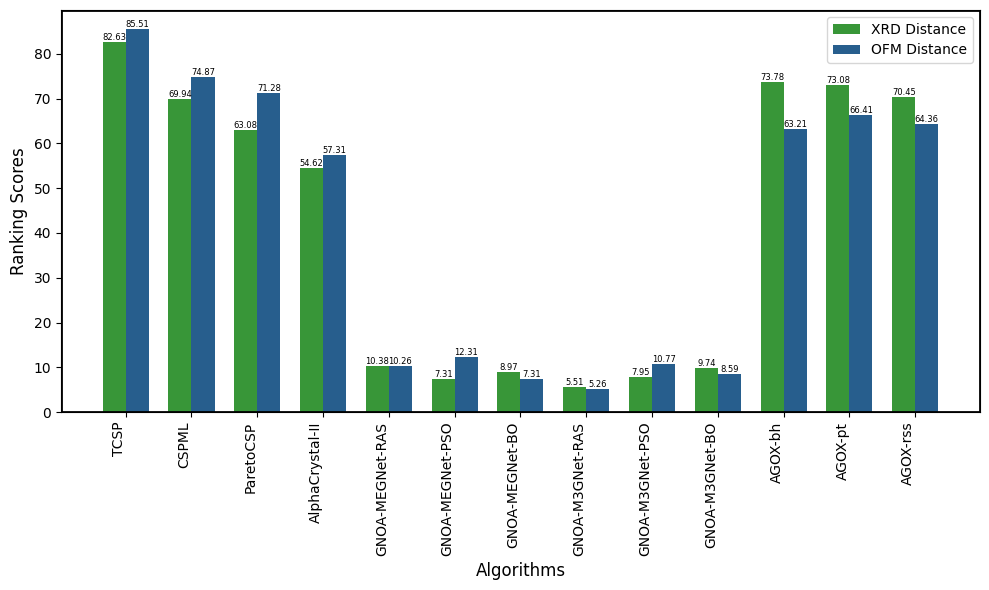}
  \caption{Comparison of CSP algorithms by their ranking scores based on the average of XRD distance and OFM distance across 60 quaternary predicted structures against the ground truth structures.}
  \label{fig:quaternary}
\end{figure}

\FloatBarrier

\section{Case studies}

\begin{figure}[!htb]
  \includegraphics[width=1\linewidth]{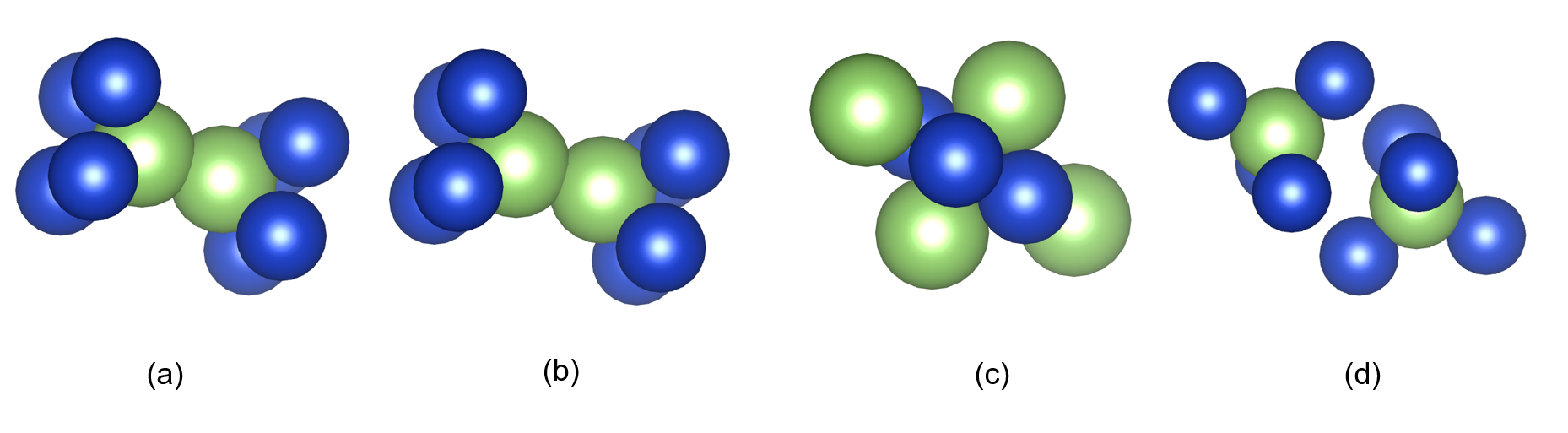}
  \caption{Comparison of the ground truth and predicted crystal structures of Ga$_2$Cu. (a). Ground truth structure. (b). Predicted by ParetoCSP (c). Predicted by TCSP. (d). Predicted by CSPML.}
  \label{fig:nontemplate}
\end{figure}

\begin{table}[htb!]
\caption{The metrics for seven distances of Ga$_2$Cu as evaluated by the ParetoCSP, TCSP and CSPML algorithms. This includes M3GNet Energy Distance (eV/atom); Sinkhorn Distance, Chamfer Distance, Superpose RMSD, and Fingerprint Distance (all in Å); XRD Spectrum Distance (counts/sec); and OFM Distance (measured in valence electrons).}
\centering
\begin{tabular}{|l|l|l|l|l|l|l|l|}
\hline
\textbf{Algorithm} & \textbf{\begin{tabular}[c]{@{}l@{}}M3gnet \\ Energy \\ Distance\end{tabular}} & \textbf{\begin{tabular}[c]{@{}l@{}}Sinkhorn \\ Distance\end{tabular}} & \textbf{\begin{tabular}[c]{@{}l@{}}Chamfer \\ Distance\end{tabular}} & \textbf{\begin{tabular}[c]{@{}l@{}}Superpose \\ RMSD\end{tabular}} & \textbf{\begin{tabular}[c]{@{}l@{}}Fingerprint \\ Distance\end{tabular}} & \textbf{\begin{tabular}[c]{@{}l@{}}XRD \\ Distance\end{tabular}} & \textbf{\begin{tabular}[c]{@{}l@{}}OFM \\ Distance\end{tabular}} \\ \hline
\textbf{ParetoCSP} & \textbf{0.012}                                                                & \textbf{0.316}                                                        & \textbf{0.211}                                                       & \textbf{0.075}                                                     & \textbf{0.445}                                                           & \textbf{1.764}                                                   & \textbf{0.109}                                                   \\ \hline
\textbf{TCSP}      & 0.265                                                                         & 13.891                                                                & 5.485                                                                & 1.946                                                              & 2.150                                                                    & \textbf{2.433}                                                   & 1.503                                                            \\ \hline
\textbf{CSPML}     & \textbf{0.006}                                                                & 20.104                                                                & 5.792                                                                & 1.118                                                              & 2.202                                                                    & 2.506                                                            & 0.682                                                            \\ \hline
\end{tabular}
\label{tab:Ga2Cu}
\end{table}

\begin{figure}[!htb]
  \includegraphics[width=1\linewidth]{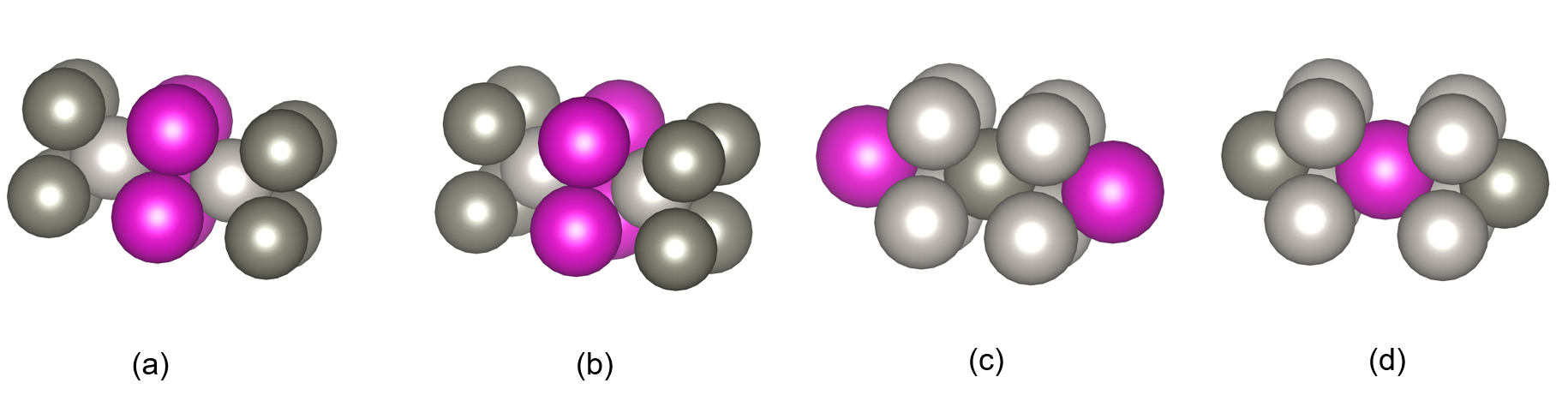}
  \caption{Comparison of the ground truth and predicted crystal structures of ZnCdPt$_2$. (a). Ground truth structure. (b). Predicted by CALYPSO. (c). Predicted by ParetoCSP.  (d). Predicted by AlphaCrystal}
  \label{fig:ZnCdPt2}
\end{figure}

\begin{table}[th]
\caption{The metrics for seven distances of ZnCdPt$_2$ as evaluated by the CALYPSO, ParetoCSP and AlphaCrystal-II algorithms. This includes M3GNet Energy Distance (eV/atom); Sinkhorn Distance, Chamfer Distance, Superpose RMSD, and Fingerprint Distance (all in Å); XRD Spectrum Distance (counts/sec); and OFM Distance (measured in valence electrons).}
\begin{tabular}{|l|l|l|l|l|l|l|l|}
\hline
\textbf{Algorithm}       & \textbf{\begin{tabular}[c]{@{}l@{}}M3gnet \\ Energy \\ Distance\end{tabular}} & \textbf{\begin{tabular}[c]{@{}l@{}}Sinkhorn \\ Distance\end{tabular}} & \textbf{\begin{tabular}[c]{@{}l@{}}Chamfer \\ Distance\end{tabular}} & \textbf{\begin{tabular}[c]{@{}l@{}}Superpose \\ RMSD\end{tabular}} & \textbf{\begin{tabular}[c]{@{}l@{}}Fingerprint \\ Distance\end{tabular}} & \textbf{\begin{tabular}[c]{@{}l@{}}XRD \\ Distance\end{tabular}} & \textbf{\begin{tabular}[c]{@{}l@{}}OFM \\ Distance\end{tabular}} \\ \hline
\textbf{CALYPSO}         & \textbf{0.008}                                                                & \textbf{0.290}                                                        & \textbf{0.145}                                                       & \textbf{0.049}                                                     & 0.490                                                           & \textbf{1.632}                                                   & 0.102                                                   \\ \hline
\textbf{ParetoCSP}       & 0.010                                                                         & 7.273                                                                 & 3.554                                                                & 1.528                                                              & \textbf{0.064}                                                                    & 2.147                                                            & \textbf{0.036}                                                            \\ \hline
\textbf{AlphaCrystal-II} & 0.009                                                                & 7.295                                                                 & 3.636                                                                & 1.347                                                              & 0.325                                                                    & 2.113                                                            & 0.073                                                            \\ \hline
\end{tabular}
\label{tab:ZnCdPt2}
\end{table}

\begin{figure}[!htb]
  \includegraphics[width=1\linewidth]{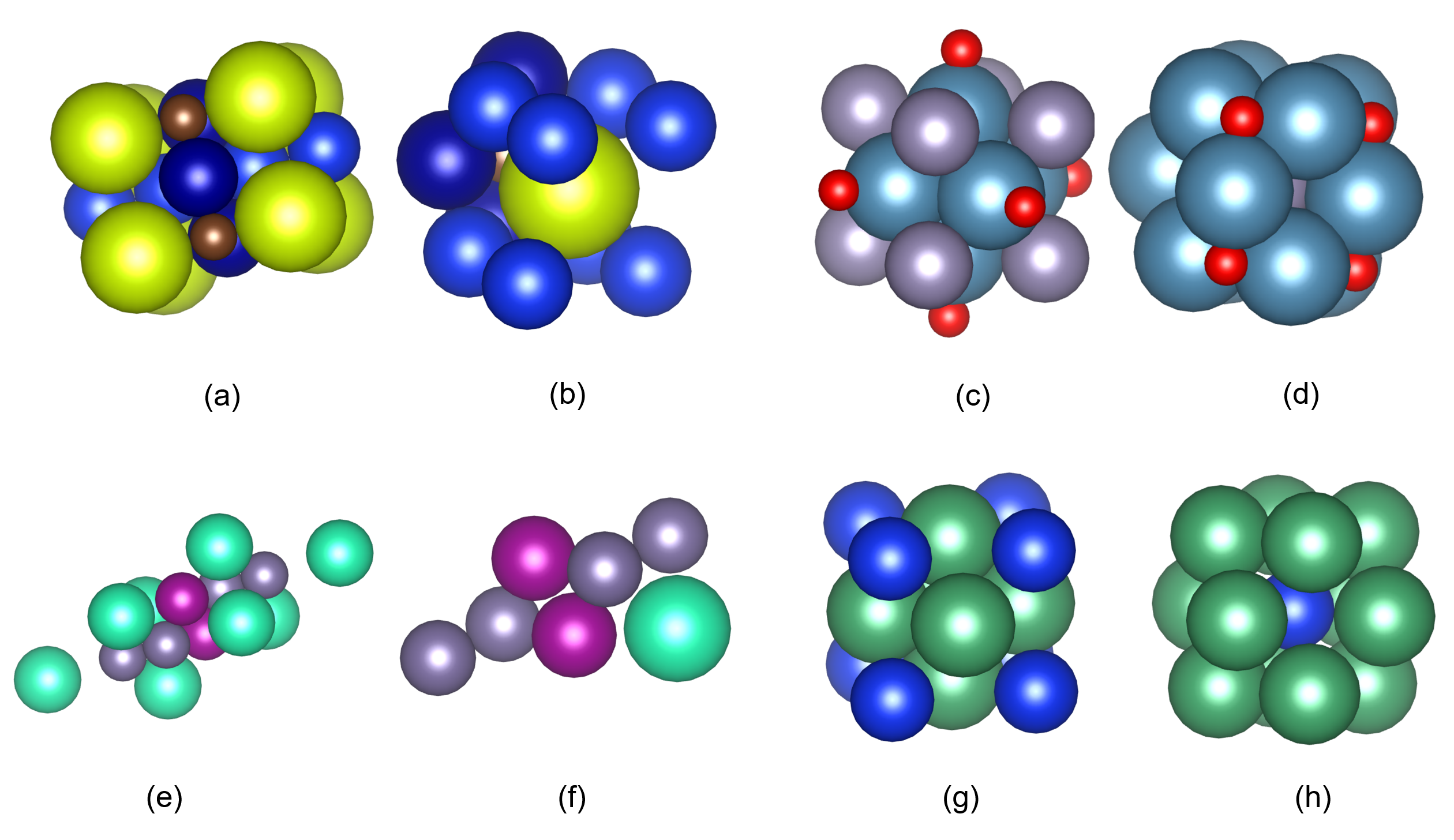}
  \caption{Comparison of the ground truth and predicted crystal structures based on StructureMatcher, showing similar materials found in the MP database. (a). Ground truth structure of CeCr$_2$Si$_2$C. (b). CeCr$_2$Si$_2$C predicted by AlphaCrystal-II. (c). Ground truth structure of Ca$_3$SnO. (d). Ca$_3$SnO predicted by CALYPSO. (e). Ground truth structure of LuMn$_2$Ge$_2$. (f). LuMn$_2$Ge$_2$ predicted by CSPML. (g).Ground truth structure of Nb$_3$Si. (h). Nb$_3$Si predicted by ParetoCSP.}
  \label{fig: fail_results}
\end{figure}

\begin{table}[]
\caption{ Predicted structures and MP structures reported by StructureMatcher (ltol=0.2, stol=0.3, angle\_tol=5) as similar, but have different space group numbers.}
\centering
\begin{tabular}{|l|l|l|l|l|l|}
\hline
\textbf{Algorithm} & \textbf{Formula} & \textbf{Target space group} & \textbf{Predicted space group} & \textbf{Fit} & \textbf{RMS} \\ \hline
AlphaCrystal-II       & CeCr$_2$Si$_2$C        & 123                         & 99                             & TRUE         & 0.004895     \\ \hline
CALYPSO            & Ca$_3$SnO           & 221                         & 123                            & TRUE         & 1.40E-32     \\ \hline
CSPML              & LuMn$_2$Ge$_2$         & 139                         & 2                              & TRUE         & 0.015918     \\ \hline
ParetoCSP          & Nb$_3$Si           & 221                         & 123                            & TRUE         & 7.45E-17     \\ \hline
\end{tabular}
\label{tab:fail_results}
\end{table}

\begin{table}[]
\centering
\caption{ Cif files of the predicted structures and MP structures reported by StructureMatcher as similar, but have different space group numbers.}
\begin{tabular}{|l|l|}
\hline
\textbf{Ground Truth Structure of CeCr$_2$Si$_2$C}                                                                                                                                  & \textbf{CeCr$_2$Si$_2$C predicted by AlphaCrystal-II}                                                                                                                                       \\ \hline
\begin{tabular}[c]{@{}l@{}}data\_CeCr2Si2C\\ \_symmetry\_space\_group\_name\_H-M   'P 1'\\ \_cell\_length\_a   3.98559300\\ \_cell\_length\_b   3.98559300\\ \_cell\_length\_c   5.26171600\\ \_cell\_angle\_alpha   90.00000000\\ \_cell\_angle\_beta   90.00000000\\ \_cell\_angle\_gamma   90.00000000\\ \_symmetry\_Int\_Tables\_number   1\\ \_chemical\_formula\_structural   CeCr2Si2C\\ \_chemical\_formula\_sum   'Ce1 Cr2 Si2 C1'\\ \_cell\_volume   83.58210379\\ \_cell\_formula\_units\_Z   1\\ loop\_\\  \_symmetry\_equiv\_pos\_site\_id\\  \_symmetry\_equiv\_pos\_as\_xyz\\   1  'x, y, z'\\ loop\_\\  \_atom\_site\_type\_symbol\\  \_atom\_site\_label\\  \_atom\_site\_symmetry\_multiplicity\\  \_atom\_site\_fract\_x\\  \_atom\_site\_fract\_y\\  \_atom\_site\_fract\_z\\  \_atom\_site\_occupancy\\   Ce  Ce0  1  0.00000000  0.00000000  0.00000000  1.0\\   Cr  Cr1  1  0.00000000  0.50000000  0.50000000  1.0\\   Cr  Cr2  1  0.50000000  0.00000000  0.50000000  1.0\\   Si  Si3  1  0.50000000  0.50000000  0.77139900  1.0\\   Si  Si4  1  0.50000000  0.50000000  0.22860100  1.0\\   C  C5  1  0.00000000  0.00000000  0.50000000  1.0\end{tabular} & \begin{tabular}[c]{@{}l@{}}data\_CeCr2Si2C\\ \_symmetry\_space\_group\_name\_H-M   'P 1'\\ \_cell\_length\_a   3.99187396\\ \_cell\_length\_b   3.99187399\\ \_cell\_length\_c   5.25647435\\ \_cell\_angle\_alpha   89.99999827\\ \_cell\_angle\_beta   90.00000196\\ \_cell\_angle\_gamma   90.00000024\\ \_symmetry\_Int\_Tables\_number   1\\ \_chemical\_formula\_structural   CeCr2Si2C\\ \_chemical\_formula\_sum   'Ce1 Cr2 Si2 C1'\\ \_cell\_volume   83.76222276\\ \_cell\_formula\_units\_Z   1\\ loop\_\\  \_symmetry\_equiv\_pos\_site\_id\\  \_symmetry\_equiv\_pos\_as\_xyz\\   1  'x, y, z'\\ loop\_\\  \_atom\_site\_type\_symbol\\  \_atom\_site\_label\\  \_atom\_site\_symmetry\_multiplicity\\  \_atom\_site\_fract\_x\\  \_atom\_site\_fract\_y\\  \_atom\_site\_fract\_z\\  \_atom\_site\_occupancy\\   Ce  Ce0  1  0.50000000  0.50000000  0.63277947  1.0\\   Cr  Cr1  1  0.50000000  0.00000000  0.13525183  1.0\\   Cr  Cr2  1  0.00000000  0.50000000  0.13525183  1.0\\   Si  Si3  1  0.00000000  0.00000000  0.86833114  1.0\\   Si  Si4  1  0.00000000  0.00000000  0.40863335  1.0\\   C  C5  1  0.50000000  0.50000000  0.13409138  1.0\end{tabular} \\ \hline
\end{tabular}
\end{table}

\begin{table}[]
\centering
\begin{tabular}{|l|l|}
\hline
\textbf{Ground Truth Structure of Ca$_3$SnO}                                                                                                                      & \textbf{Ca$_3$SnO predicted by CALYPSO}                                                                                                                         \\ \hline
\begin{tabular}[c]{@{}l@{}}data\_Ca3SnO\\ \_symmetry\_space\_group\_name\_H-M   'P 1'\\ \_cell\_length\_a   4.82648645\\ \_cell\_length\_b   4.82648645\\ \_cell\_length\_c   4.82648645\\ \_cell\_angle\_alpha   90.00000000\\ \_cell\_angle\_beta   90.00000000\\ \_cell\_angle\_gamma   90.00000000\\ \_symmetry\_Int\_Tables\_number   1\\ \_chemical\_formula\_structural   Ca3SnO\\ \_chemical\_formula\_sum   'Ca3 Sn1 O1'\\ \_cell\_volume   112.43286407\\ \_cell\_formula\_units\_Z   1\\ loop\_\\  \_symmetry\_equiv\_pos\_site\_id\\  \_symmetry\_equiv\_pos\_as\_xyz\\   1  'x, y, z'\\ loop\_\\  \_atom\_site\_type\_symbol\\  \_atom\_site\_label\\  \_atom\_site\_symmetry\_multiplicity\\  \_atom\_site\_fract\_x\\  \_atom\_site\_fract\_y\\  \_atom\_site\_fract\_z\\  \_atom\_site\_occupancy\\   Ca  Ca0  1  0.50000000  0.00000000  0.50000000  1.0\\   Ca  Ca1  1  0.00000000  0.50000000  0.50000000  1.0\\   Ca  Ca2  1  0.50000000  0.50000000  0.00000000  1.0\\   Sn  Sn3  1  0.00000000  0.00000000  0.00000000  1.0\\   O  O4  1  0.50000000  0.50000000  0.50000000  1.0\end{tabular} & \begin{tabular}[c]{@{}l@{}}data\_Ca3SnO\\ \_symmetry\_space\_group\_name\_H-M   'P 1'\\ \_cell\_length\_a   4.86831000\\ \_cell\_length\_b   4.86831000\\ \_cell\_length\_c   4.71799000\\ \_cell\_angle\_alpha   90.00000000\\ \_cell\_angle\_beta   90.00000000\\ \_cell\_angle\_gamma   90.00000000\\ \_symmetry\_Int\_Tables\_number   1\\ \_chemical\_formula\_structural   Ca3SnO\\ \_chemical\_formula\_sum   'Ca3 Sn1 O1'\\ \_cell\_volume   111.81844956\\ \_cell\_formula\_units\_Z   1\\ loop\_\\  \_symmetry\_equiv\_pos\_site\_id\\  \_symmetry\_equiv\_pos\_as\_xyz\\   1  'x, y, z'\\ loop\_\\  \_atom\_site\_type\_symbol\\  \_atom\_site\_label\\  \_atom\_site\_symmetry\_multiplicity\\  \_atom\_site\_fract\_x\\  \_atom\_site\_fract\_y\\  \_atom\_site\_fract\_z\\  \_atom\_site\_occupancy\\   Ca  Ca1  1  0.00000000  0.50000000  0.00000000  1.0\\   Ca  Ca2  1  0.00000000  0.00000000  0.50000000  1.0\\   Ca  Ca3  1  0.50000000  0.00000000  0.00000000  1.0\\   Sn  Sn1  1  0.50000000  0.50000000  0.50000000  1.0\\   O  O1  1  0.00000000  0.00000000  0.00000000  1.0\end{tabular} \\ \hline
\end{tabular}
\end{table}

\begin{table}[]
\centering
\begin{tabular}{|l|l|}
\hline
\textbf{Ground Truth Structure of LuMn$_2$Ge$_2$}                                                            & \textbf{LuMn$_2$Ge$_2$ predicted by CSPML}                                                                   \\ \hline
\begin{tabular}[c]{@{}l@{}}data\_Lu(MnGe)2\\ \_symmetry\_space\_group\_name\_H-M   'P 1'\\ \_cell\_length\_a   6.12371019\\ \_cell\_length\_b   6.12371019\\ \_cell\_length\_c   6.12371019\\ \_cell\_angle\_alpha   141.71630620\\ \_cell\_angle\_beta   141.71630620\\ \_cell\_angle\_gamma   55.25636711\\ \_symmetry\_Int\_Tables\_number   1\\ \_chemical\_formula\_structural   Lu(MnGe)2\\ \_chemical\_formula\_sum   'Lu1 Mn2 Ge2'\\ \_cell\_volume   87.50433646\\ \_cell\_formula\_units\_Z   1\\ loop\_\\  \_symmetry\_equiv\_pos\_site\_id\\  \_symmetry\_equiv\_pos\_as\_xyz\\   1  'x, y, z'\\ loop\_\\  \_atom\_site\_type\_symbol\\  \_atom\_site\_label\\  \_atom\_site\_symmetry\_multiplicity\\  \_atom\_site\_fract\_x\\  \_atom\_site\_fract\_y\\  \_atom\_site\_fract\_z\\  \_atom\_site\_occupancy\\   Lu  Lu0  1  0.00000000  0.00000000  0.00000000  1.0\\   Mn  Mn1  1  0.75000000  0.25000000  0.50000000  1.0\\   Mn  Mn2  1  0.25000000  0.75000000  0.50000000  1.0\\   Ge  Ge3  1  0.61238208  0.61238208  0.00000000  1.0\\   Ge  Ge4  1  0.38761792  0.38761792  0.00000000  1.0\end{tabular} & \begin{tabular}[c]{@{}l@{}}data\_Lu(MnGe)2\\ \_symmetry\_space\_group\_name\_H-M   'P 1'\\ \_cell\_length\_a   3.89603496\\ \_cell\_length\_b   3.90082675\\ \_cell\_length\_c   6.06380844\\ \_cell\_angle\_alpha   108.78424278\\ \_cell\_angle\_beta   108.74365528\\ \_cell\_angle\_gamma   90.01193034\\ \_symmetry\_Int\_Tables\_number   1\\ \_chemical\_formula\_structural   Lu(MnGe)2\\ \_chemical\_formula\_sum   'Lu1 Mn2 Ge2'\\ \_cell\_volume   82.06637886\\ \_cell\_formula\_units\_Z   1\\ loop\_\\  \_symmetry\_equiv\_pos\_site\_id\\  \_symmetry\_equiv\_pos\_as\_xyz\\   1  'x, y, z'\\ loop\_\\  \_atom\_site\_type\_symbol\\  \_atom\_site\_label\\  \_atom\_site\_symmetry\_multiplicity\\  \_atom\_site\_fract\_x\\  \_atom\_site\_fract\_y\\  \_atom\_site\_fract\_z\\  \_atom\_site\_occupancy\\   Lu  Lu0  1  0.99271167  0.00258012  0.00062319  1.0\\   Mn  Mn1  1  0.24486748  0.75726560  0.50064173  1.0\\   Mn  Mn2  1  0.74103641  0.24668739  0.49995412  1.0\\   Ge  Ge3  1  0.37355053  0.38990140  0.76504097  1.0\\   Ge  Ge4  1  0.61217791  0.61415549  0.23546598  1.0\end{tabular} \\ \hline
\end{tabular}
\end{table}

\begin{table}[]
\centering
\begin{tabular}{|l|l|}
\hline
\textbf{Ground Truth Structure of Nb$_3$Si}                                                                                & \textbf{Nb$_3$Si predicted by ParetoCSP}                                                                                                                                             \\ \hline
\begin{tabular}[c]{@{}l@{}}data\_Nb3Si\\ \_symmetry\_space\_group\_name\_H-M   'P 1'\\ \_cell\_length\_a   4.08386900\\ \_cell\_length\_b   4.08386900\\ \_cell\_length\_c   4.08386900\\ \_cell\_angle\_alpha   90.00000000\\ \_cell\_angle\_beta   90.00000000\\ \_cell\_angle\_gamma   90.00000000\\ \_symmetry\_Int\_Tables\_number   1\\ \_chemical\_formula\_structural   Nb3Si\\ \_chemical\_formula\_sum   'Nb3 Si1'\\ \_cell\_volume   68.11071005\\ \_cell\_formula\_units\_Z   1\\ loop\_\\  \_symmetry\_equiv\_pos\_site\_id\\  \_symmetry\_equiv\_pos\_as\_xyz\\   1  'x, y, z'\\ loop\_\\  \_atom\_site\_type\_symbol\\  \_atom\_site\_label\\  \_atom\_site\_symmetry\_multiplicity\\  \_atom\_site\_fract\_x\\  \_atom\_site\_fract\_y\\  \_atom\_site\_fract\_z\\  \_atom\_site\_occupancy\\   Nb  Nb0  1  0.00000000  0.50000000  0.50000000  1.0\\   Nb  Nb1  1  0.50000000  0.50000000  0.00000000  1.0\\   Nb  Nb2  1  0.50000000  0.00000000  0.50000000  1.0\\   Si  Si3  1  0.00000000  0.00000000  0.00000000  1.0\end{tabular} & \begin{tabular}[c]{@{}l@{}}data\_Nb3Si\\ \_symmetry\_space\_group\_name\_H-M   'P 1'\\ \_cell\_length\_a   4.43420316\\ \_cell\_length\_b   4.43420316\\ \_cell\_length\_c   3.50747317\\ \_cell\_angle\_alpha   90.00000000\\ \_cell\_angle\_beta   90.00000000\\ \_cell\_angle\_gamma   90.00000000\\ \_symmetry\_Int\_Tables\_number   1\\ \_chemical\_formula\_structural   Nb3Si\\ \_chemical\_formula\_sum   'Nb3 Si1'\\ \_cell\_volume   68.96449047\\ \_cell\_formula\_units\_Z   1\\ loop\_\\  \_symmetry\_equiv\_pos\_site\_id\\  \_symmetry\_equiv\_pos\_as\_xyz\\   1  'x, y, z'\\ loop\_\\  \_atom\_site\_type\_symbol\\  \_atom\_site\_label\\  \_atom\_site\_symmetry\_multiplicity\\  \_atom\_site\_fract\_x\\  \_atom\_site\_fract\_y\\  \_atom\_site\_fract\_z\\  \_atom\_site\_occupancy\\   Nb  Nb0  1  0.00000000  0.50000000  0.00000000  1.0\\   Nb  Nb0  1  0.50000000  0.00000000  0.00000000  1.0\\   Nb  Nb1  1  0.00000000  0.00000000  0.50000000  1.0\\   Si  Si2  1  0.50000000  0.50000000  0.50000000  1.0\end{tabular} \\ \hline
\end{tabular}
\end{table}

\FloatBarrier

\begin{figure}[!htb]
  \centering
  \includegraphics[width=1\linewidth]{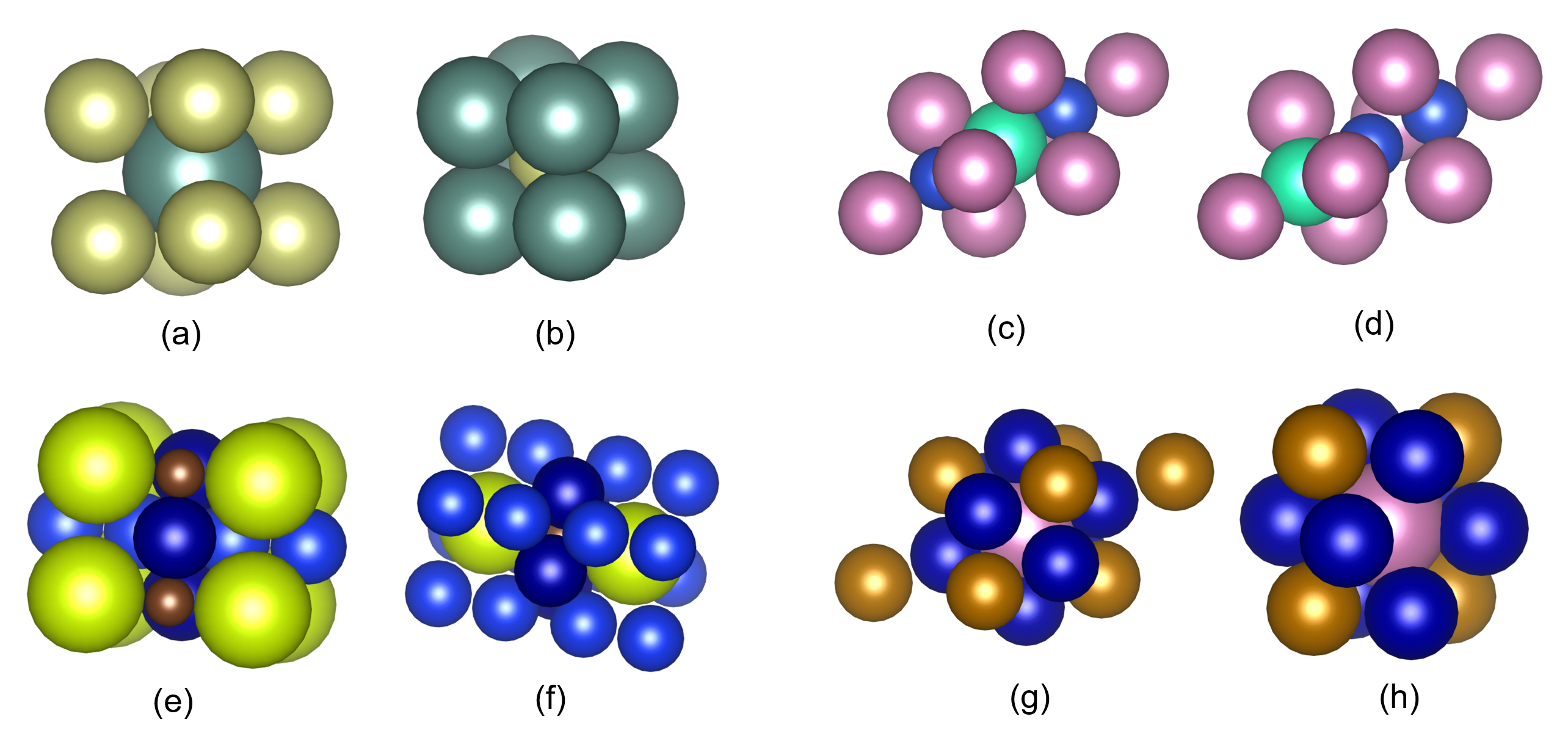}
  \caption{Comparison of the ground truth and predicted crystal structures based on Fingerprint distance, showing similar materials found in the MP database. (a). Ground truth structure of YIr. (b). YIr predicted by CSPML. (c). Ground truth structure of LuInCu$_2$. (d). LuInCu$_2$ predicted by CSPML. (e). Ground truth structure of CeCr$_2$Si$_2$C. (f). CeCr$_2$Si$_2$C predicted by ParetoCSP. (g).Ground truth structure of InFeCo$_2$. (h). InFeCo$_2$ predicted by ParetoCSP.}
  \label{fig: fingerprint_case}
\end{figure}

\begin{table}[]
\centering
\caption{Predicted structures and MP structures reported by Fingerprint distance as similar, but have large geometric distances and different space group numbers. GT represents ground truth structure, SG \# represents space gourp number.}
\centering
\begin{tabular}{|
>{\columncolor[HTML]{FFFFFF}}l |
>{\columncolor[HTML]{FFFFFF}}l |
>{\columncolor[HTML]{FFFFFF}}l |
>{\columncolor[HTML]{FFFFFF}}l |
>{\columncolor[HTML]{FFFFFF}}l |
>{\columncolor[HTML]{FFFFFF}}l |
>{\columncolor[HTML]{FFFFFF}}l |
>{\columncolor[HTML]{FFFFFF}}l |
>{\columncolor[HTML]{FFFFFF}}l |}
\hline
\textbf{Algorithm} & \textbf{Formula} & \textbf{\begin{tabular}[c]{@{}l@{}}Sinkhorn \\ Distance\end{tabular}} & \textbf{\begin{tabular}[c]{@{}l@{}}Chamfer \\ Distance\end{tabular}} & \textbf{\begin{tabular}[c]{@{}l@{}}Hausdorff \\ Distance\end{tabular}} & \textbf{\begin{tabular}[c]{@{}l@{}}Superpose \\ RMSD\end{tabular}} & \textbf{\begin{tabular}[c]{@{}l@{}}Fingerprint\\ Distance\end{tabular}} & \textbf{\begin{tabular}[c]{@{}l@{}}GT \\ SG \#\end{tabular}} & \textbf{\begin{tabular}[c]{@{}l@{}}Predicted\\  SG\#\end{tabular}} \\ \hline
CSPML              & YIr              & 6.127                                                                 & 2.966                                                                & 2.972                                                                  & 1.479                                                              & 0.060                                                                   & 123                                                          & 221                                                                \\ \hline
CSPML              & LuInCu$_2$          & 4.693                                                                 & 1.608                                                                & 3.000                                                                  & 1.195                                                              & 0.000                                                                   & 216                                                          & 225                                                                \\ \hline
ParetoCSP          & CeCr$_2$Si$_2$C        & 5.327                                                                 & 1.761                                                                & 1.436                                                                  & 0.891                                                              & 0.110                                                                   & 99                                                           & 123                                                                \\ \hline
ParetoCSP          & InFeCo$_2$          & 23.920                                                                & 7.278                                                                & 8.725                                                                  & 1.699                                                              & 0.007                                                                   & 123                                                          & 225                                                                \\ \hline
\end{tabular}

\end{table}

\FloatBarrier

\bibliographystyle{ieeetr}
\bibliography{references}